\documentclass[11pt,draftcls,onecolumn]{IEEEtran}

\usepackage{epsfig}
\usepackage{amsmath, amssymb, tabularx}
\usepackage{graphicx}

\newcommand{\tagarray}{\mbox{}\refstepcounter{equation}$(\theequation)$}
\newcommand{\beq}{\begin{equation}}
\newcommand{\eeq}{\end{equation}}
\def\bt{\mbox{\boldmath $t$}}
\def\bv{\mbox{\boldmath $v$}}
\def\bx{\mbox{\boldmath $x$}}
\def\bz{\mbox{\boldmath $z$}}
\def\bg{\mbox{\boldmath $g$}}

\def\bw{\mbox{\boldmath $w$}}
\def\by{\mbox{\boldmath $y$}}
\def\bL{\mbox{\boldmath $L$}}
\def\bC{\mbox{\boldmath $C$}}

\def\bD{\mbox{\boldmath $D$}}

\def\bgamma{\mbox{\boldmath $\gamma$}}

\def\bA{\mbox{\boldmath $A$}}

\def\bI{\mbox{\boldmath $I$}}

\def\bW{\mbox{\boldmath $W$}}

\def\bB{\mbox{\boldmath $B$}}

\def\bdelta{\mbox{\boldmath $\delta$}}

\newcommand{\ds}{\displaystyle}

\newcommand{\E}{{\cal E}}

\def\bA{\mbox{\boldmath $A$}}
\def\bB{\mbox{\boldmath $B$}}
\def\bC{\mbox{\boldmath $C$}}
\def\bD{\mbox{\boldmath $D$}}

\def\bI{\mbox{\boldmath $I$}}

\def\bL{\mbox{\boldmath $L$}}

\def\bT{\mbox{\boldmath $T$}}
\def\bU{\mbox{\boldmath $U$}}

\def\bW{\mbox{\boldmath $W$}}

\def\b0{\mbox{\boldmath $0$}}
\def\ba{\mbox{\boldmath $a$}}

\def\bd{\mbox{\boldmath $d$}}
\def\bee{\mbox{\boldmath $e$}}

\def\bg{\mbox{\boldmath $g$}}

\def\bp{\mbox{\boldmath $p$}}
\def\bq{\mbox{\boldmath $q$}}

\def\bs{\mbox{\boldmath $s$}}
\def\bu{\mbox{\boldmath $u$}}
\def\bw{\mbox{\boldmath $w$}}
\def\bx{\mbox{\boldmath $x$}}
\def\by{\mbox{\boldmath $y$}}
\def\bz{\mbox{\boldmath $z$}}
\def\b1{\mbox{\boldmath $1$}}
\def\buno{\mbox{\boldmath $1$}}
\def\buno{\mbox{\boldmath $1$}}
\def\bzero{\mbox{\boldmath $0$}}
\def\bmu{\mbox{\boldmath $\mu$}}

\def\blambda{\mbox{\boldmath $\lambda$}}

\def\btheta{\mbox{\boldmath $\theta$}}

\def\bv{\mbox{\boldmath $v$}}

\def\tr{\, \mbox{trace} }

\newtheorem{theorem}{Theorem}
\newtheorem{proposition}{Proposition}

\newcommand{\qedsymbol}{\hspace{\fill}\rule{1.5ex}{1.5ex}}

\title{Distributed detection and estimation \\in wireless sensor networks}

\author{Sergio~Barbarossa,~\IEEEmembership{Fellow,~IEEE}, Stefania Sardellitti,~\IEEEmembership{Member,~IEEE},\\ and Paolo~Di Lorenzo,~\IEEEmembership{Member,~IEEE}\\
\vspace{.3cm}
Department of Information, Electronics, and Telecommunications \\ ``Sapienza'' University of Rome, Via Eudossiana 18, 00184 Rome, Italy.\\ e-mail: {\tt sergio@infocom.uniroma1.it}, {\tt Stefania.Sardellitti@uniroma1.it}, {\tt dilorenzo@infocom.uniroma1.it}
\thanks{This work has been supported by TROPIC Project, Nr. 318784. To appear in E-Reference Signal Processing, R. Chellapa and S. Theodoridis, Eds., Elsevier, 2013.}}

\begin{document}

\maketitle

\vspace{-1.5cm}


\section{Introduction}
\label{Introduction}
Wireless sensor networks (WSN) are receiving a lot of attention from both the theoretical and
application sides, in view of the many applications spanning from environmental monitoring,
as a tool to control physical parameters such as temperature, vibration, pressure, or pollutant
concentration, to the monitoring of civil infrastructures, such as roads, bridges, buildings, etc. \cite{Akyildiz}.
Some new areas of applications are emerging rapidly and have great potentials.
A field that is gaining more and more interest is the use of WSN's as a support for
{\it smart grids}. In such a case, a WSN is useful to: i) monitor and predict energy production
from renewable sources of energy such as wind or solar energy, ii)
monitor energy consumption; iii) detect anomalies in the network.
A further area of increasing interest is {\it vehicular sensor networks}. In such a case,
the vehicles are nodes of an ad hoc network. The sensors onboard the vehicle can measure
speed and position of the vehicle and forward this information to nearby vehicles
or to the road side units (RSU). This information enables the construction
of dynamic spatial traffic maps, which can be exploited to reroute traffic in case
of accidents or to minimize energy consumption.
A relatively recent and interesting application of WSNs is {\it cognitive radio (CR)}.
In such a case, opportunistic (or secondary) users are allowed to access temporally unoccupied
spectrum holes, under the constraint of not interfering with licensed (primary) users,
and to release the channels as soon as they are requested by licensed users. The basic step
enabling this dynamic access is sensing. The problem is that if sensing is carried out at
a single location, it might be severely degraded by shadowing phenomena:
If the sensor is in shadowed area, it might miss the presence of a primary user and
then transmit by mistake over occupied slots, thus generating an undue interference. To overcome
shadowing, it is useful to resort to a WSN whose nodes sense the channels and exchange information
with each other in order to mitigate the effect of local shadowing phenomena. The goal of
the WSN in such an application is to build a spatial {\it map} of
channel occupancy. An opportunistic user willing to access radio resources
within a confined region could then interrogate the closest sensor of a WSN and get
a reliable information about which channels are temporarily available and when this
utilization has to be stopped.\\
The plethora of applications raises a series of challenging technical issues, which
may be seen as sources of opportunities for engineers. Probably the first most important question
concerns {\it energy supply}. In many applications, in fact, the sensors are battery-operated
and it may be difficult or costly to recharge the batteries or to substitute them.
As a consequence, energy consumption is a basic constraint that should be properly
taken into account.
A second major concern is {\it reliability} of the whole system. In many cases, to allow for an economy of
scale, the single sensors are devices with limited accuracy and computational capabilities.
Nevertheless, the decision taken by the network as a whole must be very reliable, because
it might affect crucial issues like security, safety, etc. The question is then how to build
a reliable system out of the combination of many potentially unreliable nodes. Nature exhibits many
examples of such systems. Human beings are capable of solving very sophisticated tasks
and yet they are essentially built around basic unreliable chemical reactions occurring
within cells whose lifetime is typically much smaller than the lifetime of a human being.
Clearly, engineering is still far away from approaching the skills of living systems, but
important inspirations can be gained by observing biological systems. Two particular features
possessed by biological systems are self-organization and self-healing capabilities.
Introducing these capabilities within a sensor network is the way to
tackle the problem of building a reliable system out of the cooperation of
many potentially unreliable units. In particular, self-organization is a key tool
to enable the network to reconfigure itself, in terms of acquisition and transfer of
information from the sensing nodes to the control centers, responsible for taking decisions,
launching alarms or activating actuators aimed to counteract adverse phenomena.
The network architecture plays a fundamental role in terms of reliability of the whole system.
In conventional WSNs, there is typically one or a few sink nodes that collect the observations taken by the sensor nodes and process them, in a centralized fashion, to produce the desired decision about the observed phenomenon. This architecture arises a number of critical issues, such as: a) potential congestion around the sink nodes; b) vulnerability of the whole network to attacks or failure of sink nodes; c) efficiency of the communication links established to send data from the sensor nodes to the sink.
For all these reasons, a desirable characteristic of a WSN is to be designed
in such a way that decisions are taken in a decentralized manner. Ideally, every node should
be able, in principle, to achieve the final decision, thanks to the exchange of information with
the other nodes, either directly or through multiple hops. In this way, vulnerability would be
strongly reduced and the system would satisfy a scalability property. In practice, it is not
necessary to make every single node to be able to take decisions as reliably as in a
centralized system. But what is important to emphasize is that proper interaction
among the nodes may help to improve reliability of single nodes, reduce vulnerability
and congestion events, and make a better usage of radio resource capabilities. This last
issue points indeed to one of the distinctive features of decentralized decision systems, namely
the fact that {\it sensing and communicating are strictly intertwined} with each
other and a proper system design must consider them jointly.
The first important constraint inducing a strict link between sensing and communicating is
that the transmission of the measurements collected by the nodes to the decision points
occurs over realistic channels, utilizing standard communication protocols.
For example, adopting common digital communication systems, the data
gathered by the sensors need to be quantized and encoded before transmission.
In principle, the number of bits used in each sensor should depend on the
accuracy of the data acquisition on that sensor. At the same time, the number of bits
transmitted per each channel use is upper bounded by the channel capacity, which
depends on the transmit power and on the channel between sensor and sink node.
This suggests that the number of bits to be used in each node for data quantization
should be made dependent on both sensor accuracy and transmission channel.
A further important consequence of the network architecture and of the resulting
flow of information from peripheral sensing nodes to central decision nodes is
the latency with which a global decision can be taken. In a centralized decision system, the
flow of information proceeds from the sensing nodes to the central control nodes, usually
through multiple hops. The control node collects all the data, it carries out the computations,
and takes a decision.
Conversely, in a decentralized decision system, there is typically an iterated
exchange of data among the nodes. This determines an increase of the time
necessary to reach a decision. Furthermore, an iterated exchange of data implies
an iterated energy consumption. Since in WSN's
energy consumption is a fundamental concern, all the means to minimize
the overall energy consumption necessary to reach a decision within a maximum latency
are welcome. At a very fundamental level, we will see how an efficient design of the
network requires a global cross layer design where the physical and the routing layers
take explicitly into account the specific application for which the network has been
built.\\
This article is organized as follows. In Section \ref{General framework}, we provide a
general framework aimed to show how an efficient design of a sensor network requires
a joint organization of in-network processing and communication. We show
how the organization of the flow of information from the sensing nodes to the
decision centers should depend not only on the WSN topology, but also on the statistical
model of the observation. Finally, we briefly recall some fundamental information
theoretical issues showing how in a multi-terminal decision
network source and channel coding are strictly related to each other.
In Section \ref{Consensus algorithms for distributed detection and estimation} we
introduce the graph model as the formal tool to describe the interaction among the nodes.
Then, we illustrate the so called consensus algorithm as a basic tool to reach
globally optimal decisions through a decentralized approach. Since the interaction
among the nodes occurs through a wireless channel, we also consider
the impact of realistic channel models on consensus algorithm and show how consensus
algorithms can be made robust against channel impairments.
In Section \ref{Distributed estimation} we address the distributed estimation problem.
We show first an entirely decentralized approach, where observations and estimations
are performed without the intervention of a fusion center. In such a case, we show
how to achieve a globally optimal estimation through the local exchange of information
among nearby nodes. Then, we consider the case where the estimation is
performed at a decision center. In such a case, we show how to allocate quantization bits
and transmit powers in the links between the sensing nodes and the fusion center,
in order to accommodate the requirement on the maximum estimation variance,
under a constraint on the global transmit power.
In Section \ref{Decentralized detection} we extend the approach to the detection problem.
Also in this case, we consider the entirely distributed approach, where every node
is enabled to achieve a globally optimal decision, and the case where the decision is
taken at a central control node.
In such a case, we show how to allocate coding bits and transmit power in order to
maximize the detection probability, under constraints on the false alarm rate and the global transmit power.
Then, in Section \ref{Beyond consensus: Distributed projected algorithms}, we generalize
consensus algorithms illustrating a distributed procedure that does not force all the nodes
to reach a common value, as in consensus algorithms, but rather to converge to
the projection of the overall observation vector onto a signal subspace. This algorithm
is especially useful, for example, when it is required to smooth out the effect of noise,
but without destroying valuable information present in the spatial variation
of the useful signal.
In wireless sensor networks, a special concern is energy consumption.
We address this issue in Section \ref{Minimum energy consensus},
where we show how to optimize the network topology in order to minimize the energy necessary
to achieve a global consensus. We show how to convert this, in principle,
combinatorial problem, into a convex problem with minimal performance losses.
Finally, in Section \ref{Matching communication network topology to statistical dependency graph}
we address the problem of matching the topology of the observation
network to the graph describing the statistical dependencies among the observed variables.
Finally, in Section \ref{Conclusions} we draw some conclusions and we try to highlight
some open problems and possible future developments.

\section{General framework}
\label{General framework}
The distinguishing feature of a decentralized detection or estimation system is that
the measurements are gathered by a multiplicity of sensors dispersed over space, while
the decision about what is being sensed is taken at one or a few
fusion centers or sink nodes. The information gathered by the sensors has then
to propagate from the peripheral nodes to the central control nodes.
The challenge coming from this set-up is that in a WSN, information propagates through
wireless channels, which are inherently broadcast, affected by fading and prone to interference.
Installing a WSN requires then to set up a proper medium access control protocol (MAC) able to handle
the communications among the nodes, in order to avoid interference and to ensure that
the information reaches the final destination in a reliable manner. But what is decidedly
specific of a WSN is that the sensing and communication aspects are strictly related
to each other.
In designing the MAC of a WSN, there are some fundamental aspects that distinguish a WSN
from a typical telecommunication (TLC) network.
The main difference stems from the analysis of goal and constraints of these two kinds of networks.
A TLC network must make sure that every source packet reaches the final destination,
perhaps through retransmission in case of errors or packet drop, irrespective of the
packet content. In a WSN, what is really important is that the decision about what is being sensed
be taken in the most reliable way, without necessarily implying the successful delivery
of all source packets.
Moreover, one of the major constraints in WSN's is energy consumption,
because the nodes are typically battery operated and recharging the batteries is sometimes
troublesome, especially when the nodes are installed in hard to reach places. Conversely,
in a TLC network, energy provision is of course important, but it is not
the central issue. At the same time, the trend in TLC networks is to support
higher and higher data rates to accommodate for ever more demanding applications,
while the data rates typically required in most WSN's are not so high.
These considerations suggest that an efficient design of a WSN should take into account
the application layer directly. This means, for example, that it is not really necessary that
every packet sent by a sensor node reaches the final destination. What is important is
only that the correct decision is taken in a reliable manner, possibly with low latency and
low energy consumption. This enables data aggregation or in-network processing
to avoid unnecessary data transmissions. It is then important to formulate this change
of perspective in a formal way to envisage ad hoc information transmission and processing
techniques.

\subsection{Computing while communicating}
\begin{figure}[t]
\centering
\includegraphics[scale=0.6]{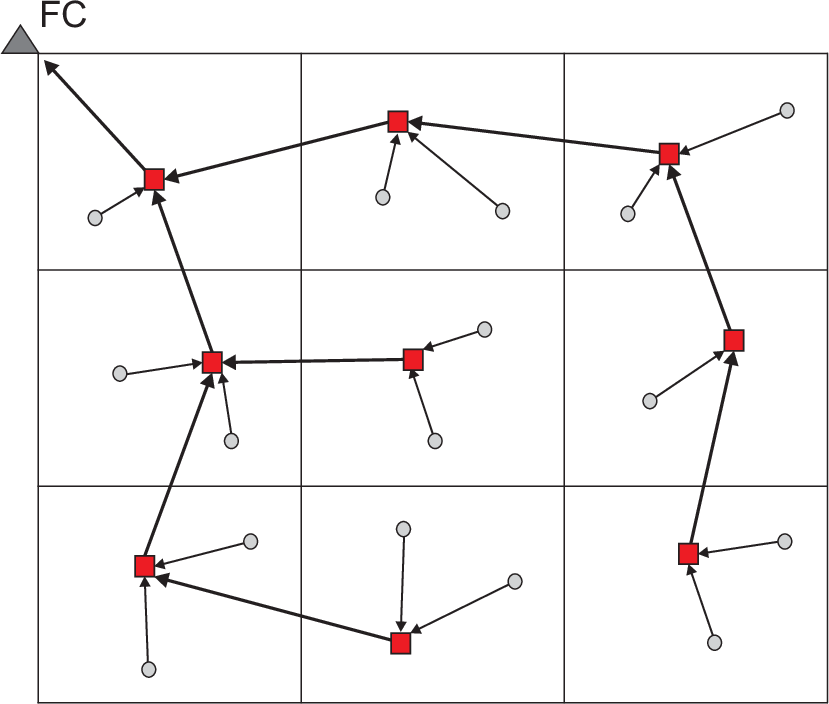}
\caption{Hierarchical organization of information flow from peripheral nodes
to fusion center.}\label{tessellation}
\end{figure}

In a very general setting, taking a decision based on the data collected by the sensors
can be interpreted as computing a function of these data.
Let us denote by $x_i$, with $i=1, \ldots, N$, the measurements collected by the $i$-th node of the network,
and by $f(\bx)=f(x_1, \ldots, x_N)$ the function to be computed. The straightforward
approach for computing this function consists in sending all the measurements $x_i$
to a fusion center through a proper communication network and then implement
the computation of $f(\bx)$ at the fusion center. However, if $f(\bx)$ possesses
a structure, it may be possible to take advantage of such a structure
to better organize the flow of data from the sensing nodes to the fusion center.
The idea of mingling computations and communications to make an efficient use
of the radio resources, depending on the properties that the function $f(\bx)$ might possess,
was proposed in \cite{Giridhar-Kumar}. Here, we will first recall the main results of
\cite{Giridhar-Kumar}. Then, we will show how the interplay between computation and
communication will be further affected by the structure of the probabilistic model underlying the
observations.\\
To exploit the structure of the function $f(x_1, \ldots, x_N)$ to be computed, it is necessary to
define some relevant structural properties. One important property
is {\it divisibility}. Let ${\cal C}$ be a subset of $\{ 1, 2, \ldots, N\}$ and let
$\pi:=\{C_1, \ldots, C_s\}$ be a partition of ${\cal C}$. We denote by $\bx_{C_i}$
the vector composed by the set of measurements collected by the nodes
whose indices belong to $C_i$. A function $f(x_1, \ldots, x_N)$ is said to be divisible if,
for any ${\cal C}\subset \{ 1, 2, \ldots, N\}$
and any partition $\pi$, there exists a function $g^{(\pi)}$
such that
\begin{equation}
\label{divisibility}
f(\bx_{C})=g^{(\pi)}\left(f(\bx_{C_1}), f(\bx_{C_2}), \ldots, f(\bx_{C_s})\right).
\end{equation}
In words, (\ref{divisibility}) represents a sort of
``divide and conquer'' property: A function $f(\bx)$ is divisible if it is
possible to split its computation into partial computations
over subsets of data and then recombine the partial results to
yield the desired outcome.\\
Let us suppose now that the $N$ sensing nodes are randomly distributed
over a circle of radius $R$. We assume a simple propagation
model, such that two nodes are able to send information to each other
in a reliable way if their distance is less than a coverage radius $r_0(N)$.
At the same time, the interference between two links is considered
negligible if the interfering transmitter is at a distance greater than
$\alpha r_0(N)$ from the receiver, where $\alpha$ is chosen according to the
propagation model. For any random deployment of the nodes, the choice of $r_0(N)$
induces a network topology, such that there is a link
between two nodes if their distance is less than $r_0(N)$. The resulting
graph having the nodes as vertices and the edges as links, is a random graph,
because the positions of the nodes are random. This kind of graph
is known as a {\it Random Geometric Graph} (RGG)\footnote{A basic
review of graph properties is reported in Appendix A.}.
To make an efficient use of the radio resources, it is useful to take
$r_0(N)$ as small as possible, to save local transmit power and make
possible the reuse of radio resources, either frequency or time slots.
However, $r_0(N)$ should not be too small to loose connectivity.
In other words, we do not want the network to split in subnetworks
that do not interact with each other. Since the node location is random,
network connectivity can only be guaranteed in probability.
It has been proved in \cite{Gupta-Kumar98} that, if $r_0(N)$ is chosen as follows
\begin{equation}
\label{theor_r0(N)}
r_0(N)=R\sqrt{\frac{\log N+c(N)}{\pi N}}
\end{equation}
with $c(N)$ going to infinity, as $N$ goes to infinity,
the resulting RGG is asymptotically connected with
high probability, as $N$ goes to infinity.
For instance, if we take $c(N)=(\pi-1)\log N$, the coverage radius can be
expressed simply as
\begin{equation}
r_0(N)=R\sqrt{\frac{\log N}{N}}.
\label{cover_radius}
\end{equation}
A further property of a node is the number of neighbors
of that node. For an undirected graph, the number of neighbors of a node is
known as the degree of the node.
Denoting by $d(N)$ the degree of an RGG with $N$
nodes, it was proved in \cite{Giridhar-Kumar-05} that, choosing the
coverage radius as in (\ref{theor_r0(N)}), $d(N)$ is (asymptotically)
upper bounded by a function that behaves as $\log N$.
More specifically,
\begin{equation}
\lim_{N\rightarrow \infty}\mathbb{P}\left\{d(N)\le c \log N\right\}=1
\end{equation}
In \cite{Giridhar-Kumar-05} it was established an interesting link
between the properties of the function $f(\bx)$ to be computed by the network
and the topology of the communication network. In particular, assuming as usual
that the measurements are quantized in order to produce a value belonging to a
finite alphabet, let us denote by ${\cal R}(f, N)$ the range of $f(\bx)$ and by $|{\cal R}(f, N)|$
the cardinality of ${\cal R}(f, N)$. In \cite{Giridhar-Kumar-05}, it was proved
that, under the following assumptions:\\
\noindent{\bf A.1} $f(\bx)$ is divisible;\\
\noindent{\bf A.2} the network is connected;\\
\noindent{\bf A.3} the degree of each node is chosen as $d(N)\le k_1 \log|{\cal R}(f, N)|$; \\
\noindent then, the rate for computing $f(\bx)$ scales with $N$ as
\begin{equation}\label{rate-scaling-law}
R(N)\ge\frac{c_1}{\log|{\cal R}(f, N)|}.
\end{equation}
This is an important result that has practical consequences. It states, in fact,
that, whenever $\log|{\cal R}(f, N)|$ scales with a law that increases more
slowly than $N$, we can have an increase of efficiency if we organize
the local computation and the flow of partial results properly.
For instance, if the sensors communicate to the sink node through
a Time Division Multiplexing Access (TDMA) scheme, with a standard approach
it is necessary to allocate $N$ time slots to send all the data to the sink node.
Conversely, Eqn. (\ref{rate-scaling-law}) suggests that, to compute the function
$f(\bx)$, it is sufficient to allocate $\log|{\cal R}(f, N)|/c_1$ slots.
The same result would apply in a Frequency Division Multiplexing Access (FDMA) scheme,
simply reverting the role of time slots and frequency subchannels. This is indeed
a paradigm shift, because it suggests that an efficient radio resource allocation in
a WSN should depend on the cardinality of ${\cal R}(f, N)$. This implies a sort of cross-layer
approach that involves physical, MAC and application layers jointly.
The next question is how to devise an access protocol that enables such an efficient design.
To this regard, the theorem proved in \cite{Giridhar-Kumar-05} contains
a constructive proof, which suggests how to organize the flow of information from
the sensing nodes to the control center. In particular, the strategy consists
in making a tessellation of the area monitored by the sensor network, similarly
to a cellular network, as pictorially described in Fig. \ref{tessellation}.
Furthermore, the information flows from the peripheral nodes to the fusion center
through a tree-like graph, having the fusion center as its root.
In each cell, the nodes (circles) identify a node as the relay node (square).
The relay node collects data from the nodes within its own cell and from relay nodes
of its leaves, performs local computations and communicates the result to the parent
relay nodes, with the goal of propagating these partial results towards the root (sink node).
To handle interference, a graph coloring scheme is used to avoid interference
among adjacent cells. This allows spatial reuse of radio resources, e.g. frequency
or time slots, which can be used in parallel
without generating an appreciable interference.
The communication structure is conceptually similar to a cellular network,
with the important difference that now the flow of information is directly
related to the computational task.
A few examples are useful to better grasp the possibilities of this approach.\\
\noindent{\it Data uploading}: Suppose it is necessary to convey all the data to the sink node.
If each observed vector belongs to an alphabet ${\cal X}$, with cardinality $|{\cal X}|$,
the cardinality of the whole data set is $|{\cal R}(f, N)|=|{\cal X}|^N$. Hence,
$\log |{\cal R}(f, N)|=N\, \log |{\cal X}|$. This means that, according to (\ref{rate-scaling-law}),
the capacity of the network scales as $1/N$.
This is a rather disappointing result, as it shows that there is no real benefit with respect to the
simplest communication case one could envisage: The nodes have to split
the available bandwidth into a number of sub-bands equal to the number of nodes,
with a consequent rate reduction per node.\\
\noindent{\it Decision based on the histogram of the measurements}: Let us suppose now that
the decision to be taken at the control node can be based on the histogram of the
data collected by the nodes, with no information loss. In this case, the function $f(\bx)$ is the histogram.
It can be verified that the histogram is a divisible function. Furthermore, the cardinality
of the histogram is
\begin{equation}
|{\cal R}(f, N)|=
\left(
\begin{array}{c}
N+|{\cal X}|-1\\
|{\cal X}|-1
\end{array}
\right).
\end{equation}
Furthermore, it can be shown that
\begin{equation}
\left(N/|{\cal X}|\right)^{|{\cal X}|}\le
\left(
\begin{array}{c}
N+|{\cal X}|-1\\
|{\cal X}|-1
\end{array}
\right)\le \left(N+1\right)^{|{\cal X}|}.
\end{equation}
Hence, in this case $\log |{\cal R}(f, N)|$ behaves as $\log N$ and then
the rate $R(N)$ in (\ref{rate-scaling-law}) scales as $1/\log N$.
This is indeed an interesting result, showing that if the decision
can be based on the histogram of the data, rather than on each
single measurement, adopting the right communication scheme,
the rate per node behaves as $1/\log N$, rather than $1/N$,
with a rate gain $N/\log N$, which increases as the number of nodes increases.\\
\noindent{\it Symmetric functions}: Let us consider now the case where
$f(\bx)$ is a symmetric function. We recall that a function $f(\bx)$ is
symmetric if it is invariant to permutations of its arguments, i.e., $f(\bx)
= f(\Pi \bx)$ for any permutation matrix $\Pi$ and any argument vector $\bx$.
This property reflects the so called {\it data-centric}
view, where what it important is the measurement {\it per se}, and not which node has
taken which measurement. Examples of symmetric functions include the mean,
median, maximum/minimum, histogram, and so on.
The key property of symmetric functions is that it can be shown that they depend on
the argument $\bx$ only through the histogram of $\bx$. Hence,
the computation of symmetric functions is a particular case
of the example examined before. Thus, the rate scales again as $1/\log N$. \\

\subsection{Impact of observation model}
\label{Observation model}
Having recalled that the efficient design of a WSN requires an information flow
that depends on the scope of the network, more specifically, on the structural
properties of the function to be computed by the network, it is now time to
be more specific on the decision tasks that are typical of WSN's, namely
detection and estimation.
Let us consider for example the simple hypothesis testing problem.
In such a case, an ideal centralized detector having
error-free access to the measurements collected by the
nodes, should compute the likelihood ratio and compare it with a
suitable threshold \cite{SB-kay-II}. We denote with ${\cal H}_0$ and
${\cal H}_1$ the two alternative hypotheses, i.e. absence or
presence of the event of interest, and with $\bx_i$ the set of
measurements collected by node $i$. If we indicate with
$p(\bx_1, \ldots, \bx_N; {\cal H}_i)$ the joint probability density function of the
whole set of observed data, under the hypothesis ${\cal H}_i$, with $i= 0, 1,$,
the likelihood ratio test amounts to comparing the likelihood
ratio (LR) with a threshold $\gamma$, and decide for ${\cal H}_1$
if the threshold is exceeded or for ${\cal H}_0$, otherwise. In
formulas
\begin{equation}
\label{LR}
\Lambda\left(\bx\right):=\Lambda\left(\bx_1, \ldots, \bx_N\right)
=\frac{p(\bx_1, \ldots, \bx_N; {\cal H}_1)}{p(\bx_1, \ldots, \bx_N; {\cal H}_0)}
\overset{\mathcal{H}_{1}}{\underset{\mathcal{H}_{0}}{\gtreqless}} \gamma
\end{equation}
The LR test (LRT) is optimal under a Bayes or a Neyman-Pearson
criterion, the only difference being that the threshold $\gamma$
assumes different values in the two cases \cite{SB-kay-II}.
In principle, to implement the LRT at the fusion center,
every node should send its observation vector $\bx_i$ to the fusion center,
through a proper MAC protocol. The fusion center, after having collected all the data,
should then implement the LRT, as indicated in (\ref{LR}). However, the computation of
the LR in (\ref{LR}) does not necessarily imply the transmission of the single
vectors $\bx_i$. Conversely, according to the theory recalled above, the transmission
strategy should depend on the structural properties of the LR function, if any.
Let us see how to exploit the structure of the LR function in two
cases of practical interest.\\

\subsubsection{Statistically independent observations}
Let us start assuming that the observations taken by different sensors are statistically
independent, conditioned to each hypothesis. This is an assumption valid in many cases.
Under such an assumption, the LR can be factorized as follows
\begin{equation}
\label{LR_cond_ind}
\Lambda\left(\bx\right):=\frac{\prod_{n=1}^{N}p(\bx_n; {\cal H}_1)}{\prod_{n=1}^{N}p(\bx_n; {\cal H}_0)}:=\prod_{n=1}^{N}\Lambda_n(\bx_n)\overset{\mathcal{H}_{1}}{\underset{\mathcal{H}_{0}}{\gtreqless}} \gamma
\end{equation}
where $\Lambda_n(\bx_n)=p(\bx_n; {\cal H}_1)/p(\bx_n; {\cal H}_0)$
denotes the local LR at the $n$th node. In this case, the global function $\Lambda\left(\bx\right)$ in (\ref{LR_cond_ind}) possesses a clear structure: It is factorizable in the product of
the local LR functions. Then, since a factorizable function is divisible, it is possible to implement
the efficient mechanisms described in the previous section to achieve an efficient design.
The network nodes should cluster as in Fig. \ref{tessellation}. Every relay node should
compute the local LR, multiply it to the data received from the relays pertaining to the lower clusters
and send the partial result to the relay of the upper cluster, until the result reaches the fusion center.
The efficiency comes from the fact that many transmissions can occur in parallel,
exploiting spatial reuse of radio resources.
This result suggests also that the proper source encoding to be implemented at
each sensor node consists in the computation of the local LR.\\

\subsubsection{Markov observations}
\begin{figure}[t]
\centering
\includegraphics[scale=0.6]{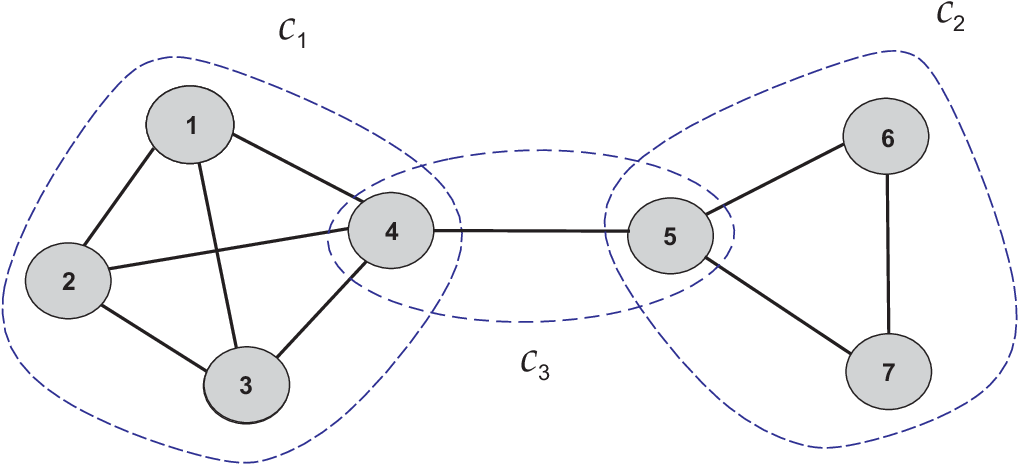}
\caption{Example of Markov graph.}\label{Markov-graph}
\end{figure}

\begin{figure}[t]
\centering
\includegraphics[scale=0.6]{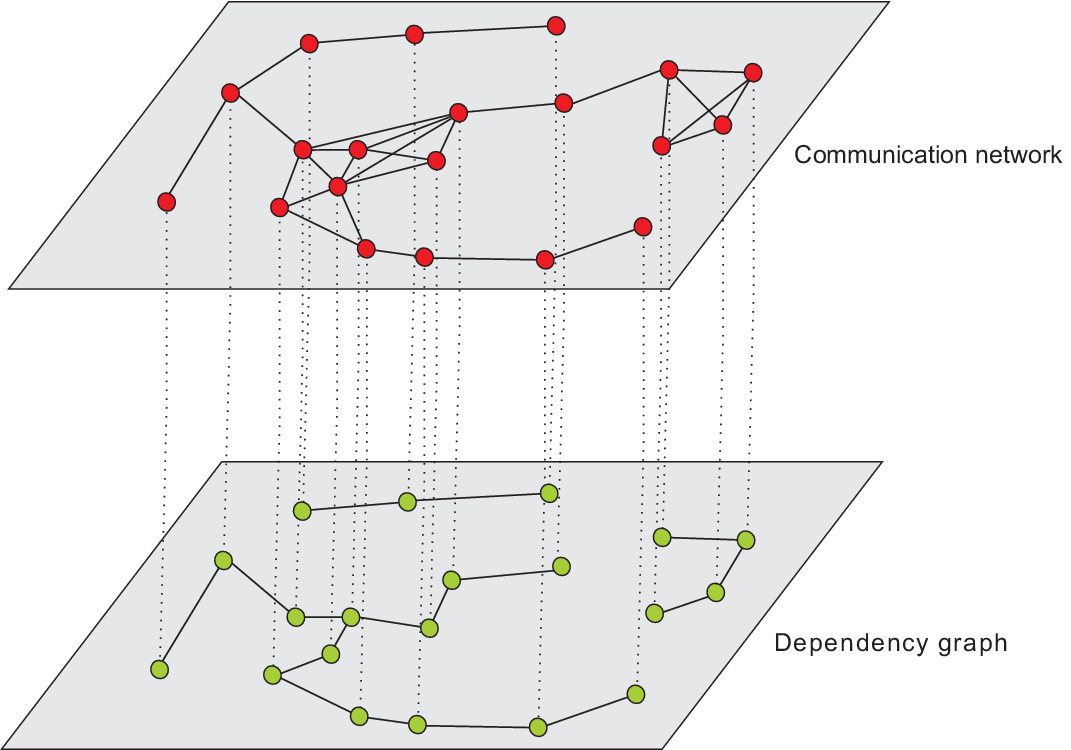}
\caption{Superposition of communication layer (top) over a Markov statistical dependency graph (bottom).}
\label{WSN_MRF}
\end{figure}

The previous result is appealing, but it pertains to the simple situation where the observations
are statistically independent, conditioned to the hypotheses. In some circumstances, however, this
assumption is unjustified. This is the case, for example, when the sensors monitor
a field of spatially correlated values, like a temperature or atmospheric pressure field.
In such cases, nearby nodes sense correlated values and then the statistical independence assumption is
no longer valid. It is then of interest, in such cases, to check whether the statistical properties
of the observations can still induce a structure on the function to be computed that can
be exploited to improve network efficiency.\\
There is indeed a broad class of observation models where the joint pdf cannot be factorized into
the product of the individual pdf's pertaining to each node, but it can still be factorized
into functions of subsets of variables. This is the case of Bayes networks or Markov random fields.
Here we will recall the basic properties of these models, as relevant to our problem. The interested reader
can refer to many excellent books, like, for example, \cite{Whittaker-book} or \cite{Lauritzen-book}.\\
In the Bayes network's case, the statistical dependency among the random variables is
described by an acyclic directed graph, whose vertices represent the random variables,
while the edges represent local conditional probabilities. In particular, given a node $x_i$,
whose parent nodes are identified by the set of indices ${\rm pa}(i)$, the joint probability density
function (pdf) of a Bayes network can be written as
\begin{equation}
\label{Bayes-network}
p(x_1, \ldots, x_N)=\prod_{i=1}^{N}p(x_i/\bx_{{\rm pa}(i)}),
\end{equation}
where $\bx_{{\rm pa}(i)}$ collects all the variables corresponding to the
parents of node $i$. If a node in (\ref{Bayes-network}) does not have parents,
the corresponding probability is unconditional.\\
Alternatively, a Markov random field is represented through an undirected graph.
More specifically, a Markov network consists of:
\begin{enumerate}
\item An undirected graph $G = (V,E)$, where each vertex $v \in V$ represents a
random variable and each edge $\{u,v\} \in E$ represents statistical dependency
between the random variables $u$ and $v$;
\item
A set of potential (or compatibility) functions $\psi_c(\bx_c)$ (also called clique potentials),
that associate a non-negative number to the cliques \footnote{A clique is a subset
of nodes which are fully connected and maximal, i.e. no additional node can be added
to the subset so that the subset remains fully connected.} of $G$.
\end{enumerate}
Let us denote by ${\cal C}$ the set of all cliques present in the graph. The random vector $\bx$ is
Markovian if its joint pdf admits the following factorization
\begin{equation}
\label{Markov_factor}
p(\bx)=\frac{1}{Z}\prod_{c\in {\cal C}}\psi_c(\bx_c),
\end{equation}
where $\bx_c$ denotes the vector of variables belonging
to the clique $c$. The functions $\psi_c(\bx_c)$ are called {\it compatibility functions}.
The term $Z$ is simply a normalization factor necessary to guarantee that
$p(\bx)$ is a valid pdf.
A node $p$ is conditionally independent of another node $q$ in the Markov network,
given some set $S$of nodes, if every path from $p$ to $q$ passes through a
node in $S$. Hence, representing a set of random variables
by drawing the correspondent Markov graph is a meaningful pictorial
way to identify the conditional dependencies occurring across the random variables.
As an example, let us consider the graph reported in Fig. \ref{Markov-graph}.
The graph represents conditional independencies among seven random variables.
The variables are grouped into $3$ cliques. In this case, for example, we can say that nodes $1$ to $4$ are
statistically independent of nodes $6$ and $7$, conditioned to the knowledge of node $5$.
In this example, the joint pdf can be written as follows
\begin{equation}
\label{Markov_factor_ex}
p(\bx)=\frac{1}{Z}\psi_1(x_1, x_2, x_3, x_4)\psi_2(x_5, x_6, x_7)\psi_3(x_4, x_5).
\end{equation}
If the product in (\ref{Markov_factor}) is strictly positive for any $\bx$, we can introduce the functions
\begin{equation}
V_c(\bx_c)=-\log \psi_c(\bx_c)
\end{equation}
so that (\ref{Markov_factor}) can be rewritten in exponential form as
\begin{equation}
\label{Markov_factor_exp}
p(\bx)=\frac{1}{Z}\exp\left(-\sum_{c\in {\cal C}}V_c(\bx_c)\right).
\end{equation}
This distribution is known, in physics, as the Gibbs (or Boltzman) distribution with interaction
{\it potentials} $V_c(\bx_c)$ and {\it energy} $\sum_{c\in {\cal C}}V_c(\bx_c)$.\\
The independence graph conveys the key probabilistic information through absent
edges: If nodes $i$ and $j$ are not neighbors, the random variables $x_i$ and $x_j$ are statistically independent, conditioned to the other variables. This is the so called {\it pairwise Markov property}.
Given a subset $a \subset V$ of vertices, $p({\bx})$ factorizes as
\begin{equation}
\label{Markov_factor_2}
p(\bx)=\frac{1}{Z}\prod_{c:c\cap a\neq\emptyset}\psi_c(\bx_c)\,\prod_{c:c\cap a = \emptyset}\psi_c(\bx_c)
\end{equation}
where the second factor does not depend on $a$. As a consequence, denoting by $S-a$ the set of all nodes except the nodes in $a$ and by ${\cal N}_a$ the set of neighbors of the nodes in $a$, $p(\bx_a/\bx_{S-a})$ reduces to $p(\bx_a/{\cal N}_a)$. Furthermore,
\begin{equation}
\label{Markov_factor_3}
p(\bx_a/{\cal N}_a)=\frac{1}{Z_a}\prod_{c:c\cap a\neq\emptyset}\psi_c(\bx_c)=\frac{1}{Z_a}\,
\exp\left(-\sum_{c:c\cap a\neq\emptyset}\,V_c(\bx_c)\right).
\end{equation}
This property states that the joint pdf factorizes in terms that contain only variables
whose vertices are neighbors.\\
An important example of jointly Markov random variables is the Gaussian Markov Random Field (GMRF),
characterized by having a pdf expressed as in (\ref{Markov_factor_exp}), with the additional property
that the energy function is a quadratic function of the variables. In particular, a
vector $\bx$ of random variables is a GMRF if its joint pdf can be written as
\begin{equation}
\label{gmrf}
p(\bx)=\frac{1}{\sqrt{(2\pi)^N |\bC|}}\,e^{-\frac{1}{2}(\bx-\bmu)^T\bC^{-1}(\bx-\bmu)}=\sqrt{\frac{|\bA|}{(2\pi)^N }}\,e^{-\frac{1}{2}(\bx-\bmu)^T\bA(\bx-\bmu)},
\end{equation}
where $\bmu=\mathbb{E}\{\bx\}$ is the expected value of $\bx$,
$\bC=\mathbb{E}\{(\bx-\bmu)(\bx-\bmu)^T\}$ is the
covariance matrix of $\bx$ and $\bA=\bC^{-1}$ is the so called {\it precision} matrix.
In this case, the {\it Markovianity} of $\bx$ manifests itself
through the {\it sparsity} of the precision matrix.
As a particular case of (\ref{Markov_factor_3}), the coefficient $a_{ij}$ of $\bA$
is different from zero if and only if nodes $i$ and $j$ are neighbors.\\
Having recalled the main properties of GMRF's, let us now go back to the problem
of organizing the flow of information in a WSN aimed at deciding between two
alternative hypotheses of GMRF. Let us consider for example the decision about
the two alternative hypotheses:
\begin{equation}
\label{mrf0}
\mathcal{H}_{0} : \bx \sim p(\bx; \mathcal{H}_{0})= \frac{1}{Z_0}\prod_{c\in {\cal C}}\psi_c(\bx_c; \mathcal{H}_{0})
\end{equation}
\begin{equation}
\label{mrf1}
\mathcal{H}_{1} : \bx \sim p(\bx; \mathcal{H}_{1})= \frac{1}{Z_1}\prod_{c'\in {\cal C'}}\psi_{c'}(\bx_{c'}; \mathcal{H}_{1})
\end{equation}
where the sets of cliques involved in the two cases are, in general, different.
The factorizations in (\ref{mrf0}, \ref{mrf1}) suggest how to implement the computation of the LRT:\\
\noindent
{\it 1) } Each cluster in the WSN should be composed of the nodes associated to the random variables
pertaining to the same clique in the statistical dependency graph;\\
\noindent{\it 2) } The observations gathered by the nodes pertaining to a clique $c$ are locally encoded
into the clique potential $\psi_c(\bx_c; \mathcal{H}_{i})$. This is the value
that has to be transmitted by each cluster towards upper layers or to the FC;\\
\noindent{\it 3) } As in Fig. \ref{tessellation}, each relay in the lowest layer compute the local potentials
and forward these results to the upper layers. The relays of the intermediate clusters receive
the partial results from the lower clusters, multiply these values by the local potential and forward
the results to the relay of the upper cluster, until reaching the FC.\\
In general, different grouping may occur depending on the hypothesis.
This organization represents a generalization of the distributed computation
observed in the conditionally independent case, where the groups are simply
singletons, i.e. sets composed by exactly one element. In that case,
the clustering among nodes is only instrumental to the communication
purposes, i.e. to enable spatial reuse of radio resources. In the more general Markovian case,
the organization of the communication network in clusters (cells) should take into account,
{\it jointly}, the grouping suggested by the cliques of the underlying dependency graph
and the spatial grouping of nodes to enable concurrent transmission over the
same radio resources without incurring in undesired interference.
To visualize this general perspective, it is useful to have in mind two
superimposed graphs, as depicted in Fig. \ref{WSN_MRF}: the communication
graph (top), whose vertices are the network nodes while the edges are
the radio links; the dependency graph (bottom), whose vertices represent random
variables, while the arcs represent statistical dependencies.
Each communication cluster should incorporate at least one clique.
Furthermore, in each cluster there is a relay node that is responsible for the exchange
of data with nearby clusters.
The whole communication network has a hierarchical tree-structure. Each node in the tree is a relay node
belonging to a cluster. This node collects the measurements from the nodes belonging to its cluster,
computes the potential (or the product of potentials if more cliques belong to the same cluster)
and forwards this value to its relay parents.
While we have depicted the two graphs as superimposed in Fig. \ref{WSN_MRF},
it is useful to clarify that the nodes of the communication network are located in space and their
relative position is well defined in a metric space. Conversely, the nodes of the Markov graph
represent random variables for which there is no well defined notion of distance or,
even if we define one, it is a notion that in general does not have a correspondence
with distance in space. In other words, while the neighborhood of nodes in the
top graph has to do with the concept of spatial distance among the nodes, the
neighborhood of the nodes in the Markov graph has only to do with
statistical dependencies. Nevertheless, it is also true that in the observation of physical
entities like a temperature field, for example, it is reasonable to expect higher
correlation among nearby (in the spatial sense) nodes (variables).
An example of GMRF where the statistical dependencies incorporate the spatial distances
was suggested in \cite{Anandkumar-Tong-Swami}.
In summary, the previous considerations suggest that an efficient design
of the communication network topology should keep into account the
structure, if any, of dependency graph describing the observed variables.
At the same time, the design of the network topology
should keep into account physical constraints like the power consumption necessary
to maintain the links with sufficient reliability (i.e., to insure the sufficient signal-to-noise
ratio at the receiver). This is indeed an interesting line of research: How to match
the network topology to the dependency graph, under physical constraints
dictated by energy consumption, delay, etc. Some works have already addressed this issue.
For example, in \cite{Anandkumar-Yukich-Tong-Swami} the authors addressed the problem
of implementing data fusion policies with minimal energy consumption, assuming a
Markov random field observation model, and established the scaling laws for
optimal and suboptimal fusion policies.
An efficient message-passing algorithm taking into account the communication
network constraints was recently proposed in \cite{Kreidl-Willsky}.

\subsection{Fundamental information-theoretical issues}
In this section, we recall very briefly some of the fundamental information-theoretic limits
of multi-terminal decision networks. We will not go into the details of this challenging
fundamental problem. The interested reader can refer to \cite{Gastpar-chapter} and the
references therein.
In a WSN, each sensor is observing a physical phenomenon, which can be regarded as
a source of information, and the goal of the network is to take decisions about what is being sensed.
In some cases, the decision is taken by a fusion center; in others, the decision is distributed
across the nodes. In general, the data gathered by the nodes
has to travel through realistic channels, prone to additive noise,
channel fading and interference. This requires source and channel coding.
In a point-to-point communication,
when there is only one sensor transmitting data to the fusion center, the
encoding of the data gathered by the sensor follows well known rules.
In particular, the observation is first time-sampled and
each sample is encoded in a finite number, let us say $R$, of bits per symbol.
This converts an analog source of information into a digital source. In this
analog-to-digital (AD) conversion, there is usually a distortion that can be properly quantified.
More precisely, the source coding rate $R$ depends on the constraint on the
mean-square distortion level $D$. At the same time, given a constraint
on the power budget (cost) $P$ available at the transmit side, the maximum rate
that can be transmitted with arbitrarily low error probability is the channel
capacity $C(P)$, which depends on the transmit power constraint. A
rate-distortion pair $(D, P)$ is achievable if and only if
\begin{equation}
R(D)\le C(P).
\end{equation}
The source-channel coding separation theorem \cite{cover-book} states that
the encoding operation necessary to transmit information through a noisy channel
can be split, without loss of optimality,
into the cascade of two successive {\it independent} operations: i) source coding, where each symbol
emitted by the source is encoded in a finite number of bits per symbol;
ii) channel coding, where a string of $k$ bits are encoded into a codeword of
length $n$ bits, to make the codeword error probability arbitrarily low.
This theorem has been a milestone in digital communications, as it
allows system designers to concentrate, separately, on source coding and channel
coding techniques, with no loss of optimality.
However, when we move from the point-to-point link to the multipoint-to-multipoint
case, there is no equivalent of the source-channel coding separation theorem.
This means that in the multi-terminal setting, splitting coding into source and channel coding does not come
without a cost, anymore. Rephrasing the source/channel coding theorem
in the multi-terminal context, denoting by ${\cal R}(D)$ the {\it rate region},
comprising all the source codes that satisfy the distortion constraint $D$,
and by ${\cal C}(P)$ the {\it capacity region}, containing all the transmission rates
satisfying the transmit power constraint $P$, a pair $(D, P)$ is achievable if
\begin{equation}
\label{RD-multiT}
{\cal R}(D) \cap {\cal C}(P) \neq \emptyset.
\end{equation}
However, Equation (\ref{RD-multiT}) is no longer a necessary condition, meaning
that there may exist a code that achieves the prescribed distortion $D$
at a power cost $P$, which cannot be split into a source compression encoder
{\it followed} by a channel encoder. In general, in the multiterminal case,
a {\it joint} source/channel encoding is necessary. This suggests, from a
fundamental theoretical perspective, that, again, in a distributed WSN
local processing and communication have to be considered jointly.

\subsection{Possible architectures}
\begin{figure}[h]
\centering
\includegraphics[scale=0.45]{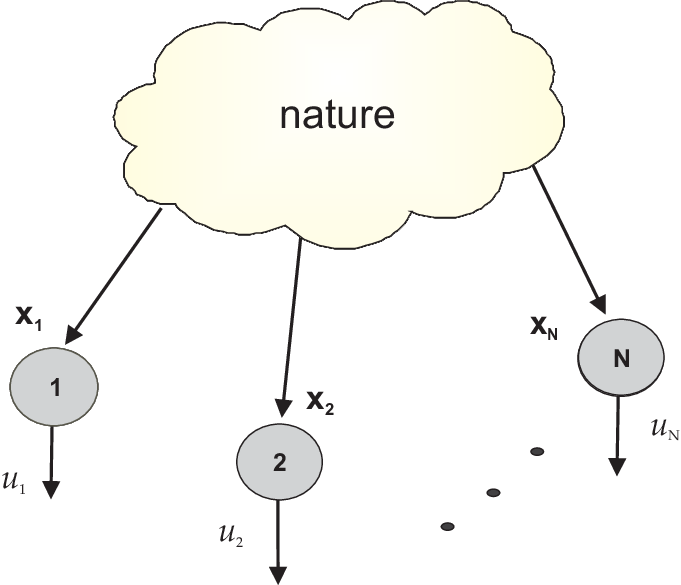}\hspace{2cm}
\includegraphics[scale=0.45]{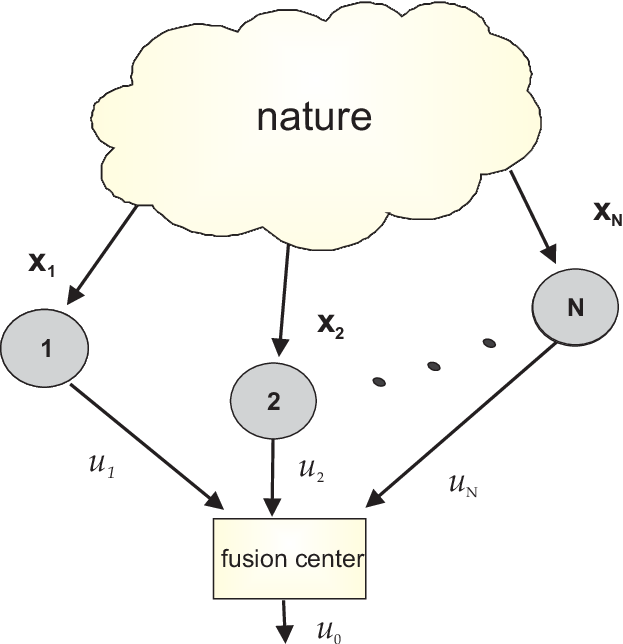}\\
a)\hspace{6cm} b)\vspace{.5cm}\\
\hspace{-.5cm}\includegraphics[scale=0.45]{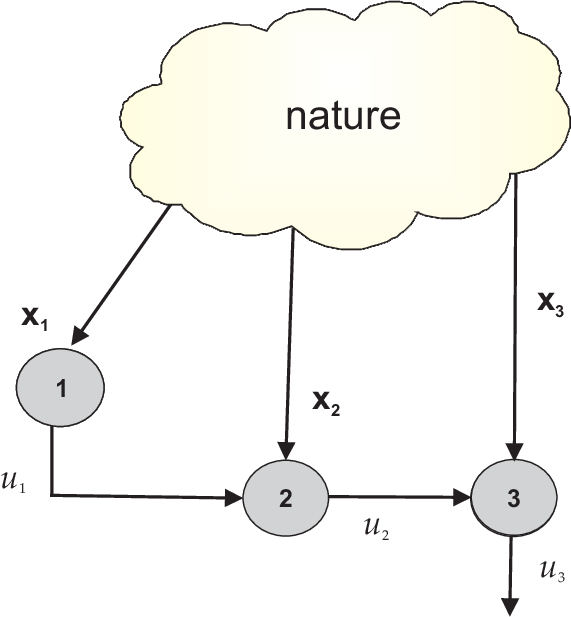}\hspace{2.5cm} \includegraphics[scale=0.45]{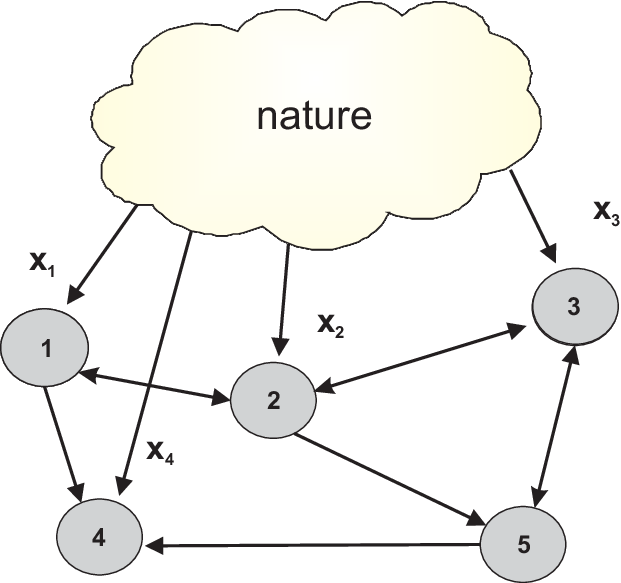}\\
c)\hspace{6cm} d)
\caption{Alternative communication architectures between peripheral nodes and fusion center.}
\label{fig_arch}
\end{figure}

Alternative networks architectures may be envisaged depending on how the nodes take
decision and exchange information with each other. A few examples are shown in Fig. \ref{fig_arch}
where there is a set of $N$ nodes observing a given phenomenon, denoted
as ``nature'' for simplicity.
The measurements made by node $i$ are collected into the vector $\by_i$, with
$i=1, \ldots, N$. In Fig. \ref{fig_arch} a), each node takes an individual decision, which is
represented by the variable $u_i$: $u_i=1$ if node $i$ decides for the presence of the event,
otherwise $u_i=0$. More generally, $u_i$ could also be the result of a local source encoder, whose
aim is to reduce the redundancy present in the observed data.
The simplest case is sketched in Fig. \ref{fig_arch} a), where a set of nodes observes a
state of nature and each node takes a decision. Even if this is certainly the simplest
form of monitoring, if the local decisions are taken according to a global optimality criterion,
even in the case of statistically independent observations, the local decisions are coupled
in a non trivial form.
The next step, in terms of complexity, is to combine all the observations collected by the sensing nodes in a
centralized node, called fusion center or sink node. This strategy is depicted in the architecture
of Fig. \ref{fig_arch} b).
In such a case, each node takes a local decision and sends this information to the fusion center,
which combines the local decision according to a globally optimum criterion.
What is important, in a practical setting, is that the limitations occurring in the transmission
of information from the sensing nodes to the fusion center are properly taken into account.
An alternative approach is reported in Fig. \ref{fig_arch} c), where node $1$
takes a local decision and it notifies node $2$ about this decision.
Node $2$, on its turn, based on the decision of node $1$ and on its
own measurements as well, takes a second decision, and so on.
A further generalization occurs in the example of Fig. \ref{fig_arch} d), where the nodes
take local decisions and exchange information with the other nodes. In such a case,
there is no fusion center and the final decision can be taken, in principle, by every node.\\
Besides the architecture describing the flow of information through the network,
a key aspect concerns the constraint imposed by the communication links.
Realistic channels are in fact affected by noise, fading, delays, and so on.
Hence, a globally optimal design must incorporate the decision and
communication aspects jointly in a common context. The first step
in this global design passes through a formal description of the
interaction among the nodes.

\section{Graphical models and consensus algorithm}
\label{Consensus algorithms for distributed detection and estimation}
The proper way to describe the interactions among the network nodes
is to introduce the graph model of the network.
Let us consider a  network composed of $N$ sensors.
The flow of information across the sensing nodes implementing some form of distributed
computation can be properly described by introducing a graph model whose
vertices are the sensors and there is an edge between two nodes if they exchange information
with each other\footnote{ We refer the reader to Appendix A for a review of the basic notations and properties of graphs.}. Let us denote the graph as
$\mathcal{G}=\{\mathcal{V},\mathcal{E}\}$ where $\mathcal{V}$ denotes the
set of $N$ vertices (nodes) $v_i$  and  $\mathcal{E}\subseteq
{\mathcal{V} \times \mathcal{V}}$ is the set of   edges
$e_{i j}(v_i,v_j)$. The most powerful tool to grasp the properties of a graph
is {\it algebraic graph theory} \cite{SB-godsil}, which is based on the description of the graph
through appropriate matrices, whose definition we recall here below.
Let $\bA\in \mathbb{R}^{N\times N}$ be the \textit{adjacency} matrix of the graph $\mathcal{G}$,
whose elements $a_{ij}$ represent the weights associated to each edge with $a_{ij}>0$ if
$e_{ij}\in \mathcal{E} $ and $a_{ij}=0$ otherwise.
According to this notation  and assuming no self-loops, i.e., $a_{ii}=0$, $\;   \forall i=1,\ldots,N$,
the out-degree of node $v_i$ is defined as $ \mbox{deg}_{out}(v_i)=\ds
\sum_{j=1}^{N}a_{ji}$. Similarly, the in-degree of node  $v_i$ is $ \mbox{deg}_{in}(v_i)=\ds
\sum_{j=1}^{N}a_{ij}$. The \textit{degree} matrix $\bD$ is defined as the diagonal matrix
whose $i$-th diagonal entry is $d_{ii}=\mbox{deg}(v_i)$. Let ${\cal N}_i$
denote the set of neighbors of node $i$, so that $|{\cal N}_i| =
\mbox{deg}_{in}(v_i)$\footnote{By $\mid \cdot \mid$ we denote the
cardinality of the set.}. The \textit{Laplacian} matrix $\bL \in \mathbb{R}^{N\times N}$
of the graph $\mathcal{G}$ is defined as $\bL:=\bD-\bA$.
Some properties of the Laplacian will be used in the
distributed algorithms to be presented later on, and then it is
useful to recall them.\\

\noindent{\it Properties of the Laplacian matrix}\\

\noindent{\it P.1:} $\bL$ has, by construction, a null eigenvalue
with associated eigenvector the vector $\buno$ composed by all ones.

This property can be easily checked verifying that $\bL \buno=\bzero$ since by
construction, $\ds \sum_{j=1}^{N} a_{ij}=d_{ii}$.\\

\noindent{\it P.2:} The multiplicity of the null eigenvalue is equal to the number
of connected components of the graph. Hence, the null eigenvalue
is {\it simple} (it has multiplicity one) if and only if the graph is connected.\\

\noindent{\it P.3:} If we associate a state variable $x_i$ to each node of the graph,
if the graph is undirected, the disagreement between the values assumed by the
variables is a quadratic form built on the Laplacian \cite{SB-godsil}:
\begin{equation}
J(\bx):= \frac{1}{4}\sum_{i=1}^N \sum_{j\in \mathcal{N}_i } a_{ij} (x_i-x_j)^2
=\frac{1}{2}\,\bx^T\bL\bx,
\label{SB-cons-disag}
\end{equation}
where $\bx=[x_1,\ldots, x_N]^T$ denotes the network state vector.\\

\subsection{Consensus algorithm}
Given a set of measurements $x_{i}(0)$, for $i=1, \ldots, N$, collected by the network nodes, the goal of
consensus algorithm is to minimize the disagreement among the nodes. This can be useful, for example,
when the nodes are measuring some common variable and their measurement is affected by error.
The scope of the interaction among the nodes is to reduce the effect of local errors on the final estimate.
In fact, consensus is one of fundamental tools to design distributed decision algorithms that
satisfy a global optimality principle, as corroborated by many works on distributed optimization,
see, e.g., \cite{Jadbabaie-Morse, Xiao-Boyd, SB-Ren-Beard-Control-Magazine, SB-olfati, Nedic-Ozdaglar,
SB-barbarossa_ML, SB-barbarossa_Pesc}.
We recall now the consensus algorithm as this will form the basis of the distributed estimation and detection
algorithms developed in the ensuing sections.

Let us consider, for simplicity, the case where the nodes are measuring a temperature and
the goal is to find the average temperature. In this case,  reaching  a consensus over the
average temperature can be seen as the minimization of the disagreement, as defined in (\ref{SB-cons-disag}),
between the states $x_i(0)$ associated to the nodes.
The minimization of the disagreement can be obtained  by using a simple gradient-descent algorithm.
More specifically, using a continuous-time system, the minimum of  (\ref{SB-cons-disag}) can be
achieved by running the following dynamical system \cite{SB-olfati}
\begin{equation}
\label{ct-consensus}
\dot{\bx}(t)=-\bL\,\bx(t),
\end{equation}
initialized with $\bx(0)=\bx_0$, where $\bx_0$ is the vector
containing all the initial measurements collected by the network nodes.
This means that the state of each node evolves in time
according to the first order differential equation
\begin{equation}
\dot{x}_i(t)=\sum_{j\in {\cal N}_i}\,a_{ij}(x_j(t)-x_i(t)) \label{ct2}
\end{equation}
where ${\cal N}_i$ indicates the set of neighbors of node $i$. Hence,
every node updates its own state only by interacting with its neighbors.

Equation (\ref{ct-consensus}) assumes the form of a diffusion equation.
Let us consider for example the evolution of a diffusing physical quantity  $\psi(z; t)$
as a function of the spatial variable $z$ and of time $t$ ($\psi(z; t)$ could
represent, for instance, the heat distribution), the diffusion
equation assumes the form
\begin{equation}
\label{diffusion}
\frac{\partial \psi(z; t)}{\partial t}=D\,\frac{\partial^2 \psi(z; t)}{\partial z^2}
\end{equation}
where $D$ is the diffusion coefficient. If we discretize the space variable
and approximate the second order derivative with a discrete-time second order
difference,  the diffusion equation (\ref{diffusion}) can be written as in
(\ref{ct-consensus}), where the Laplacian matrix represents the
discrete version of the Laplacian operator.  This conceptual link between
consensus equation and diffusion equation has been exploited in
\cite{SB-sard_barb_giona} to derive a fast consensus algorithm,
mimicking the effect of advection. The interesting result derived in
\cite{SB-sard_barb_giona} is that to speed up the consensus (diffusion)
process, it is necessary to use a {\it directed} graph, with time-varying
adjacency matrix coefficients $a_{ij}$.

The solution of (\ref{ct-consensus}) is given by
\begin{equation}
\label{expLx}
\bx(t)= \exp(- \bL \, t)\,\,\bx(0) \;.
\end{equation}
In the case analyzed so far, since the consensus algorithm has been
deduced from the minimization of the disagreement and the
disagreement has been defined for undirected graphs, the matrix
$\bL$ is symmetric. Hence, its eigenvalues are real.
The convergence of (\ref{expLx}) is guaranteed
because all the eigenvalues of $\bL$ are non-negative, by construction.
If the graph is connected, according to property {\it P.2},
the eigenvalue zero has multiplicity
one. Furthermore, the eigenvector associated to the zero eigenvalue
is the vector $\buno$. Hence, the system (\ref{ct-consensus}) converges to
the consensus state:
\begin{equation}
\lim_{t\rightarrow \infty}\bx(t)=\frac{1}{N} \buno\,\buno^T\,\bx(0).
\end{equation}
This means that every node converges to the average value
of the measurements collected by the whole network, i.e.,
\begin{equation}
\lim_{t\rightarrow \infty}\,x_i(t)=\frac{1}{N} \,\,\sum_{i=1}^{N}x_{i}(0)=x^{*}.
\end{equation}
The convergence rate of system (\ref{ct2})  is lower bounded by the slowest decaying mode
of the dynamical system (\ref{ct-consensus}), i.e.
by the second smallest eigenvalue of $\bL$, $\lambda_2(\bL)$, also known as the {\it algebraic
connectivity} of the graph \cite{SB-fiedler}.
More specifically, if the graph is connected or, equivalently, if  $\lambda_2(\bL)>0$, then the dynamical system
(\ref{ct-consensus}) converges to consensus exponentially \cite{SB-olfati}, i.e.
$\parallel \bx(t)-x^{*} \buno\parallel \leq \parallel \bx(0)-x^{*} \buno\parallel O(e^{-rt})$
with $r=\lambda_2(\bL)$.

In some applications, the nodes are required to converge to a {\it weighted} consensus,
rather than average consensus.
This can be achieved with a slight modification of the consensus algorithm. If we premultiply the left side
of (\ref{ct2}) by a positive coefficient $c_i$, the resulting equation
\begin{equation}
c_i\dot{x}_i(t)=\sum_{j\in {\cal N}_i}\,a_{ij}(x_j(t)-x_i(t))
\end{equation}
converges to the weighted average
\begin{equation}\label{weighted average}
\lim_{t \rightarrow \infty}\,x_i(t)= \frac{\sum_{i=1}^{N}\,c_i\,x_{i}(0)}{\sum_{i=1}^{N}\,c_i}.
\end{equation}
This property will be used in deriving distributed estimation mechanisms
in the next section.

Alternatively, the minimization of (\ref{SB-cons-disag})
can be achieved in discrete-time through the following iterative algorithm
\begin{equation}
\label{x[k+1]N}
\bx[k+1]=\bx[k]-\epsilon \bL \bx[k]:=\bW \bx[k],
\end{equation}
where we have introduced the so called {\it transition} matrix $\bW=\bI-\epsilon \bL$.
Also in this case, the discrete time equation is initialized with the measurements taken by the
sensor nodes at time $0$, i.e., $\bx[0]:=\bx_0$.
This time, to guarantee convergence of the system (\ref{x[k+1]N}), we need to choose the coefficient $\epsilon$
properly. More specifically, the discrete time equation  (\ref{x[k+1]N}) converges if the eigenvalues of
$\bW$ are bounded between $-1$ and $1$.  This can be seen very easily considering
that reiterating (\ref{x[k+1]N}) $k$ times, we get
\begin{equation}
\label{x[k+1]}
\bx[k]=\bW^k\bx[0]\;.
\end{equation}
Let us denote by $\bu_k$ the eigenvectors of $\bW$ associated to the
eigenvalues $\lambda_k(\bW)$, with $k=1, \ldots, N$. The eigenvalues of $\bW$ are real
and we consider them ordered in increasing sense, so that $\lambda_N(\bW)\ge \lambda_{N-1}(\bW)\ge
\cdots \lambda_1(\bW)$. Hence,  the evolution of system (\ref{x[k+1]}) can be written as
\begin{equation}
\label{x_k_vs_lambda}
\bx[k]=\sum_{n=1}^N \lambda_n^k(\bW)\,\, \bu_n\, \bu_n^T\,\bx[0].
\end{equation}
The matrix $\bW$ has an eigenvector equal to $\b1/\sqrt{N}$, associated to the eigenvalue
$1$ by construction. In fact, $\bW\, \b1/\sqrt{N}=\b1/\sqrt{N}-\epsilon\, \bL\, \b1/\sqrt{N}=\b1/\sqrt{N}$.
If the graph is connected, the eigenvalue $1$ of $\bW$ has multiplicity one. Furthermore, if
$\epsilon$ is chosen such that $\epsilon(\bL)<2/\lambda_N(\bL)$, all
other eigenvalues are less than $1$. Hence, for a connected
graph, the system (\ref{x_k_vs_lambda}) converges to
\begin{equation}
\label{x_infty}
\lim_{k\rightarrow \infty}\bx[k]= \frac{1}{N}\b1\, \b1^T\,\bx[0].
\end{equation}
Again, this corresponds to having every node converging to the average consensus.

The consensus algorithm can be extended to the case of directed graphs.
This case is indeed much richer of possibilities than the undirected case, because the consensus
value ends up to depend more strictly on the graph topology.
In the directed case, in fact, $\bL$ is an  asymmetric
matrix. The most important difference is that the graph connectivity
turns out to depend on the orientation of the edges. Furthermore,
each eigenvalue of $\bL$ gives rise to a pair of
left and right eigenvectors which do not coincide with each other.
These differences affect the final consensus state and induce
different forms of consensus, as shown below.

The convergence of the system in (\ref{x[k+1]N}) can be proved by exploiting the properties
of non-negative matrices.
A nonnegative matrix is row (or column) stochastic if
all its row (or column) sums are equal to one. Furthermore
if the graph associated to the network is strongly connected, i.e. the zero eigenvalue associated to $\bL$ has multiplicity one (see Appendix A),  $\bW$
is called an {\it irreducible} matrix. An irreducible stochastic matrix is primitive if it has only one eigenvalue
with maximum modulus. Primitive nonnegative matrices, often named
\emph{Perron matrices}, satisfy the Perron-Frobenius theorem \cite{SB-horn}.
\begin{theorem}
Let $\bgamma_l$ and $\bgamma_r$, respectively, the left and right eigenvectors associated to the
unit eigenvalue of the primitive nonnegative matrix
$\bW$, i.e. $\bW \bgamma_r=\bgamma_r$ and $\bgamma_l^T\bW =\bgamma_l^T$ with $\bgamma_r^T \bgamma_l=1$,
then $\lim_{k\rightarrow \infty}\bW^k=\bgamma_r\bgamma_l^T $. \label{SB-ther_frobenius}
\end{theorem}
Let us now apply to a sensor network modeled by the graph $\mathcal{G}$ with adjacency matrix $\bA$  the distributed consensus algorithm
 \beq
x_i[k+1]= x_i[k] - \epsilon \ds \sum_{j \in \mathcal{N}_i} a_{ij} (x_i[k]-x_j[k]) \label{SB-discrete-time1}
\eeq
with $0<\epsilon< 1/d_{max}$.

Interestingly, different forms of consensus can be achieved in a directed graph, depending on
the graph connectivity properties \cite{SB-barbarossa_Pesc}:
\begin{itemize}
  \item [a)]\emph{If the graph is strongly connected,  the dynamical system in (\ref{SB-discrete-time1})
converges to a weighted consensus}, for any initial state vector $\bx[0]$, i.e.,
\beq
\lim_{k\rightarrow \infty} \mathbf{W}^k \mathbf{x}[0]=\mathbf{x}^{\star}=\mathbf{1} \mbox{\boldmath $\gamma$}_l^T \mathbf{x}[0]\label{SB-conv-discrete}
\eeq
where $\gamma_l(i)>0$, $\forall i$, and $\sum_{i=1}^{N} \gamma_l(i)=1$.
In this case, since the graph is strongly connected,  $\bW$
is an  irreducible matrix. Then, applying Gershgorin theorem \cite{SB-horn},
it can be deduced that there exists a single eigenvalue $\mu_1(\bW)=1$ with maximum modulus.
Then  $\bW$ is a primitive nonnegative matrix and from
Theorem $\ref{SB-ther_frobenius}$  the convergence in (\ref{SB-conv-discrete})
is straightforward. In this case, every node contributes to the final consensus value.
Furthermore, the consensus value is a weighted combination of the initial observations,
where the weights are the entries of the left eigenvector associated to the null
eigenvalue of $\bL$ (or the unit eigenvalue of $\bW$).
  \item  [b)] \emph{If the digraph is strongly connected and balanced, i.e. $\b1^T \bL=\mathbf{0}$ and $\bL \b1=\mathbf{0}$, the systems achieves an average consensus or}
$\bx^{\star}=\ds \frac{\b1 \b1^T}{N} \bx[0]$. In fact, for balanced graphs, $\bW$ is  a double stochastic matrix
 with $\bgamma_l=\bgamma_r=\b1/\sqrt{N}$;
\item [c)]  \emph{If the digraph ${\mathcal{G}}$ is
weakly connected (WC)}, but not strongly connected, and it contains a forest with $K$ strongly
connected root components, the graph splits in $K$ disjoint clusters ${%
\mathcal{C}%
}_{1},\ldots ,{%
\mathcal{C}%
}_{K}\subseteq \{1,\ldots ,N\}$,\footnote{%
In general, the clusters $\mathcal{C}_{1},\ldots ,{\mathcal{C}}_{K}$ are not
a partition of the set of nodes $\{1,\cdots, N\}$.} and all the nodes pertaining to each cluster
converge to the consensus values
\begin{equation}
{x}_{q}^{\star }=\frac{\sum_{i\in {%
\mathcal{C}}_{k}}\gamma _{i} x_i[0]}{\sum_{i\in {%
\mathcal{C}}_{k}}\gamma _{i}},\quad \quad
\forall q\in {\mathcal{C}}_{k},\quad k=1,\ldots ,K.  \label{Eq-Corollary-cluster}
\end{equation}
In words, there is no single consensus, in this case, but there is a
local consensus within each cluster. Different clusters typically converge
to different consensus values.
\item [d)]  \emph{If the digraph ${\mathcal{G}}$ is composed of a single spanning tree},
every node converges to the value assumed by the root node.
\end{itemize}
As far as the convergence rate, instead, in  \cite{SB-olfati} it has been shown for undirected connected
graphs that the dynamical system in
(\ref{x[k+1]N}) converges exponentially to the average consensus with a rate at least equal to
$\mu_2(\bW)=1-\epsilon \lambda_2(\bL)$ where $\mu_2(\bW)$ is the second largest eigenvalue of the Perron matrix $\bW$.
 In fact by defining the disagreement vector $\bdelta=\bx-\bx^{\star}$, it can be easily verified \cite{SB-olfati}
that $\bdelta$  evolves according to the disagreement dynamic given by $\bdelta[k+1]=\bW \bdelta[k]$.
Hence $\psi[k]:=\mbox{\boldmath $\delta$}[k]^T \mbox{\boldmath $\delta$}[k]$ represents a candidate Lyapunov function for the disagreement dynamics  so that
\beq
\psi[k+1]=\bdelta[k+1]^T \bdelta[k+1]=\parallel \bW \bdelta[k]\parallel^2 \leq \mu_2(\bW)^2 \parallel  \bdelta[k]\parallel^2=\mu_2(\bW)^2\psi[k]
\eeq
with $0<\mu_2(\bW)<1$ since $\bW$ is a symmetric and primitive matrix.
 As  a consequence the algorithm converges exponentially to consensus
with \emph{a rate at least equal to $\mu_2(\bW)$}.

\subsection{Consensus algorithms over realistic channels}
\label{Consensus algorithms over realistic channels}
So far, we have recalled the basic properties of consensus algorithm assuming that
the exchange of information across the nodes occurs with no errors. In this section we study
what happens to consensus algorithms when the communications among the nodes
are affected by quantization errors, noise, packet drops, etc.  The problem of consensus protocols affected by stochastic disturbance has been considered in a series of previous papers \cite{PDL-Hatano}-\cite{PDL-Kar-Moura}. In \cite{PDL-Hatano}, the authors use a decreasing sequence of weights to prove the convergence of consensus protocols to an agreement space in the presence of additive noise under a fixed network topology. The works in \cite{Tahbaz-Salehi-Jadb1}-\cite{Tahbaz-Salehi-Jadb2} consider consensus algorithms in the presence of link failures, which are modeled as i.i.d. Laplacian matrices of a directed graph. The papers present necessary and sufficient conditions for consensus exploiting the ergodicity of products of stochastic matrices. A distributed consensus algorithm in which the nodes utilize probabilistically quantized information to communicate with each other was proposed in \cite{PDL-Aysal-Coates-Rabbat}. As a result, the expected value of the consensus is equal to the average of the original sensor data. A stochastic approximation approach was followed in \cite{PDL-Huang-Manton}, which considered a stochastic consensus problem in a strongly connected directed graph where each agent has noisy measurements of its neighboring states. Finally, the study of a consensus protocol that is affected by both additive channel noise and a random topology was considered in \cite{PDL-Kar-Moura}. The resulting algorithm relates to controlled Markov processes and the convergence analysis relies on stochastic approximation techniques.

In the study of consensus mechanisms over realistic channels, we consider the following sources of randomness:\\

\noindent {\textit{1) Node positions}}: The first randomness is related to the spatial positions of the nodes, which are in general unknown. We model the spatial distribution of nodes as a random geometric graph composed of $N$ nodes. In graph theory, a random geometric graph (RGG) is a random undirected graph drawn on a bounded region, eg. the unit disk, generated by:
\begin{enumerate}
  \item Placing vertices at random uniformly and independently on the region,
  \item Connecting two vertices, $u$, $v$ if and only if the distance between them is inside a threshold radius $r_0$, i.e. $d (u, v) \leq r_0$.
\end{enumerate}
Several probabilistic results are known about RGG's.  In particular, as shown in \cite{Gupta-Kumar98}, if $N$ nodes are placed in a disc of unit area in $\mathbb{R}^2$ and each node transmits with a power
scaling with $N$ as in (\ref{theor_r0(N)}),
the resulting network is asymptotically connected with probability one, as $N \rightarrow \infty$.\\

\noindent {\textit{2) Random link failures model}}: In a realistic communication scenario, the packets exchanged among sensors may be received with errors, because of channel fading or noise. The retransmission of erroneous packets can be incorporated into the system, but packet retransmission introduces
a nontrivial additional complexity in decentralized implementations and, most important, it introduces an unknown delay and delay jitter.
It is then of interest to examine simple protocols where erroneous packets are simply dropped. Random packet dropping can be taken into account by modeling the coefficient $a_{ij}$ describing the network
topology as random variables that assume the value $1$ or $0$, if the packet is correctly delivered or not, respectively. In this case, the Laplacian varies with time
as a sequence of i.i.d. matrices $\{\bL[k]\}$, which can be written, without any loss of generality, as
\begin{equation}\label{PDL-RandomLaplacian}
 \bL[k]=\bar{\bL}+\tilde{\bL}[k]
\end{equation}
where $\bar{\bL}$ denotes the mean matrix and $\tilde{\bL}[k]$ are i.i.d. perturbations around the mean.
We do not make any assumptions about the link failure model. Although the link failures and the Laplacians are independent over time, during the same iteration, the link failures can still be spatially correlated. It is important to remark that we do not require the random instantiations $G[k]$ of the graph be connected for all $k$. We only require the graph to be connected on average. This condition is captured by requiring $\lambda_2(\bar{\bL})>0$.

\noindent {\textit{3) Dithered quantization} :} We assume that each node encodes the message to be exchanged with the other nodes using a uniform quantizer, with a finite number of bits $n_b$, defined by the following vector mapping, $\bq(\cdot):\mathbb{R}^L \rightarrow Q^L$,
\begin{eqnarray}\label{PDL-Quantizer}
\bq(\by)=[b_1\Delta,\ldots,b_L\Delta]^T=\by+\bee_q(\by) ,
\end{eqnarray}
where the entries of the vector $\by$, the quantization step $\Delta>0$, and the error $\bee_q$ satisfy
\begin{eqnarray}
(b_m-1/2)\Delta \quad \leq &y_m& \leq \quad (b_m+1/2)\Delta, \quad 1\leq m\leq L,\\
-\Delta/2\hspace{.1cm} \mathbf{1}_L \quad \leq&\bee_q(\by)&\leq \quad \Delta/2 \hspace{.1cm}\mathbf{1}_L, \quad \mbox{for all $\by$}.
\end{eqnarray}
The quantization alphabet is
\begin{equation}
Q^L=\{[b_1\Delta,\ldots,b_L\Delta]^T|b_m\in\mathbb{Z},\forall m\}.
\end{equation}
Conditioned on the input, the quantization error $\bee_q(\by)$ is deterministic.
This induces a correlation among the quantization errors resulting at different nodes
and different times, which may affect the convergence properties of the distributed algorithm.
To avoid undesired error correlations, we introduce dithering, as in  \cite{PDL-Lip-Wan-Vander,PDL-Wan-Lip-Vander-Wright}.
In particular, the dither added to randomize the quantization effects satisfies a special condition,
namely the Schuchman conditions, as in subtractively dithered systems, \cite{PDL-Schuchmann}.
Then, at every time instant $k$, adding to each component $y_m[k]$ a dither sequence $\{d_m[k]\}_{k\geq0}$ of i.i.d. uniformly distributed random variables on $[-\Delta/2,\Delta/2)$ independent of the input sequence, the resultant error sequence $\{e_m[k]\}_{k\geq0}$ becomes
\begin{equation}\label{PDL-Dithering}
e_m[k]=q(y_m[k]+d_m[k])-(y_m[k]+d_m[k]).
\end{equation}
The sequence $\{e_m[k]\}_{k\geq 0}$ is now an i.i.d. sequence of uniformly distributed random variables on $[-\Delta/2,\Delta/2)$, which is independent of the input sequence.\\ 

The convergence of consensus algorithm in the presence of random disturbance can be proved by
exploiting results from supermartingale theory \cite{PDL-Polyak}. In an ideal communication case, by selecting the step-size of the algorithm to be sufficiently small (smaller than $2/\lambda_N(\bL)$, where $\lambda_N(\bL)$ is the maximum eigenvalue of the Laplacian matrix of the graph), the discrete-time consensus algorithm will asymptotically converge to the agreement subspace. However, in a realistic communication scenario, the links among the sensors may fail randomly and the exchanged data is corrupted by quantization noise. Under these nonideal conditions, the consensus algorithm needs to be properly adjusted to guarantee convergence. A discrete time consensus algorithm that accounts for random link failures and dithered quantization noise can be written as:
\begin{eqnarray}\label{PDL-x_punto_disc}
\bx_i[k+1]=\bx_i[k]+\alpha[k]\sum_{j=1}^{N}a_{ij}[k]\hspace{.1cm}(\bq(\bx_j[k]+\bd_{ij}[k])-\bx_i[k]), \quad i=1, \ldots, N\nonumber,
\end{eqnarray}
where $\alpha[k]$ is a positive iteration dependent step-size, and $\mathbf{d}_{ij} [k]$ is the dithered quantization vector. Now, exploiting the feature of subtractively dithered systems in (\ref{PDL-Dithering}), the previous expression can be recast as:
\begin{eqnarray}\label{PDL-x_punto_quant_disc}
\bx_i[k+1]=\bx_i[k]+\alpha[k]\sum_{j=1}^{N}a_{ij}[k]\hspace{.1cm}(\bx_j[k]-\bx_i[k]+\bd_{ij}[k]+\bee_{ij}[k]), \quad i=1, \ldots, N\nonumber.
\end{eqnarray}
Starting from some initial value, $\bx_i[0]\in\mathbb{R}^L$, each node generates via (\ref{PDL-x_punto_quant_disc}) a sequence of state variables, $\{\bx_i[k]\}_{k\geq0}$. The value $\bx_i[k+1]$ at the $i$-th node at time $k+1$ is a function of: its previous state $\bx_i[k]$ and the quantized states correctly received at time $k$ by the neighboring sensors. As described previously, the data are subtractively dithered-quantized, so that the quantized data received by the $i$-th sensor from the $j$-th sensor at time $k$ is $\bq(\bx_j[k] + \bd_{ij}[k])$. It then follows that the quantization error $\bee_{ij}[k]$ is a random vector, whose components are i.i.d., uniformly distributed on $[-\Delta/2,\Delta/2)$, and independent of $\bx_j[k]$.

One way to guarantee convergence of the previous system is to use  a positive iteration-dependent step size
$\alpha[k]$ satisfying \cite{PDL-Hatano}, \cite{PDL-Kar-Moura}
\begin{eqnarray}\label{PDL-Decreasing_step}
\lim_{k\rightarrow\infty}\alpha[k]=0, \quad \sum_{k=0}^\infty \alpha[k]=\infty, \quad  \sum_{k=0}^\infty \alpha^2[k]<\infty.
\end{eqnarray}
Exploiting results from stochastic approximation theory, this choice drives the noise variance to zero while
guaranteeing the convergence to  the consensus subspace.

\begin{figure}[t]
\centering
\includegraphics[scale=0.6]{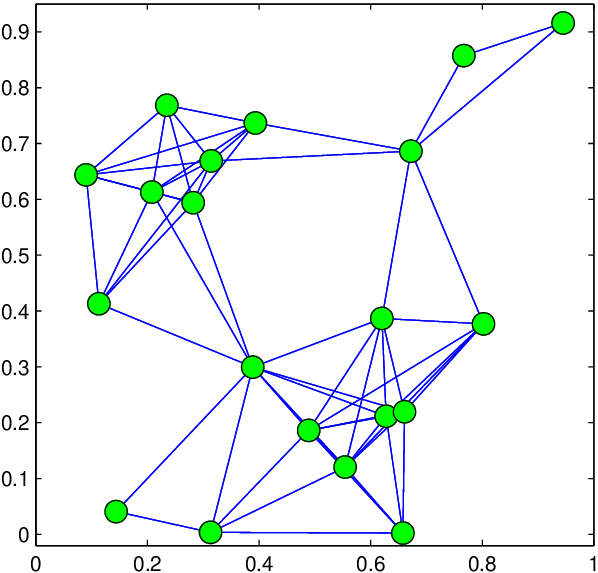} \hspace{.4cm}
\includegraphics[scale=0.6]{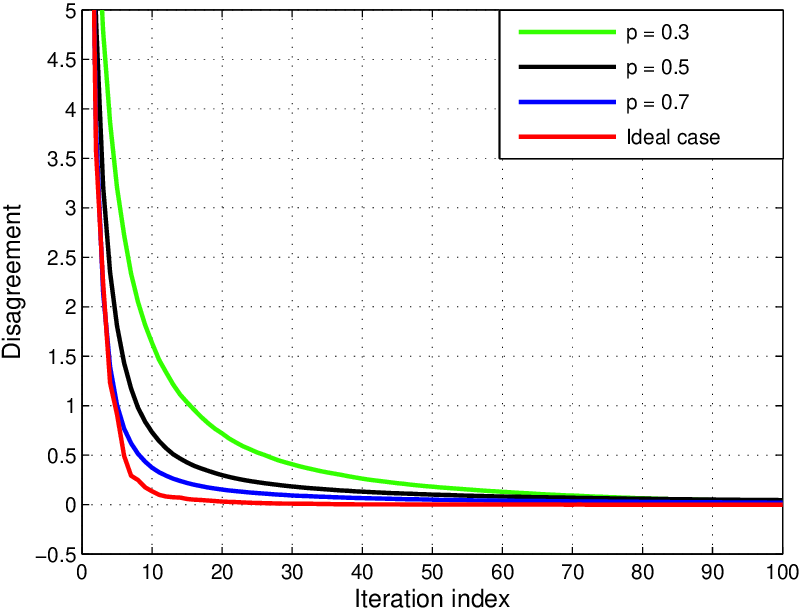}
  \caption{Network (left). Disagreement vs. time index (right), for different probabilities of correct packet reception.}\label{PDL-consensus_rand}
\end{figure}

A numerical example is useful to show the robustness of consensus algorithm in the presence of link failures and quantization noise. We consider a connected network composed of $20$ nodes as depicted on the left side of Fig. \ref{PDL-consensus_rand}. The initial value of the state variable at each node is randomly chosen in the interval
$[0, 1)$. At the $k$-th iteration of the updating rule (\ref{PDL-x_punto_disc}), each node communicates to its neighbors its current state, i.e., a scalar $x_i[k]$. Because of fading and additive noise, a communication link among two neighbors has a certain probability $p$ to be established correctly. The values to be exchanged are
(dither) quantized with $6$ bits. The iteration-dependent step size is chosen as $\alpha[k]=\alpha_0/k$, with $\alpha_0=1.5/\lambda_N(\bL)$, in order to satisfy (\ref{PDL-Decreasing_step}).
The right side of Fig. \ref{PDL-consensus_rand} shows the average behavior of the disagreement among the sensors in the network, versus the iteration index, for different values of the probability $p$ to establish a communication link correctly. The result is averaged over 100 independent realizations. The ideal case corresponds to $p = 1$ and it is shown as a benchmark. As we can notice from the right side of Fig. \ref{PDL-consensus_rand}, even in the presence of random disturbances, an agreement is always reached by the network for any value of $p$. The only effect of the random link failures is to slow down the convergence process, without altering the final value of the global potential function. This proves the robustness of the algorithm.

\section{Distributed estimation}
\label{Distributed estimation}
Having introduced all the tools necessary to study distributed estimation and detection mechanisms,
let us now start with the estimation problem. This problem has been the subject of an extensive literature,
see, e.g. \cite{SB-Barbarossa-Scutari-SP-Magazine}, \cite{SB-Schizas-Ribeiro-GG, Schizas-Giannakis-Roumeliotis-Ribeiro, Cattivelli-Lopes-Sayed-08, Braca-Marano-Matta, Dimakis-Kar-Moura-Rabbat-Scaglione, Bertrand-Moonen, Li-Scaglione-Manton, Kar-Moura-Ramanan-12}. Most of the algorithms proposed in these works propose a mix of local estimation and
consensus among neighbor nodes to improve upon the performance of the local estimators. In a first class
of methods, like \cite{SB-Schizas-Ribeiro-GG, Schizas-Giannakis-Roumeliotis-Ribeiro} for example,
the nodes collect all the data first, perform local estimation and then interact iteratively with their neighbors.
In alternative methods, the nodes keep interacting with each other while collecting new measurements
or, in general, receiving new information, like in \cite{Cattivelli-Lopes-Sayed-08, Braca-Marano-Matta, Kar-Moura-Ramanan-12}. These two classes of methods can be seen as assigning different time scales to the local estimation
and consensus steps.
Indeed, it can be proved that a proper combination of local estimation and consensus can bring
the whole network to a globally optimal estimate, provided that the graph describing the interaction
among the nodes is connected. This approach was pursued, for example in \cite{SB-Schizas-Ribeiro-GG}, where the so called bridge nodes fulfilled the scope of enforcing local consensus.
In the following, we will show how alternative formulations of the globally optimal estimation problem
naturally lead to a different mix of the local estimation and the consensus steps,
without the need to introduce any node having a special role nor enforcing different time scales a priori.

Let us denote with $\btheta \in \mathbb{R}^M$ the parameter vector to be estimated.
In some cases, there is no prior information about $\btheta$. In other cases,
$\btheta$ is known to belong to a given set $\mathcal{C}$:
For instance, its entries are known to be positive or to belong to a finite
interval of known limits, and so on. In some applications,
$\btheta$ may be the outcome of a random variable described by a known pdf $p_\Theta(\btheta)$.
Let us denote by $\bx_i$ the measurement vector collected by node $i$ and by $\bx:=[\bx_1^T, \ldots, \bx_N^T]^T$
the whole set of data collected by all the nodes.
In the two cases of interest, the estimation can be obtained as the solution
of the following problems:\\

\noindent{\it Arbitrary case}
\begin{eqnarray}
\max_{\btheta} p_{X;\Theta}(\bx; \btheta)\\
\mbox{s.t.} \quad \btheta \in \mathcal{C}
\end{eqnarray}
where $p_{X;\Theta}(\bx; \btheta)$ is the joint pdf of vector $\bx$, for a given
arbitrary vector $\btheta$, or\\

\noindent{\it Random case}
\begin{equation}
\label{random_parameter}
\max_{\btheta} p_{X/\Theta}(\bx/\btheta)p_{\Theta}(\btheta)
\end{equation}
where $p_{\Theta}(\btheta)$ is (known) prior pdf of the parameter vector and
$p_{X/\Theta}(\bx/\btheta)$ is the pdf of $\bx$ conditioned to $\btheta$.\\

In general, it is not necessary to reconstruct the whole joint pdf $p_{X;\Theta}(\bx; \btheta)$
(or $p_{X/\Theta}(\bx/\btheta)$) to obtain the optimal estimate. Let us consider, for example,
the case where the pdf can be factorized as
\begin{equation}
\label{suff_stat}
p_{X;\Theta}(\bx; \btheta)=g\left[\bT(\bx),\btheta\right]\,h(\bx),
\end{equation}
where $g(\cdot, \cdot)$ depends on $\bx$ only through $\bT(\bx)$,
whereas $h(\cdot)$ does not depend on $\btheta$.
The function $\bT(\bx)$ is called a {\it sufficient statistic} for $\btheta$ \cite{SB-kay}.
In general, the sufficient statistic $\bT(\bx)$ is a vector, as it may be constituted by a
set of functions.
If (\ref{suff_stat}) holds true, all is necessary to estimate $\btheta$ is not
really $p_{X;\Theta}(\bx; \btheta)$, but only $g\left[\bT(\bx),\btheta\right]$.
This means that any sensor able to evaluate $\bT(\bx)$ through an interaction with
the other sensors is able to find out the optimal parameter vector $\btheta$
as the vector that maximizes $g\left[\bT(\bx),\btheta\right]$.

A simple (yet common) example is given by the so called {\it exponential family} of pdf
\begin{equation}
\label{exp_pdf}
p(\bx; \btheta)=\exp\left[A(\btheta) B(\bx)+C(\bx)+D(\btheta)\right].
\end{equation}
Examples of random variables described by this class include the Gaussian, Rayleigh, and exponential
pdf's. Hence, this is a rather common model.
Let us assume now that the observations $\bx_i$ collected by different nodes are statistically
independent and identically distributed (i.i.d.), according to (\ref{exp_pdf}).
It is easy to check, simply applying the definition
in (\ref{suff_stat}), that a sufficient statistic in such a case is the scalar function:
\begin{equation}
T(\bx)=\sum_{i=1}^{N}B(\bx_i).
\end{equation}
This structure suggests that a simple distributed way to enable every node
in the network to estimate the vector $\btheta$ locally, without loss of
optimality with respect to the centralized approach, is to run a consensus
algorithm, where the initial state of every node is set equal to $B(\bx_i)$.
At convergence, if the network is connected, every node has a state equal to
the consensus value, i.e., $T(\bx)/N$. This enables every node to implement
the optimal estimation by simply interacting with its neighbors to achieve a consensus.
The only necessary condition for this simple method to work properly is that the network be connected.
This is indeed a very simple example illustrating how consensus can be
a fundamental step in deriving an optimal estimation through a purely decentralized
approach relying only upon the exchange of data among neighbors.\\

In the next two sections, we will analyze in more details the purely distributed case
(with no fusion center) where the global estimation can be carried out in any node
and the centralized case, where the final estimation is taken at the fusion center.

\subsection{Decentralized observations with decentralized estimation}
In the following we analyze different observation models and illustrate
alternative distributed estimation algorithms. We will start with
the conditionally independent case and then we will generalize
the approach to a conditionally dependent model.\\

\subsubsection{Conditionally independent observations}
A case amenable for finding distributed solutions
is given by the situations where the observations collected by different sensors
are conditionally independent. In such a case, the joint pdf $p_{X;\Theta}(\bx; \btheta)$
can be factorized as follows
\begin{equation}
p_{X;\Theta}(\bx; \btheta)=\prod_{i=1}^{N}p_{X_i;\Theta}(\bx_i; \btheta)
\end{equation}
where $p_{X_i;\Theta}(\bx_i; \btheta)$ is the pdf of the vector $\bx_i$
observed by node $i$. Taking the $\log$ of this expression, the
optimization problem can be cast, equivalently, as
\begin{equation}
\label{maxtheta}
\max_{\btheta} \,\sum_{i=1}^{N} \log p_{X_i;\Theta}(\bx_i; \btheta).
\end{equation}
Even if the objective function to be maximized is written as a sum
of functions depending each on a local observation vector, the
solution of the previous problem still requires a centralized approach
because the vector $\btheta$ to be estimated is common to all the terms.
A possible way to find a distributed solution to the problem in (\ref{maxtheta})
consists in introducing an instrumental common variable $\bz$ and
rewriting the previous problem in the following form
\begin{equation}
\begin{array}{lll}
\displaystyle \min_{\btheta_i} - \sum_{i=1}^{N} \log p_{X_i;\Theta}(\bx_i; \btheta_i)\\
\mbox{s.t.} \quad \quad \ds \btheta_i=\bz,\,\,\, i=1, 2, \ldots, N.
\end{array}\;
\label{min_local}
\end{equation}
This is a constrained problem,  whose Lagrangian is
\begin{equation}
\label{Lagrangian_local}
L(\btheta,\blambda, \bz):= \sum_{i=1}^{N} \left[-\log p_{X_i;\Theta}(\bx_i; \btheta_i)+\blambda_i^T (\btheta_i-\bz)\right],
\end{equation}
where $\blambda_i$ are the vectors whose entries are the Lagrange
multipliers associated to the equality constraints in (\ref{min_local}).
In many cases, it is useful to introduce the so called {\it augmented} Lagrangian \cite{SB-Bertsekas}:
\begin{equation}
\label{aug_Lagrangian_local}
L_{\rho}(\btheta,\blambda, \bz):= \sum_{i=1}^{N} \left[-\log p_{X_i;\Theta}(\bx_i; \btheta_i)+\blambda_i^T (\btheta_i-\bz)+\frac{\rho}{2}\|\btheta_i-\bz\|_2^2  \right],
\end{equation}
where $\rho$ is a {\it penalty} parameter. Minimizing the augmented Lagrangian
leads to the same solution as minimizing the original Lagrangian
because any feasible vector satisfying the linear constraint yields
a zero penalty. Nevertheless, there are some benefits in working
with the augmented Lagrangian, namely: i) the objective function is
differentiable under milder conditions than with the original Lagrangian; ii)
convergence can be achieved without requiring strict convexity of the
objective function (see \cite{SB-Bertsekas} for more insight into the
augmented Lagrangian method).

If the pdf's involved in (\ref{aug_Lagrangian_local}) are log-concave
functions of $\btheta$, the problem in (\ref{aug_Lagrangian_local})
is strongly convex and then it admits a unique solution and there are
efficient algorithms to compute the solution. Here, we are interested
in deriving decentralized solutions.\\

A possible method to find a distributed solution of the problem in (\ref{aug_Lagrangian_local})
is the {\it alternating direction method of multipliers (ADMM)} \cite{SB-Bertsekas}.
An excellent recent review of ADMM and its applications is \cite{boyd_now}.
The application of ADMM to distributed estimation problems was proposed
in \cite{SB-Schizas-Ribeiro-GG}. The method used in \cite{SB-Schizas-Ribeiro-GG}
relied on the introduction of the so called {\it bridge} nodes. Here, we will describe methods
that do not require the introduction of any special class of nodes (in principle,
every node has the same functionality as any other node). This is
useful to simplify the estimation method as well as network design and management.\\

The ADMM algorithm applied to  solve (\ref{aug_Lagrangian_local})
works through the following steps:
\begin{eqnarray}
\btheta_i[k+1]&=&{\rm arg}\min_{\btheta_i} \left\{-\log p_{X_i;\Theta}(\bx_i; \btheta_i)+\blambda_i^T[k] (\btheta_i-\bz[k])+\frac{\rho}{2}\|\btheta_i-\bz[k]\|_2^2  \right\},\nonumber\\
\bz[k+1]&=&{\rm arg}\min_{\bz} \sum_{i=1}^{N}
\left\{\blambda_i^T[k] (\btheta_i[k+1]-\bz)+\frac{\rho}{2}\|\btheta_i[k+1]-\bz\|_2^2  \right\},\nonumber\\
\blambda_i[k+1]&=&\blambda_i[k]+\rho \left( \btheta_i[k+1]-\bz[k+1]\right).
\end{eqnarray}
The first and second steps aim at minimizing the primal function (i.e., the augmented Lagrangian)
over the unknown variables $\btheta$ and $\bz$, for a given value of the
Lagrange multipliers' vectors $\blambda_i$, as computed in the previous iteration.

The third step is a dual variable update, whose goal is to maximize
the dual function, as in the dual ascent method. We recall that, in our case,
the dual function is defined as
\begin{equation}
g(\blambda)={\inf}_{\btheta, \bz} L_{\rho}(\btheta, \blambda, \bz).
\end{equation}
In ADMM, the dual ascent step uses a gradient ascent approach
to update $\blambda$ in order to maximize $g(\blambda)$,
for a given value of vectors $\btheta_i$ and $\bz$, with
the important difference that the step size used to compute
the update is exactly the penalty coefficient $\rho$.

In our case, the second step can be computed in closed form as follows
\begin{eqnarray}
\btheta_i[k+1]&=&{\rm arg}\min_{\btheta_i} \left\{-\log p_{X_i;\Theta}(\bx_i; \btheta_i)+\blambda_i^T[k] (\btheta_i-\bz[k])+\frac{\rho}{2}\|\btheta_i-\bz[k]\|_2^2  \right\},\label{a11}\\
\bz[k+1]&=&\frac{1}{N}\sum_{i=1}^{N}\left(\btheta_i[k+1]+\frac{1}{\rho}\blambda_i[k]\right),\label{a12}\\
\blambda_i[k+1]&=&\blambda_i[k]+\rho \left( \btheta_i[k+1]-\bz[k+1]\right)\label{a13}.
\end{eqnarray}
From this formulation, we can see that the first and third steps can be
run in parallel, over each node. The only step that requires an exchange
of values among the nodes is the second step that requires the
computation of an average value. But, as we know from previous section,
the average value can be computed through a distributed consensus algorithm.
The only condition for the convergence of consensus algorithm to the
average value is that the graph representing the links among the nodes
is connected.

The step in (\ref{a12}) can be further simplified
as follows. Let us denote with $\bar{x}$ the averaging operation across the nodes, i.e.
\begin{equation}
\overline{x}:=\frac{1}{N}\sum_{i=1}^{N} x_i.
\end{equation}
Using this notation, the $\bz$-update can be written as
\begin{equation}\label{bz}
\bz[k+1]=\overline{\btheta[k+1]}+\frac{1}{\rho} \overline{\blambda[k]}.
\end{equation}
Similarly, averaging over the $\blambda$-update yields
\begin{equation}\label{blambda}
\overline{\blambda[k+1]}=\overline{\blambda[k]}+\rho \left(\overline{\btheta[k+1]}-\bz[k+1]\right).
\end{equation}
Substituting (\ref{bz}) in (\ref{blambda}), it is easy to check that, after the first iteration,
$\overline{\blambda[k+1]}=0$. Hence, using $\bz[k]=\overline{\btheta[k]}$, the overall algorithm
proceeds as indicated in Table \ref{TableA1}.

\begin{table}[h]
\centering
\begin{tabular}{|l|}
 \hline
  \textbf{A.1}\\
  \hline
   \vspace{-.15cm}\\
   STEP 1: Set $k=0$, $\epsilon$ equal to a small positive value and initialize $\btheta_i[0]$,  $\blambda_i[0]$, $\forall i$, and $\bz$ randomly;\\
   \vspace{-.15cm}\\
   STEP 2: Compute $\btheta_i[1]$, $\forall i$ using (\ref{a11});\\
   \vspace{-.15cm}\\
   STEP 3: Run consensus over $\btheta_i[1]$ and $\blambda_i[0]$ to get $\overline{\btheta[1]}$ and $\overline{\blambda[0]}$;\\
   \vspace{-.15cm}\\
   STEP 4: Set  $\displaystyle\bz[1]=\overline{\btheta[1]}+\frac{1}{\rho}\,\overline{\blambda[0]}$;\\
   \vspace{-.15cm}\\
   STEP 5: Compute $\blambda_i[1]$, $\forall i$, using (\ref{a13});\\
   \vspace{-.15cm}\\
   STEP 6: Set $k=1$;\\
   \vspace{-.15cm}\\
   STEP 7: Repeat until convergence\\
   \vspace{-.15cm}\\
  \hspace{1.3cm}$\displaystyle\btheta_i[k+1]={\rm arg}\min_{\btheta_i} \left\{-\log p_{X_i;\Theta}(\bx_i; \btheta_i)+
    \blambda_i^T[k] (\btheta_i-\overline{\btheta[k]})+\frac{\rho}{2}\|\btheta_i-\overline{\btheta[k]}\|_2^2  \right\}$ \hspace{1.1cm}\tagarray\label{primal_dual}\\
  \vspace{-.15cm}\\
  \hspace{1.3cm}Run consensus over $\btheta_i[k+1]$ until convergence;\\
  \vspace{-.15cm}\\
  \hspace{1.3cm}$\blambda_i[k+1]=\blambda_i[k]+\rho \left( \btheta_i[k+1]-\overline{\btheta[k+1]}\right)$ \hspace{5.5cm}\tagarray\label{admm2} \\
  \vspace{-.15cm}\\
  \hspace{1.3cm}Set $k=k+1$, if convergence criterion is satisfied stop, otherwise go to step 7.\\
  \\
   \hline
\end{tabular}
\caption{Algorithm A.1}
\label{TableA1}
\end{table}

The convergence criterion used in the steps of the algorithm is based on the relative absolute difference at two successive iterations: Given a sequence $\by[k]$, the algorithm stops when $\|\by[k+1]-\by[k]\|/\|\by[k]\|\le \epsilon$, with $\epsilon$ a small positive value.\\

Equations (\ref{primal_dual})-(\ref{admm2}) give rise to an interesting interpretation: the primal update (first equation) aims at implementing a local optimization, with a penalty related to the disagreement between the local solution and
the global one; the dual update (second equation) aims at driving all the local solutions to converge to a common (consensus) value, which coincides with the globally optimal solution.\\

The straightforward implementation of (\ref{primal_dual})-(\ref{admm2}) requires running, at each step $k$ of the ADMM algorithm, a consensus algorithm. A possible alternative approach can be envisaged by reformulating
the optimization problem as follows:
\begin{eqnarray}
\min_{\btheta_i} & & \left\{-\sum_{i=1}^{N} \log p_{X_i;\Theta}(\bx_i; \btheta_i)+\sum_{i=1}^{N}\sum_{j\in{\cal N}_i}
\blambda_{ij}^T(\btheta_j-\btheta_i)+\frac{\rho}{2}\sum_{i=1}^{N}\sum_{j\in{\cal N}_i}
\|\btheta_j-\btheta_i\|^2\right\}\\
& &\mbox{s.t.} \quad \quad \ds \btheta_j=\btheta_i; \forall j \in {\cal N}_i;\,\,\, i=1, 2, \ldots, N,\nonumber
\label{min_consensus_augmented}
\end{eqnarray}
where ${\cal N}_i$ denotes the set of node $i$'s neighbors. To make more clear the
interaction among the nodes, it is useful to introduce the graph notation, as in previous section.
Using the adjacency matrix $\bA$, the previous problem can be rewritten as follows:
\begin{eqnarray}
\label{min_consensus_augmented3}
\min_{\btheta_i} & &\left\{ -\sum_{i=1}^{N} \log p_{X_i;\Theta}(\bx_i; \btheta_i)+\sum_{i=1}^{N}\sum_{j=1}^{N}a_{ij}
\blambda_{ij}^T(\btheta_j-\btheta_i)+\frac{\rho}{2}\sum_{i=1}^{N}\sum_{j=1}^{N}a_{ij}
\|\btheta_j-\btheta_i\|^2\right\}\\
& &\mbox{s.t.} \quad \quad \ds \btheta_j=\btheta_i; \forall j \in {\cal N}_i;\,\,\, i=1, 2, \ldots, N.\nonumber
\end{eqnarray}
This formulation does not require the introduction of the instrumental variable $\bz$.
We keep enforcing the constraint that all the local estimates $\btheta_i$
converge to the same value. However, the penalty is now formulated as the disagreement
between the local estimates. From consensus algorithm, we know that
nulling the disagreement is equivalent to forcing all the vectors $\btheta_i$
to reach the same value if the graph describing the interactions among the nodes is connected.
Hence, if the network is connected, at convergence, the disagreement goes to zero and there is no bias
resulting from the introduction of the disagreement penalty.

The formulation in (\ref{min_consensus_augmented}) is more amenable for
an implementation that does not require, at any step of the algorithm,
the convergence of consensus algorithms. In fact, applying ADMM
to the solution of (\ref{min_consensus_augmented3}) yields the algorithm
described in Table \ref{TableA2}.\\

\begin{table}[h]
\centering
\begin{tabular}{|l|}
 \hline
  \textbf{A.2}\\
  \hline
   \vspace{-.15cm}\\
   STEP 1: Set $k=0$, and initialize $\btheta_i[0]$,  $\blambda_{ij}[0]$, $\forall i$, $j \in {\cal N}_i$;\\
   \vspace{-.15cm}\\
   STEP 2: Repeat until convergence\\
   \vspace{-.15cm}\\
  \hspace{.5cm} $\displaystyle\btheta_i[k+1]={\rm arg}\min_{\btheta_i} \left\{ -\sum_{i=1}^{N} \log p_{X_i;\Theta}(\bx_i; \btheta_i)+\sum_{i=1}^{N}\sum_{j=1}^{N}a_{ij}\blambda_{ij}^T[k](\btheta_j[k]-\btheta_i)+\frac{\rho}{2}\sum_{i=1}^{N}\sum_{j=1}^{N}a_{ij}
  \|\btheta_j[k]-\btheta_i\|^2\right\}$\\
  \hspace{13.5cm}\tagarray\label{primal_dual2}\\
  \vspace{-.15cm}\\
  \hspace{.5cm}$\blambda_{ij}[k+1]=\blambda_{ij}[k]+\rho a_{ij}\left( \btheta_j[k+1]-\btheta_i[k+1]\right)$ \hspace{6.3cm}\tagarray\label{admm3}\\
  \vspace{-.15cm}\\
  \hspace{.5cm}Set $k=k+1$, if convergence criterion is satisfied stop, otherwise go to step 2.\\
  \\
   \hline
\end{tabular}
\caption{Algorithm A.2}
\label{TableA2}
\end{table}

Some examples of applications are useful to grasp
the main features of these algorithms.\\

\subsubsection{Distributed ML estimation under Gaussian noise}
Let us consider the common situation where the
measured vector $\bx_i \in  \mathbb{R}^{Q}$ is related
to the parameter vector $\btheta \in  \mathbb{R}^{M}$, with $Q\ge M$, through a linear observation model, as:
\begin{equation}
\label{linear_model}
\bx_i=\bA_i \btheta+ \bv_i,\,\, i=1, \ldots, N
\end{equation}
where $\bA_i \in  \mathbb{R}^{Q \times M}$ and $\bv_i$ is a vector of jointly Gaussian random variables with zero mean and covariance matrix $\bC_i$, i.e. $\bv_i \sim {\cal N}(\bzero, \bC_i)$.

In such a case, algorithm {\bf A.1} in (\ref{primal_dual}) simplifies as the first step of
(\ref{primal_dual}) can be expressed in closed form
\begin{eqnarray}
\label{primal_dual_Gaussian}
\btheta_i[k+1]&=&\left(\bA_i^T \bC_i^{-1} \bA_i+\rho \bI\right)^{-1}\,\left(\bA_i^T \bC_i^{-1} \bx_i-\blambda_i[k]+\rho \overline{\btheta[k]}\right),\nonumber\\
\blambda_i[k+1]&=&\blambda_i[k]+\rho \left( \btheta_i[k+1]-\overline{\btheta[k+1]}\right).
\end{eqnarray}
The two updates can be computed in parallel by all the nodes, after having
computed the average values through the consensus algorithm.\\

\begin{figure}[t]
\begin{tabular}{c}
\centerline{{\includegraphics[width=8cm]{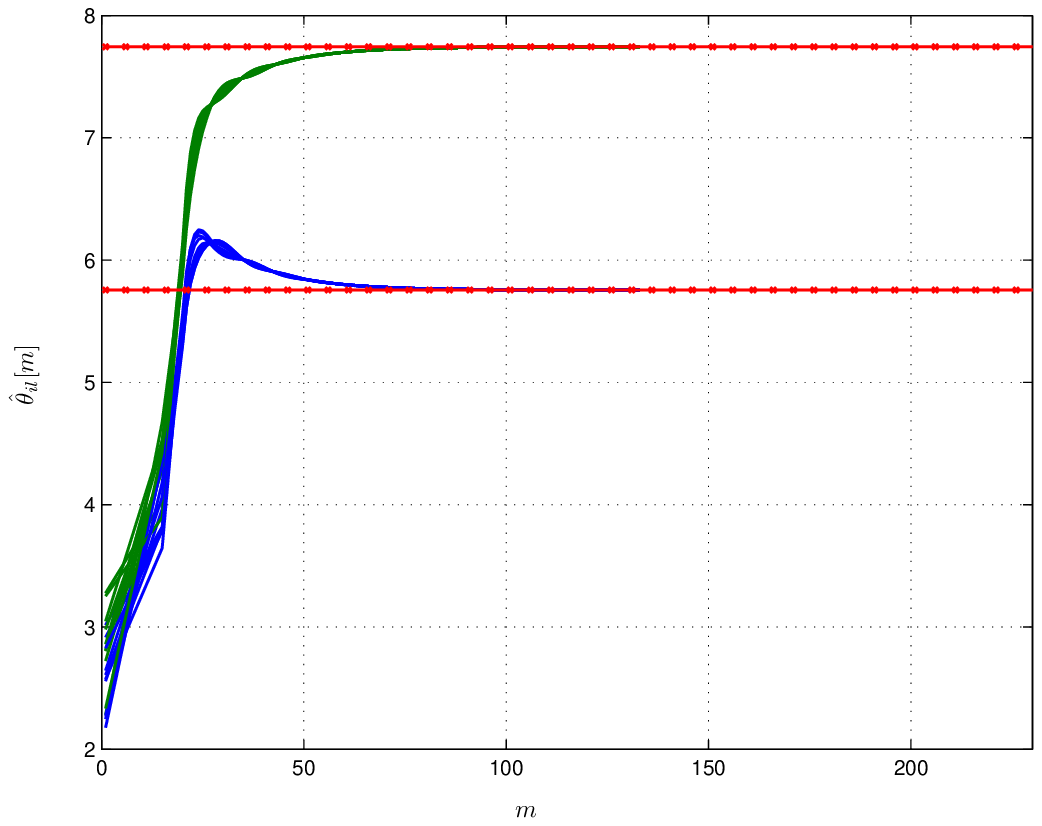}}\hspace{.5cm}{\includegraphics[width=8cm]{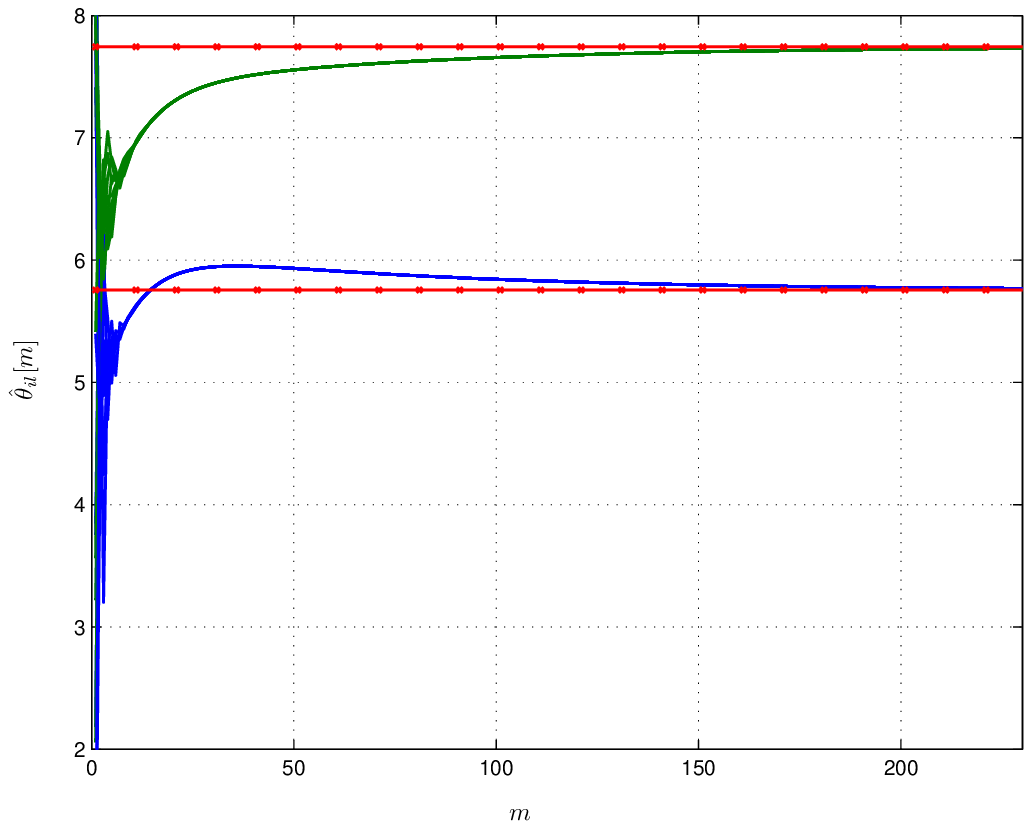}}}\\
(a)\hspace{7cm}(b)
\end{tabular}
\caption{Per node parameter estimation versus the iteration index
$m$ using algorithms {\bf A.1} (left) and {\bf A.2} (right).}
\label{fig:ADMM_local_cons}
\end{figure}

Alternatively, algorithm {\bf A.2} becomes
\begin{align}
\label{primal_dual_Gaussian_seq}
\btheta_i[k+1]&=\left(\bA_i^T \bC_i^{-1} \bA_i+ 2 \rho \sum_{j=1}^{N} a_{ij} \bI\right)^{-1}\,\left(\bA_i^T \bC_i^{-1} \bx_i+ \sum_{j=1}^{N} a_{ij} (\blambda_{ij}[k]-\blambda_{ji}[k])+2 \rho \sum_{j=1}^{N} a_{ij}\btheta_j[k]\right),\nonumber\\
\blambda_{ij}[k+1]&=\blambda_{ij}[k]+\rho a_{ij} \left( \btheta_j[k+1]-\btheta_i[k+1]\right).
\end{align}
In this case, there is no need of running the consensus algorithm for every iteration. Some numerical results are useful to compare the methods. As an example, we considered a connected network composed of $N=10$ sensors. We set $\rho=30$ and assumed an
observation vector of size $Q=30$.
In Fig. \ref{fig:ADMM_local_cons} we report the estimates
$\hat{\theta}_{i,l}$, for $l=1,2$, versus the iteration index $m$,
for the two algorithms {\bf A.1} (left plot) and {\bf A.2} (right plot). The iteration index
$m$ includes also the iterations necessary for the consensus algorithm
to converge within a prescribed accuracy (in this case, we stopped the
consensus algorithm as soon as the absolute difference between
two consecutive updates is below of $10^{-2}$ for all the nodes).
In both figures, we report, as a benchmark, the maximum likelihood estimate
(red line) achievable by a centralized node that knows all the observation vectors and
all the model parameters, i.e. $\bA_i, \bC_i, \forall i$. From Fig.\ref{fig:ADMM_local_cons}, we can see that the estimates
obtained with both methods converge to the optimal ML estimates.

In the specific case where the observation model is as in (\ref{linear_model}),
with additive Gaussian noise, and the noise vectors pertaining to different
sensors are mutually uncorrelated,  the globally optimal ML estimate is
\begin{equation}
\label{ML}
\hat{\btheta}_{ML}=\left(\sum_{i=1}^{N}\bA_i^T \bC_i^{-1} \bA_i\right)^{-1}\left(\sum_{i=1}^{N}\bA_i^T \bC_i^{-1} \bx_i\right).
\end{equation}
This formula is a vector weighted sum of the observations. Recalling that consensus
algorithms, if properly initialized, can be made to converge to a weighted sum of the
initial states, we can use the consensus algorithm directly to compute
the globally optimal ML estimate through a totally distributed mechanism.
In particular, in this case, the consensus algorithm proceeds as in Table \ref{TableA3}.\\
\begin{table}[h]
\centering
\begin{tabular}{|l|}
 \hline
  \textbf{A.3}\\
  \hline
   \vspace{-.15cm}\\
   STEP 1: Set $k=0$, and initialize $\btheta_i[0]=\left(\bA_i^T \bC_i^{-1} \bA_i\right)^{-1}\left(\bA_i^T \bC_i^{-1} \bx_i\right)$;\\
   \vspace{-.15cm}\\
   STEP 2: Repeat until convergence\\
   \vspace{-.15cm}\\
   \hspace{1.3cm}  $\displaystyle\btheta_i[k+1]=\btheta_i[k]+\epsilon \left(\bA_i^T \bC_i^{-1} \bA_i\right)^{-1}\,\sum_{j=1}^{N} a_{ij}\left(\btheta_j[k]-\btheta_i[k]\right)$ \hspace{1.85cm}\tagarray\label{vec_cons}\\
   \vspace{-.15cm}\\
   \hspace{1.3cm}Set $k=k+1$, if convergence criterion is satisfied stop, otherwise go to step 2.\\
  \\
   \hline
\end{tabular}
\caption{Algorithm A.3}
\label{TableA3}
\end{table}

Using again the basic properties of consensus algorithm, if the graph
is connected and the step size $\epsilon$ is sufficiently small, the iterations in
(\ref{vec_cons}) converge to the globally optimal estimate (\ref{ML}).\\

\subsubsection{Distributed Bayesian estimation under Gaussian noise and Laplacian prior}
Let us consider now the case where the parameter vector is a random
vector with known prior probability density function.
Following a Bayesian approach, as in (\ref{random_parameter}),
the practical difference is that in such a case
the objective function must include a term depending on the prior
probability.
Let us consider, for instance, the interesting case
where the observation is Gaussian, as in the previous example,
and the prior pdf is Laplacian, i.e.
\begin{equation}
p_{\Theta}(\btheta)=\mu\,\exp(-\mu \|\btheta\|_1)
\end{equation}
with $\mu>0$, where $\|\bx\|_1$ denotes the $l_1$ norm of vector $\bx$.
In this case, the problem to be solved is the following
\begin{equation}
\min_{\btheta} \left\{\sum_{i=1}^{N} \|\bx_i-\bA_i\btheta\|^2_{C_i^{-1}}\,+\,\mu \|\btheta\|_1\right\},
\end{equation}
where $\|\bx\|^2_{A}$ denotes the weighted $l_2$ norm of $\bx$,
i.e. $\|\bx\|^2_{A}:=\displaystyle \frac{\bx^T \bA \bx}{2}$.

Interestingly, this formulation coincides with the formulation resulting from
having no prior pdf, but incorporating an $l_1$ norm in order to drive the solution
towards a sparse vector. This is the so called {\it least-absolute shrinkage and selection operator
(lasso)} method \cite{SB-tibshirani}. A distributed algorithm to solve
a linear regression problem with sparsity constraint was proposed in
\cite{Mateos-Bazerque-GG}. Here we provide a similar approach, with
the important difference that, in each iteration, the update is computed in closed form.
A decentralized solution can be found by reformulating the
problem as follows
\begin{eqnarray}
\min_{\btheta_i} & & \left\{\sum_{i=1}^{N} \|\bx_i-\bA_i\btheta_i\|^2_{C_i^{-1}}+ \frac{\rho}{2}\sum_{i=1}^{N}
\|\btheta_i-\bz\|^2\,+\,\mu \|\bz\|_1\right\},\nonumber\\
{\rm s.t.} & & \btheta_i=\bz,\,\,\, i=1, \ldots, N.
\end{eqnarray}
Using the ADMM approach, the algorithm proceeds through the following updates
\begin{eqnarray}
\label{primal_dual_Gaussian_l1norm}
\btheta_i[k+1]&=&\left(\bA_i^T \bC_i^{-1} \bA_i+\rho \bI\right)^{-1}\,\left(\bA_i^T \bC_i^{-1} \bx_i-\blambda_i[k]+\rho \bz[k]\right),\nonumber\\
\bz[k+1]&=&{\rm arg}\min_{\bz} \left\{\mu \|\bz\|_1+\frac{\rho}{2}\sum_{i=1}^{N}\|\btheta_i[k+1]-\bz\|^2
+\sum_{i=1}^{N}\blambda_i^T(\btheta_i[k+1]-\bz)\right\}\nonumber\\
\blambda_i[k+1]&=&\blambda_i[k]+\rho \left( \btheta_i[k+1]-\bar{\bz}[k+1]\right).
\end{eqnarray}
The second equation can also be expressed in closed form. Moreover,
defining the vector threshold function $\bt_{\mu}(\bx)$ as the vector
whose entries are obtained by applying the scalar thresholding function
$t_{\mu}(x)$ to each element of vector $\bx$, where
\begin{equation}
\label{soglia}
t_{\mu}(x)=\left\{ \begin{array}{lll}
 x-\mu ,&  x>\mu\\
 0, & -\mu \le x \le \mu\\
 x+\mu, & x<-\mu
\end{array}\right.
\end{equation}
the overall algorithm is as in Table \ref{TableA4}.
\begin{table}[h]
\centering
\begin{tabular}{|l|}
 \hline
  \textbf{A.4}\\
  \hline
   \vspace{-.15cm}\\
   STEP 1: Set $k=0$, and initialize $\btheta_i[0]$,  $\blambda_i[0]$, $\forall i$, and $\bz[0]$ randomly;\\
   \vspace{-.15cm}\\
   STEP 2: Repeat until convergence\\
   \vspace{-.15cm}\\
  \hspace{1.3cm}$\displaystyle\btheta_i[k+1]=\left(\bA_i^T \bC_i^{-1} \bA_i+\rho \bI\right)^{-1}\,\left(\bA_i^T \bC_i^{-1} \bx_i-\blambda_i[k]+\rho \bz[k]\right)$ \hspace{2.92cm}\tagarray\label{primal_dual_Gaussian_l1norm_explicit}\\
  \vspace{-.15cm}\\
  \hspace{1.3cm}Run consensus over $\btheta_i[k+1]$ and $\blambda_i[k]$ to get $\overline{\btheta[k+1]}$ and $\overline{\blambda[k]}$ until $\epsilon$-convergence;\\
  \vspace{-.15cm}\\
  \hspace{1.3cm}$\displaystyle\bz[k+1]=\frac{1}{\rho N}\,t_{\mu}\left(N \overline{\blambda[k]} +\rho N \overline{\btheta[k+1]}\right)$ \hspace{5.45cm}\tagarray\label{primal_dual_Gaussian_l1norm_explicit2} \\
  \vspace{-.15cm}\\
  \hspace{1.3cm}$\blambda_i[k+1]=\blambda_i[k]+\rho \left( \btheta_i[k+1]-\bar{\bz}[k+1]\right)$ \hspace{5.02cm}\tagarray\\
  \vspace{-.15cm}\\
  \hspace{1.3cm}Set $k=k+1$, if convergence criterion is satisfied stop, otherwise go to step 2.\\
  \\
   \hline
\end{tabular}
\caption{Algorithm A.4}
\label{TableA4}
\end{table}

\begin{figure}[h]
\centering
\includegraphics[width=8.5cm]{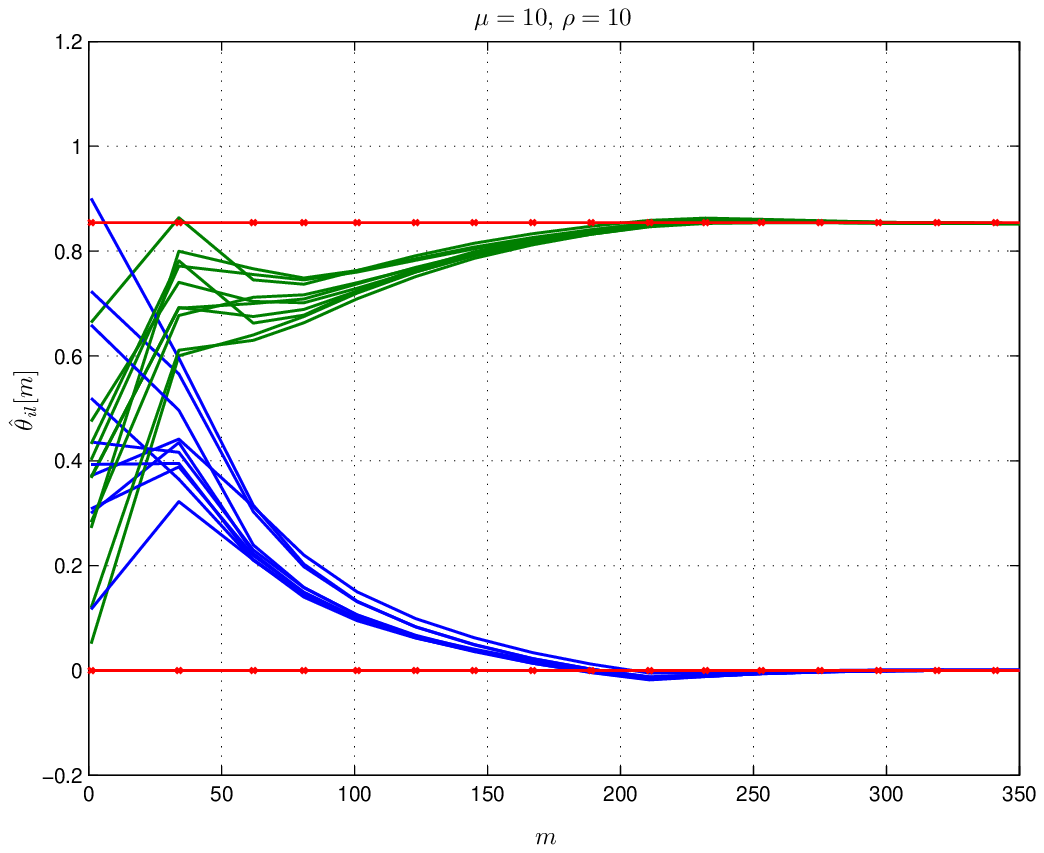}
\caption{Per node  estimated variable versus the iteration index
$m$ for distributed Bayesian estimation using the ADMM approach.}
\label{fig:ADMM_lasso}
\end{figure}

As a numerical example, in Fig. \ref{fig:ADMM_lasso}  we report
the behavior of the estimated variable obtained using
Algorithm {\bf A.4} versus the iteration index $m$, which includes the
convergence times of two consensus algorithms in the
equation (\ref{primal_dual_Gaussian_l1norm_explicit2}).
The example refers to a network of $N=10$ nodes, using $\rho=\mu=10$. The constant red line represents the centralized optimal solution.
The parameter vector of this example has two components, one of which
has been set to zero to test the capability to recover the sparsity.
We can notice from Fig. \ref{fig:ADMM_lasso} that, as expected, the
algorithm converges to the globally optimal values.

\begin{figure}[t]
\centering
\includegraphics[width=8.5cm]{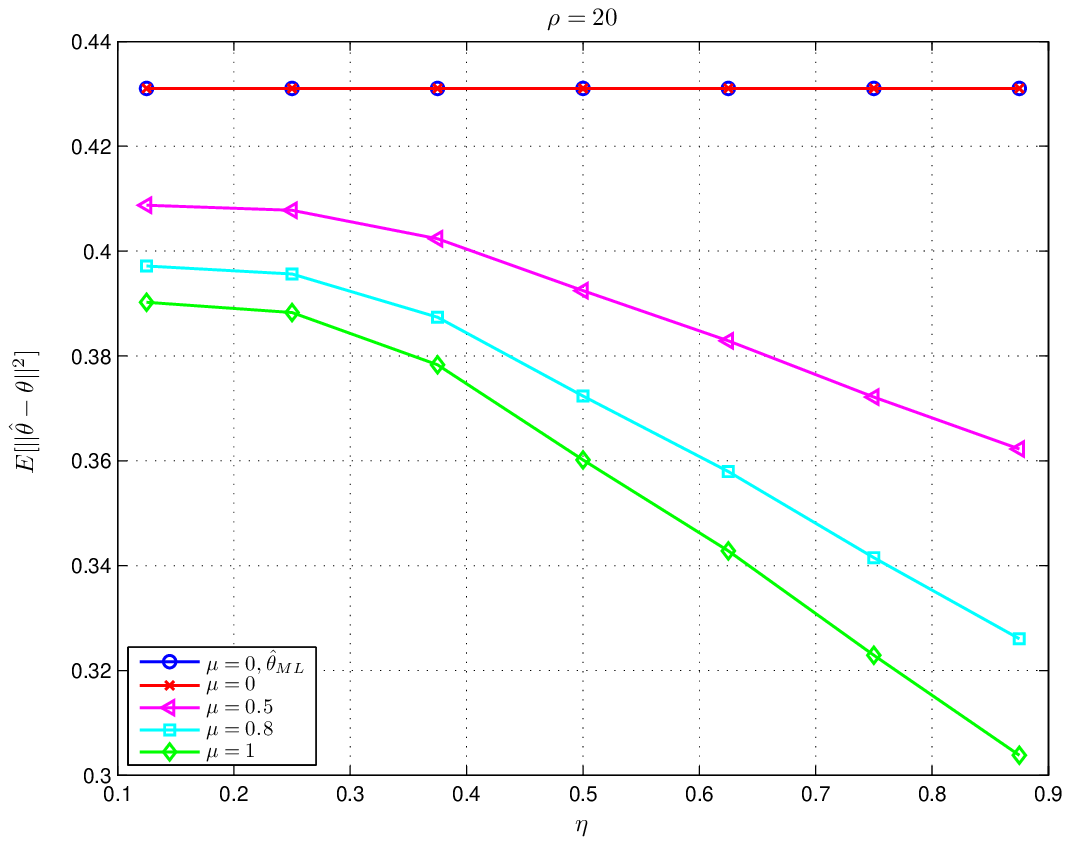}
\caption{Mean square estimation error versus the fraction $\eta$ of null entries,
for different $\mu$ values.} \label{fig:ADMM_lasso_aver}
\end{figure}

To show the impact of the penalty coefficient $\mu$ on the sparsity of the estimated vector, in Fig. \ref{fig:ADMM_lasso_aver} we have reported the average mean square estimation error versus the coefficient $\eta$, defined
as the fraction of zeros entries in the vector
$\mathbf{\theta}$ to be estimated, for different values of
$\mu$. It can be noted from Fig. \ref{fig:ADMM_lasso_aver} that for
$\mu=0$ the optimal solution is independent by $\eta$ and it
coincides with the optimal (centralized) ML solution. Furthermore, we
can observe that, as $\mu$ and  $\eta$ increase, the average
estimation error decreases thanks to the recovering sparsity
property of the ADMM approach with the lasso constraint.\\

\subsubsection{Distributed recursive least square estimation with sparsity constraint}
In some applications, the parameters to be estimated may be changing over time.
In these cases, it is more advisable to adopt recursive procedure rather than the
batch approach followed until now. We show now how to obtain a distributed
recursive least square (RLS) estimation incorporating a sparsity constraint.

Let us assume a linear observation model
\begin{equation}
\label{rls-model}
\bx_i(l)=\bA_i(l) \btheta + \bv_i(l),
\end{equation}
where $\bx_i(l)$ denotes the observation taken by node $i$ at time $l$, $\bA_i(l)$ is
a known, possibly time-varying, mixing matrix and $\bv_i(l)$
is the observation noise, supposed to have zero mean and covariance matrix $\bC_i(l)$.

In RLS estimation with a sparsity constraint, the goal is to find the parameter vector
$\btheta$, at each time instant $n$, that minimizes the following objective function
\begin{equation}
\sum_{i=1}^{N}\sum_{l=1}^{n} \beta^{n-l}\|\bx_i(l)-\bA_i(l)\, \btheta\|_{C_i^{-1}(l)}^2+\mu \|\btheta\|_1,
\end{equation}
where $0<\beta\le 1$ is a forgetting factor used to weight more
the most recent observations with respect to the older ones. The coefficient
$\mu$ weights the importance of the sparsity constraint.

Proceeding as in the previous examples, a distributed solution  can be found
by formulating the problem, at each time $n$, as a constrained problem
incorporating an instrumental variable $\bz$ to force all the nodes to converge to a common
estimate. The problem can be made explicit as
\begin{eqnarray}
\min_{\btheta_i} & & \sum_{i=1}^{N}\sum_{l=1}^{n} \beta^{n-l}\|\bx_i(l)-\bA_i(l)\, \btheta_i\|_{C_i^{-1}(l)}^2+ \frac{\rho}{2}\sum_{i=1}^{N} \|\btheta_i-\bz\|^2+\mu \|\bz\|_1,\nonumber\\
{\rm s.t.} & & \btheta_i=\bz,\,\,\, i=1, \ldots, N.
\end{eqnarray}
Again, the solution can be achieved by applying ADMM and the result is
given by the algorithm described in Table \ref{TableA5}.

\begin{table}[h]
\centering
\begin{tabular}{|l|}
 \hline
  \textbf{A.5}\\
  \hline
   \vspace{-.15cm}\\
   STEP 1: Set $n=0$ and $k=0$, and initialize $\btheta_i[0,0]$,  $\blambda_i[0,0]$, $\forall i$, and $\bz[0,0]$ randomly;\\
   \vspace{-.15cm}\\
   STEP 2: Repeat until convergence over index $k$\\
   \vspace{-.15cm}\\
  \hspace{1cm}   $\displaystyle\btheta_i[k+1, n]=\left(\sum_{l=1}^{n} \beta^{n-l}\bA_i^T(l) \bC_i^{-1}(l) \bA_i(l)+\rho \bI\right)^{-1}\,\left(\sum_{l=1}^{n} \beta^{n-l}\bA_i^T(l) \bC_i^{-1}(l) \bx_i(l)-\blambda_i[k, n]+\rho \bz[k, n]\right)$ \hspace{.5cm}\\
   \vspace{-.15cm}\\
  \hspace{1.1cm}Run consensus over $\btheta_i[k+1,n]$ and $\blambda_i[k,n]$ to get $\overline{\btheta[k+1,n]}$ and $\overline{\blambda[k,n]}$ until convergence;\\
  \vspace{-.15cm}\\
  \hspace{1.1cm}$\displaystyle\bz[k+1, n]=\frac{1}{\rho N}\,t_{\mu}\left(N \overline{\blambda[k, n]} +\rho N \overline{\btheta[k+1, n]}\right)$ \hspace{5.45cm}\tagarray\label{RLS_l1norm_explicit}\\
  \vspace{-.15cm}\\
  \hspace{1.1cm}$\blambda_i[k+1, n]=\blambda_i[k, n]+\rho \left( \btheta_i[k+1, n]-\bar{\bz}[k+1,n]\right)$\\
  \vspace{-.15cm}\\
  \hspace{1.1cm}Set $k=k+1$, if convergence criterion is satisfied set $n=n+1$ and go to step 2, otherwise \\
  \hspace{1.1cm}go to step 2.\\
  \\
   \hline
\end{tabular}
\caption{Algorithm A.5}
\label{TableA5}
\end{table}

As before, the only step requiring the interaction among the nodes
is a consensus algorithm to be run to compute the averages appearing in
(\ref{RLS_l1norm_explicit}).

{To test the convergence of Algorithm {\bf A.5},
we considered a possible application to cooperative sensing
for cognitive radio. We assumed the presence of a macro base station
transmitting using a multicarrier scheme. We considered for simplicity of representation
four channels, but the method can be easily extended to a larger number of channels.
The sensing nodes aim to recover the activity of the macro transmitter, represented by
a vector $\btheta$ composed of four entries, one for each channel.
To improve the accuracy of the local estimation, the sensors
cooperate with each other by running Algorithm {\bf A.5}.
At some time, the activity level switches from on to off or viceversa.
As an example, in Fig. \ref{fig:RLS_L_4} we report the four parameters to be estimated,
indicated by the red lines. At time $n=50$, the parameters switch to test the tracking
capability of the proposed method.

In Fig. \ref{fig:RLS_L_4} we draw also the estimated parameters
$\hat{\theta}_{l}$   for $l=1,\ldots, 4$ versus the current observation index $n$.
We used  $\beta=0.6$, $\rho=40$ and two values of the
sparsity coefficient: $\mu=0$ and $\mu=40$.
We can notice from Fig. \ref{fig:RLS_L_4} that the method is able to track
the true parameters. It is also interesting to see that,
as the penalty coefficient $\mu$ increases,
the zero coefficients are estimated with greater accuracy. Conversely, the positive coefficients
are recovered with a slightly larger bias.

\begin{figure}[h]
\begin{tabular}{c}
\centerline{{\includegraphics[width=7cm]{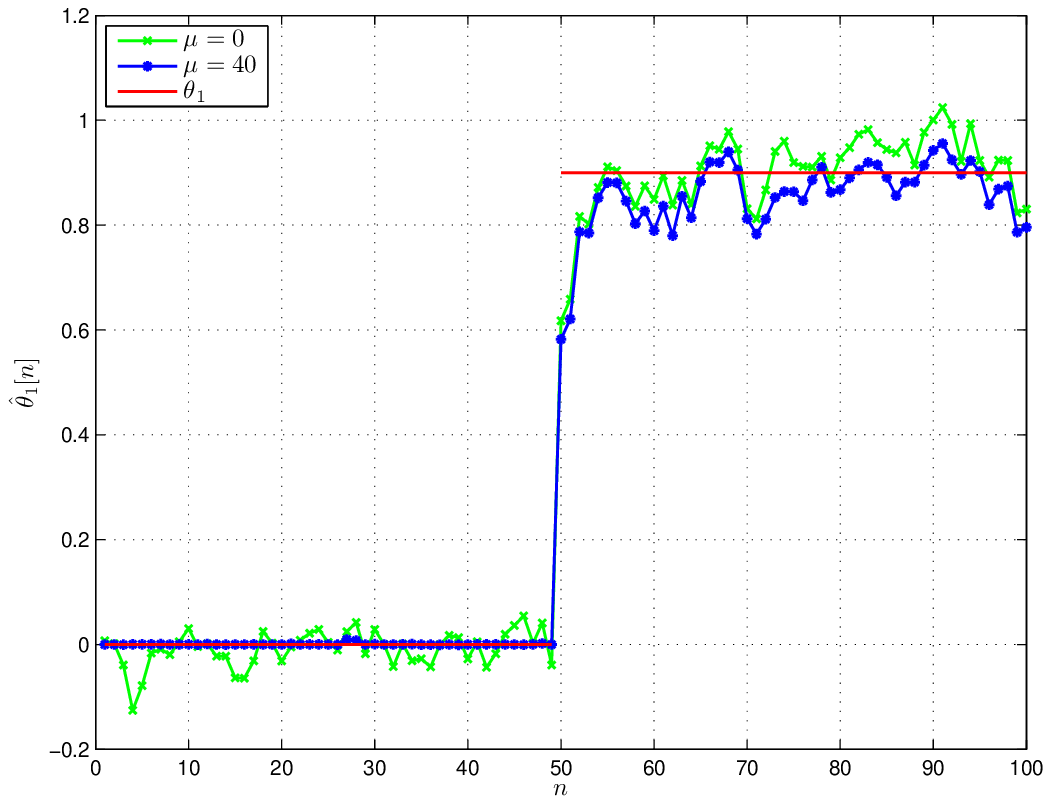}}{\includegraphics[width=7cm]{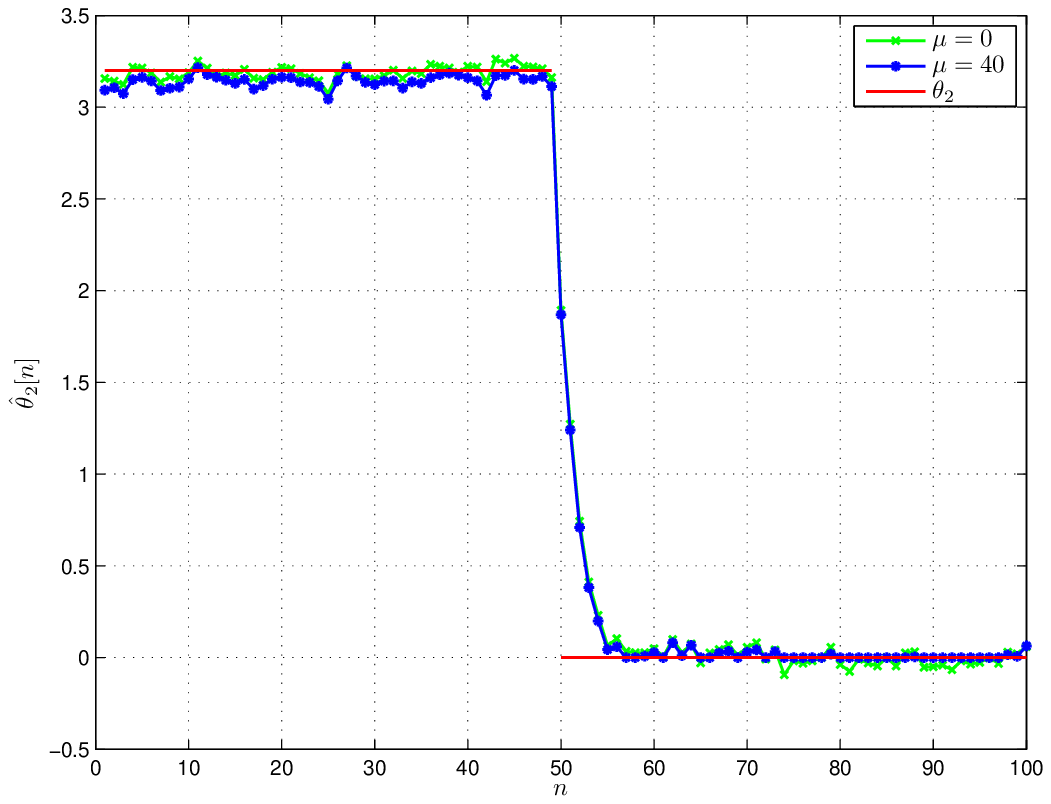}}}\\
\centerline{{\includegraphics[width=7cm]{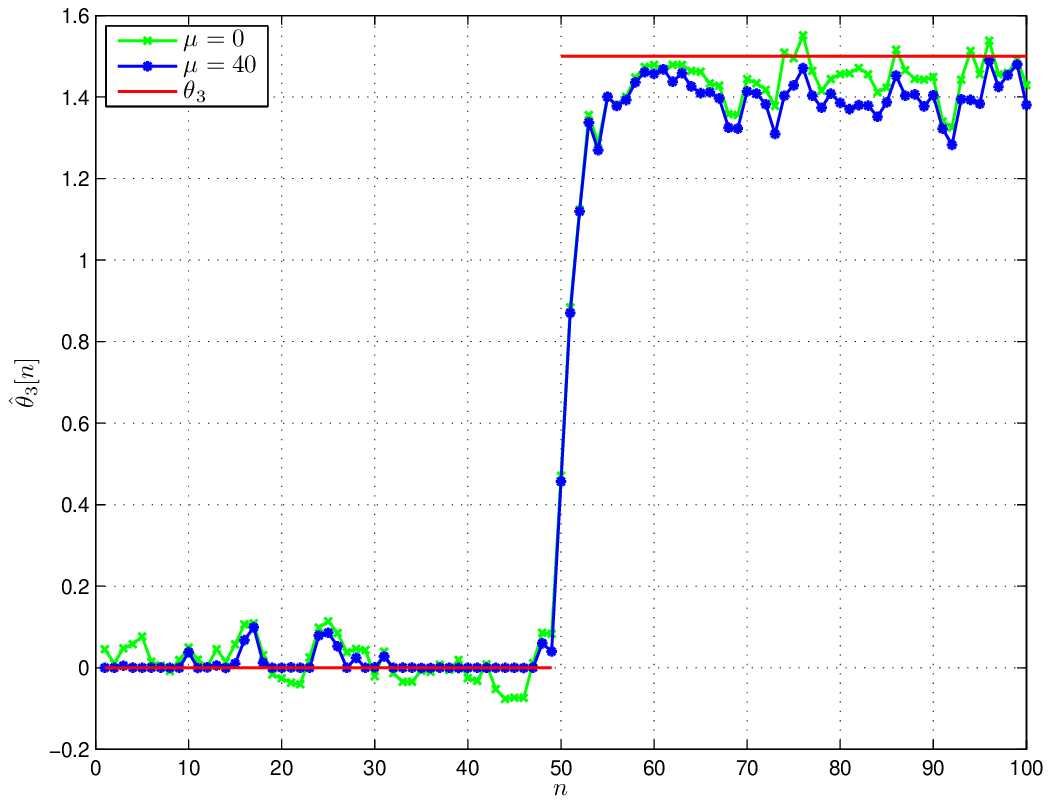}}{\includegraphics[width=7cm]{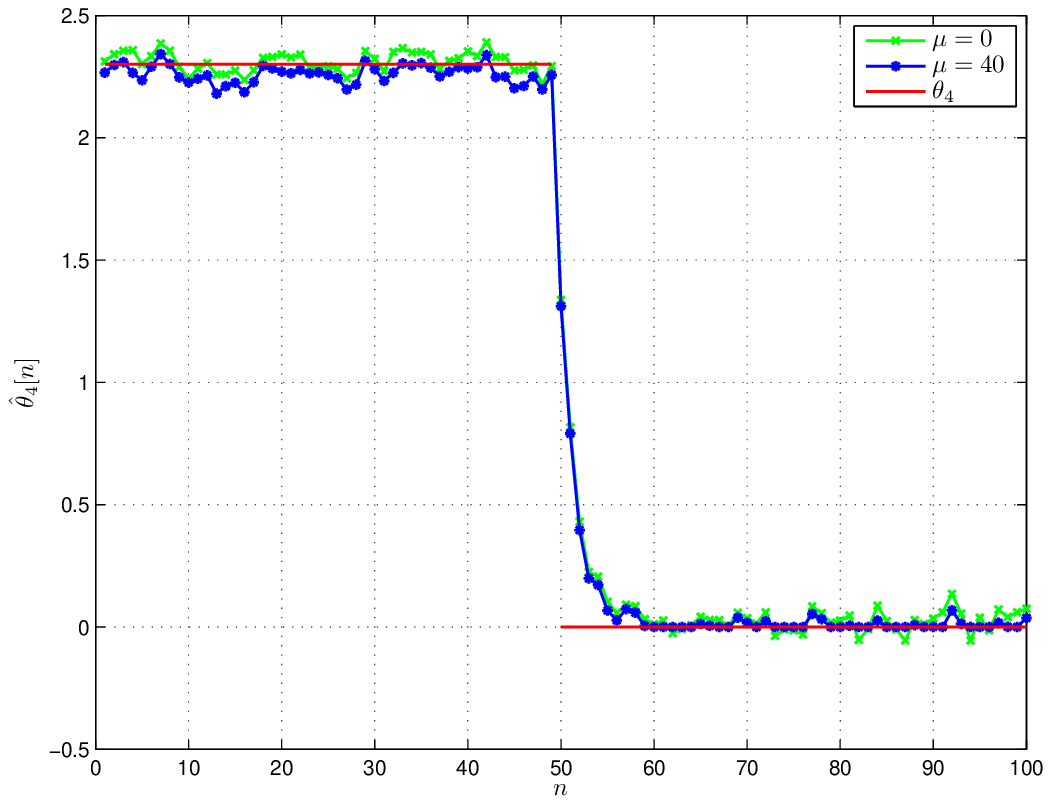}}}
\end{tabular}
\caption{ Estimated parameters versus the number of current observations $n$ for the RLS algorithmc  assuming $N=10$,
$\beta=0.6$ and $\rho=40$.} \label{fig:RLS_L_4}
\end{figure}
\begin{figure}
\centering
\includegraphics[width=10cm]{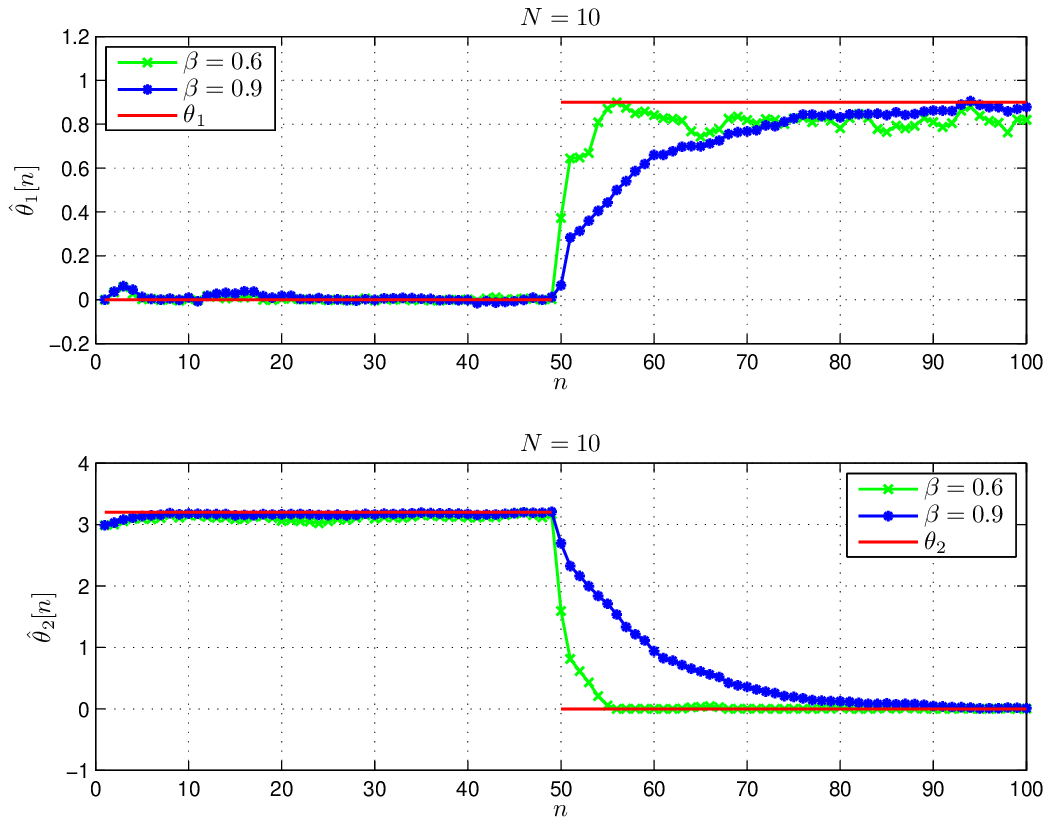}
\caption{Parameter estimation versus the number of observations
$n$ of the recursive least square estimation using the ADMM
approach, considering $N=10$ and two different values of $\beta$.}
\label{fig:ADMM_RLS1_6}
\end{figure}
\begin{figure}
\centering
\includegraphics[width=10cm]{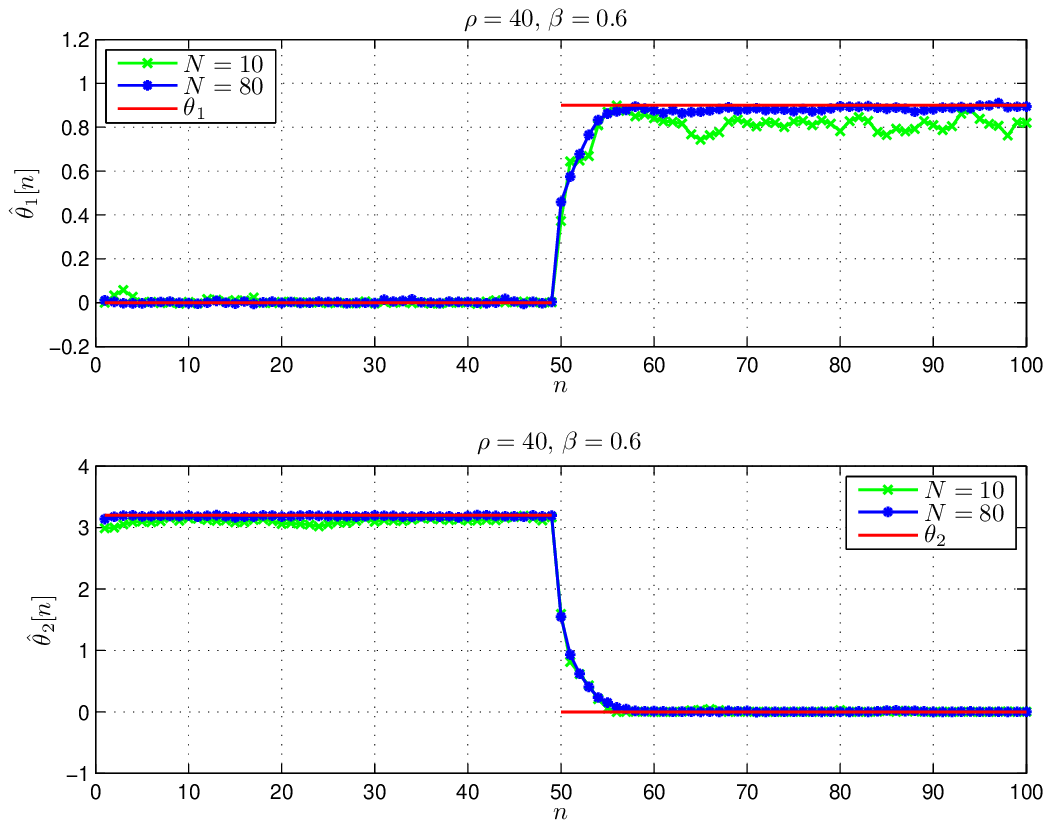}
\caption{Parameter estimation versus the number of observations
$n$ of the recursive least square estimation using the ADMM
approach, considering $\beta=0.6$, $\mu=40$ and two different values of
$N$.} \label{fig:ADMM_RLS1_7}
\end{figure}

To evaluate the impact of the forgetting
factor $\beta$ on the accuracy and tracking capability of the
distributed RLS method, in Fig. \ref{fig:ADMM_RLS1_6}  we
reported the estimated parameters using $\beta=0.6$ and $\beta=0.9$,
having set $\rho=\mu=40$. It can be noted that, as $\beta$
increases, the larger memory of the filter yields more accurate estimates.
At the same time, having a larger memory implies slower time to reaction
to the parameter switch, as evidenced in Fig.  \ref{fig:ADMM_RLS1_6}.

For any given forgetting factor $\beta$, the only possibility to
improve the estimation accuracy is to have more nodes sensing
a common macro base station. As an example, in Fig. \ref{fig:ADMM_RLS1_7}
we report the behavior of the estimates obtained with different number of nodes,
for a forgetting factor $\beta=0.6$. We can notice that,  as expected, increasing
the number of nodes, the estimation accuracy increases as well. This
reveals a trade-off between forgetting factor (time memory) and
number of nodes involved in cooperative sensing.}\\

\subsubsection{Distributed parameter estimation in spatially correlated observations}
So far, we have analyzed the case of conditionally independent observations. Let us
consider now the case where the observation noise is spatially correlated.
More specifically, we assume here the following observation model, for each sensor
\begin{equation}
x_i=\theta + v_i, \,\,\, i=1, \ldots, N,
\end{equation}
where the noise variables $v_i$ are jointly Gaussian with zero mean
and covariance matrix $\bC$, or precision matrix $\bA=\bC^{-1}$.
Furthermore, we assume that $\bv$ is a Gaussian Markov random field,
so that the precision matrix is typically a sparse matrix. The joint pdf
of the observation vector can then be written as in (\ref{gmrf}), i.e.,
\begin{equation}
\label{gmrf2}
p(\bx; \theta)=\sqrt{\frac{|\bA|}{(2\pi)^N }}\,\exp\left[-\frac{1}{2}(\bx-\theta\buno)^T\bA\, (\bx-\theta\buno)\right]:= \sqrt{\frac{|\bA|}{(2\pi)^N }}\, \exp\left[-V(\bx)\right]
\end{equation}
where $V(\bx)$ can be rewritten as follows
\begin{equation}
V(\bx)=\sum_{i=1}^{N}\phi_i(\bx_i; \theta)
\end{equation}
with $\mathbf{x}_i =[x_i, \{x\}_{j \in {\cal N}_i, j>i}]^T$, and
\begin{equation}
\phi_i(\bx_i; \theta):=\frac{1}{2}\,a_{ii}(x_i-\theta)^2+\sum_{j\in{\cal N}_i, j>i}a_{ij} (x_j-\theta)
(x_i-\theta).
\end{equation}
As in the previous cases, also here a decentralized solution
can be reached by formulating the problem as the minimization of the augmented Lagrangian
\begin{equation}
\label{aug_Lagrangian_local2}
L_{\rho}(\btheta,\blambda, \bz):= \sum_{i=1}^{N} \left\{\phi_i(\bx_i; \theta)+\lambda_i (\theta_i-z)+\frac{\rho}{2}(\theta_i-z)^2  \right\},
\end{equation}
subject to $\theta_i=z$. Applying the ADMM algorithm to this case, we
get the following algorithm\\
\begin{eqnarray}
\label{admm_corr_local}
\theta_i[k+1]&=&{\rm arg}\min_{\theta} \left[\phi_i(\bx_i;\theta_i)+\lambda_i(\theta_i-z)+\frac{\rho}{2}\left(\theta_i-z\right)^2\right],\\
z[k+1]&=&\frac{1}{N}\sum_{i=1}^{N}\left(\theta_i[k+1]+\frac{1}{\rho}\lambda_i[k]\right),\nonumber\\
\lambda_i[k+1]&=&\lambda_i[k]+\rho \left( \theta_i[k+1]-z[k+1]\right).\nonumber
\end{eqnarray}
It is important to notice that, in this case, even if the global problem
concerning the minimization of the augmented Lagrangian in (\ref{aug_Lagrangian_local2})
is certainly convex, the local problem in (\ref{admm_corr_local}) is not necessarily convex because
there is no guarantee that the term $\phi_i(\bx_i; \theta)$ is a positive definite function.
Nevertheless, the quadratic penalty present in  (\ref{admm_corr_local})
can make every local problem  in  (\ref{admm_corr_local}) convex.
At the same time, at convergence the penalty goes to zero and thus
it does not induce any undesired bias on the final result.

The first step in (\ref{admm_corr_local}) can be made explicit, so that
the algorithm assumes the form described in Table \ref{TableA6}.

\begin{table}[h]
\centering
\begin{tabular}{|l|}
 \hline
  \textbf{A.6}\\
  \hline
   \vspace{-.15cm}\\
   STEP 1: Set $k=0$, and initialize $\theta_i[0]$,  $\lambda_i[0]$, $\forall i$, and $z[0]$ randomly;\\
   \vspace{-.15cm}\\
   STEP 2: Repeat until convergence\\
   \vspace{-.15cm}\\
  \hspace{1.3cm}$\displaystyle\theta_i[k+1]=\frac{1}{a_{ii}+\rho+2\sum_{j\in{\cal N}_i, j>i}a_{ij}}\left(\rho z[k]-\lambda_i[k]+ a_{ii}x_i +  \sum_{j\in{\cal N}_i, j>i}a_{ij}(x_i+x_j)\right)$ \hspace{.5cm}\tagarray\label{admm_corr}\\
  \vspace{-.15cm}\\
  \hspace{1.3cm}Run consensus over $\theta_i[k+1]$ and $\lambda_i[k]$ to get $\overline{\theta[k+1]}$ and $\overline{\lambda[k]}$ until $\epsilon$-convergence;\\
  \vspace{-.15cm}\\
  \hspace{1.3cm}$\displaystyle z[k+1]=\overline{\theta[k+1]}+\frac{1}{\rho}\overline{\lambda[k]}$ \hspace{8cm}\tagarray \\
  \vspace{-.15cm}\\
  \hspace{1.3cm}$\lambda_i[k+1]=\lambda_i[k]+\rho \left( \theta_i[k+1]-z[k+1]\right)$ \hspace{6.05cm}\tagarray\\
  \vspace{-.15cm}\\
  \hspace{1.3cm}Set $k=k+1$, if convergence criterion is satisfied stop, otherwise go to step 2.\\
  \\
   \hline
\end{tabular}
\caption{Algorithm A.6}
\label{TableA6}
\end{table}

\begin{figure}[t]
\centering
\includegraphics[width=8cm]{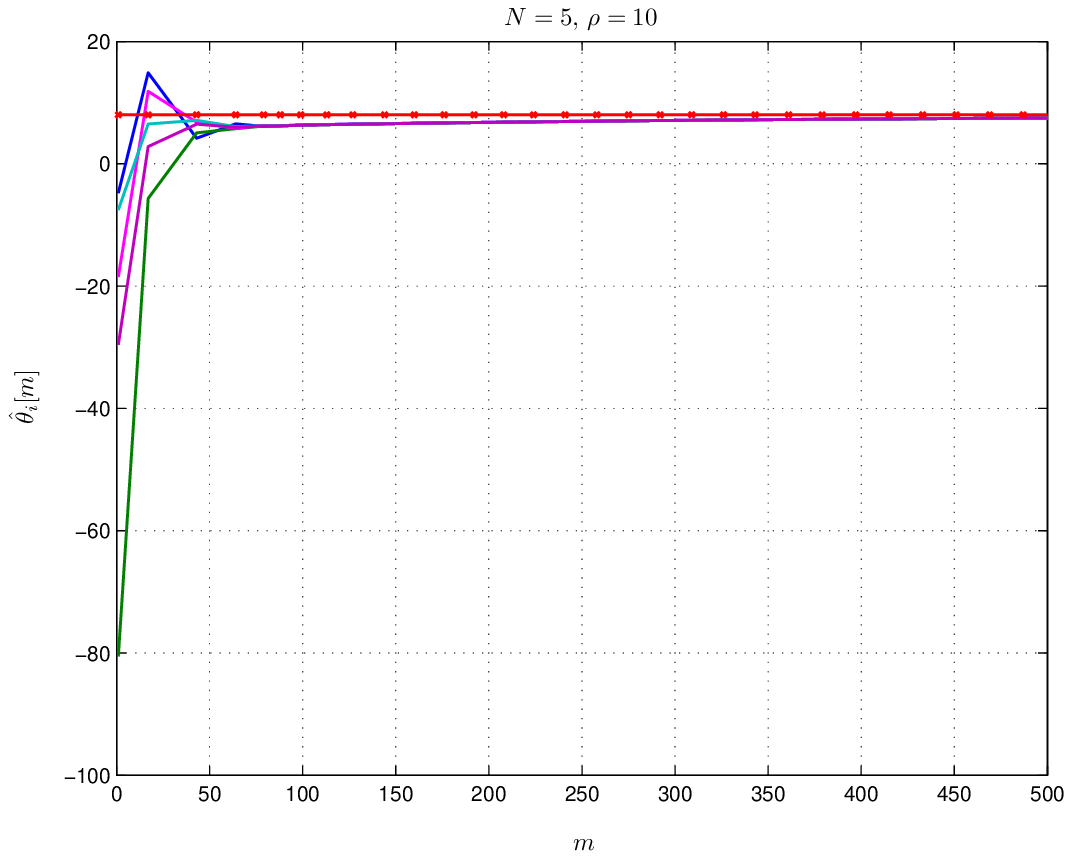}
\caption{Per node parameter estimation versus the iteration index
$m$ for spatially correlated observations using the ADMM
approach.} \label{fig:GMF}
\end{figure}

As an example, in Fig. \ref{fig:GMF} we report the
estimation versus the cumulative iteration index $m$ that
includes the consensus steps and the iterations over $k$. The results
refer to a connected network with $N=5$ nodes; $\rho$ has been chosen equal to $10$ to
guarantee that every local problem is convex. It can be
noticed from Fig. \ref{fig:GMF} that  the distributed solution
converges to the optimal centralized  solution (red line).

\subsection{Decentralized observations with centralized estimation}
In many cases, the observations are gathered in distributed form,
through sensors deployed over a certain area, but the decision
(either estimation or detection) is carried out in a central fusion center.
In this section, we review some of the problems related to
distributed estimation, with centralized decision.
In such a case, the measurements gathered by the sensors
are sent to a fusion center through rate-constrained physical channels.
The question is how to design the quantization step in each sensor in order
to optimize some performance metric related to the estimation of the
parameter of interest. Let us start with an example, to introduce the basic
issues.

Let us consider a network of $N$ sensors, each observing a value $x_k$ containing
a deterministic parameter $\theta$, corrupted by additive  noise $v_k$, i.e.
\beq
x_k=\theta+ v_k, \quad  \quad k=1,\ldots,N.
\eeq
The  noise variables  $v_k$ are supposed to be zero mean spatially uncorrelated random variables
with variance $\sigma^2_{k}$.
Suppose that the sensors transmit their observations
via some orthogonal multiple access scheme to a control center which
wishes to estimate the unknown signal $\theta$
by minimizing the estimation mean square error (MSE) $E[(\hat{\theta}-\theta)^2]$.
In the ideal case, where the observations are unquantized and received by the control center without
distortion, the best linear unbiased estimator (BLUE)  can be performed by the control center
and the estimate $\hat{\theta}$
is given by
\beq
\hat{\theta}=\left( \ds \sum_{k=1}^{N} \frac{1}{\sigma^2_{k}}\right)^{-1} \ds \sum_{k=1}^{N} \frac{x_k}{\sigma^2_{k}}
\eeq
with MSE given by
$E[\left( \hat{\theta}- \theta \right)^2 ]=\left( \ds \sum_{k=1}^{N} \frac{1}{\sigma^2_{k}}\right)^{-1}$.
This estimator coincides with the maximum likelihood estimator when the noise variables
are jointly Gaussian and uncorrelated.

Let us consider now the realistic case, where each sensor quantizes
the observation $x_k$ to generate a discrete message $m_k$ of $n_k$ bits.
Assuming an error-free transmission, the fusion center must then
provide an estimate $\hat{\theta}$ of the true parameter, based on the
messages $m_k$ transmitted by all the nodes.
More specifically, assuming a uniform  quantizer
which generates unbiased message functions,
the estimator at the control center performs a linear combination of the received
messages.
Let us suppose that the unknown signal to be estimated belongs to the range
$[-A, A]$ and each sensor uniformly divides the range  $[-A, A]$
 into $2^{n_k}$ intervals of length $W_k=2 A/2^{n_k}$ rounding $x_k$
to the midpoint  of these intervals.
In this case, the quantized value $m_k$ at the $k$-the sensor
can be written as $m_k=\theta+v_k+w_k$,
 where the quantization noise $w_k$
 is independent of $v_k$. It can be proved that $m_k$
 is an unbiased estimator of $\theta$
 with
 \beq
 \mbox{Var}\{m_k\}\leq \delta^2_k +\sigma^2_k
 \eeq
where $\delta^2_k$ denotes an upper bound on the quantization noise variance and is given by
\beq \delta^2_k=\ds \frac{W_{k}^2}{12}=\ds \frac{A^2}{3 \cdot 2^{2 n_k}}\;. \label{bound} \eeq
A linear unbiased estimator of $\theta$ is \cite{Xiao-Cui-Luo-Goldsmith}
 \beq
\hat{\theta}=\left(\ds \sum_{k=1}^{N} \ds
\frac{1}{\sigma^2_k+\delta^2_k}\right)^{-1} \ds \sum_{k=1}^{N} \ds
\frac{m_k}{\sigma^2_k+\delta^2_k}. \label{lin_est}
 \eeq
This estimate yields an MSE upper bound
 \beq
 E[\left( \hat{\theta}-\theta \right)^2]
 \leq  \left(\ds \sum_{k=1}^{N}
\ds \frac{1}{\sigma^2_k+\delta^2_k}\right)^{-1} \; . \label{upper_bound}
\eeq
As mentioned before, the previous strategy assumes that there are
no transmission errors. This property can be made as close as possible
to reality by enforcing  the transmission rate of sensor $k$ to be strictly less than
the channel capacity from sensor $k$ to the fusion center.
If we denote by $p_k$ the transmit power of sensor $k$, $h_k$
the channel coefficient between sensor $k$
and control node and $N_0$ is the noise variance at the control node receiver,
the bound on transmit rate guaranteeing an arbitrarily small error probability is
\beq
n_k \leq \ds \frac{1}{2}\log\left(1+\ds \frac{p_k h_k^2}{N_0}\right).
\label{capacity}
\eeq
The problem is then how to allocate power and bits over each channel in order
to fulfil some optimality criterion dictated by the estimation problem.
This problem was tackled in \cite{Xiao-Cui-Luo-Goldsmith} where it
was proposed the minimization of the Euclidean norm of the transmit power vector
under the constraint that the estimation variance is upper bounded by a given quantity
and that the number of bits per symbol is less than the channel capacity.
Here we formulate the problem as the minimization of the total transmit power
under the constraint that the final MSE be upper bounded by a given quantity $\epsilon >0$.
From (\ref{capacity}), defining $a_k=\ds \frac{h_k^2}{N_0}$,
we can derive the number of quantization level as a function of the transmit power,\footnote{We neglect here
the discretization of $n_{k}$, to simplify the problem and arrive at
closed form expressions.}
\beq
2^{2 n_k}=\left(1+p_k a_k\right) \;. \label{rel_n_p}
\eeq
Our aim is  to minimize the sum of powers transmitted by all the sensors
under the constraint
\beq
\left(\ds \sum_{k=1}^{N}
\ds \frac{1}{\sigma^2_k+\delta^2_k}\right)^{-1} \leq \epsilon \; . \label{constraint}
\eeq
Denoting with $\bp=[p_1,\ldots, p_N]$ the power vector,
the optimization problem can be formulated as
\beq
\begin{array}{lll}
\ds \min_{\bp} \quad \quad \ds \sum_{k=1}^{N} p_k\\
\mbox{s.t.} \quad \quad \ds \sum_{k=1}^{N} \ds \frac{1}{
\sigma^2_{k}+\ds \frac{A^2}{3 \cdot 2^{2n_k}}} \geq \ds
\frac{1}{\epsilon}\\
\quad \quad \quad \bp \geq \mathbf{0}
\end{array}\; \label{opt1} \;
\eeq
where $n_k$ is a function of $p_k$, as in (\ref{capacity}).
In practice, the values $n_k$ are integer. However, searching for the optimal
integer values $n_k$ leads to an integer programming problem. To relax the
problem, we assume that the variables $n_k$ are real.
Then, by using  (\ref{rel_n_p}), the optimization problem in (\ref{opt1})  can be formulated as
\beq
\begin{array}{lll}
\ds \min_{\bp} \quad \quad \ds \sum_{k=1}^{N} p_k\\
\mbox{s.t.} \quad \quad \ds \sum_{k=1}^{N} \ds \frac{1}{
\sigma^2_{k}+\ds \frac{A^2}{3 (1+p_k a_k)}} \geq \ds
\frac{1}{\epsilon}\\
\quad \quad \quad \bp \geq \mathbf{0}
\end{array} \quad (\mathcal{P})\;. \label{opt_app}
\eeq
Problem  $(\mathcal{P})$ is indeed a convex optimization problem
and it is feasible if $\ds \sum_{k=1}^{N} \ds \frac{1}{\sigma^2_{k}}> \frac{1}{\epsilon}$.\\

The optimal solution of the convex  problem  $(\mathcal{P})$ can be found
by imposing the KKT conditions of $(\mathcal{P})$, i.e.,
\beq
\begin{array}{lll}
1-\mu_k-\lambda \ds \frac{3 A^2 a_k}{[3 \sigma^2_{k}(1+p_k a_k)+A^2]^2}=0  \quad \forall \, k=1,\ldots, N\\
0 \leq \lambda \perp \ds \sum_{k=1}^N \frac{3(1+p_k a_k)}{3 \sigma^2_{k}(1+p_k a_k)+A^2} -\ds \frac{1}{\epsilon}\geq 0\\
0 \leq \mu_k \perp p_k \geq 0 \quad \forall \, k=1,\ldots, N\\
\end{array} \;  \label{KKT}
\eeq
where $\lambda$ and $\mu_k$ denote the Lagrangian multipliers associated to the $N+1$
constraints. The solution for the optimal powers turns out to be
 \beq
 p^{*}_k=\left[\frac{1}{\sigma^2_{k}}\sqrt{\ds \frac{\lambda A^2}{3 a_k}}
 -\ds \frac{1}{a_k}-\ds \frac{A^2}{3 a_k \sigma^2_{k} }\right]^{+} \label{final_power}
\eeq
where $(x)^{+}=\max(0,x)$ and $\lambda>0$
is found by imposing the MSE constraint to be valid with equality.\\

\begin{figure}
\centering
\includegraphics[width=8.5cm]{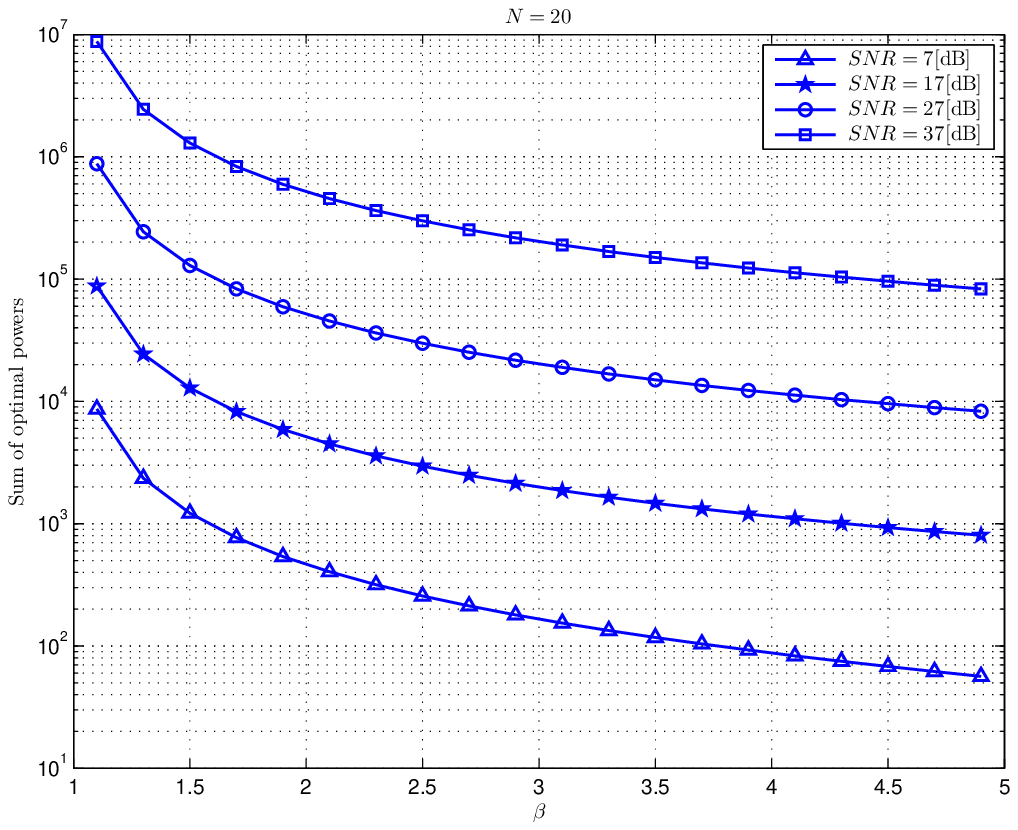}
\caption{Sum of the optimal powers for  problem $(\mathcal{P})$
 versus $\beta$ for several values of $\sigma^2_k$.}
\label{fig:fig1cap}
\end{figure}

It is now useful to present  some numerical results. To guarantee the
existence of a solution, we set the bound $\epsilon=\beta \epsilon_{min}$
with $\beta>1$ and $\epsilon_{min}=\left(\ds \sum_{k=1}^{N}
\ds \frac{1}{\sigma^2_k}\right)^{-1}$.
In Fig. \ref{fig:fig1cap} we report the sum of the optimal transmit powers vs. $\beta$, for different SNR values.
The number of sensors is $N=20$.
\begin{figure}
\centering
\includegraphics[width=8.5cm]{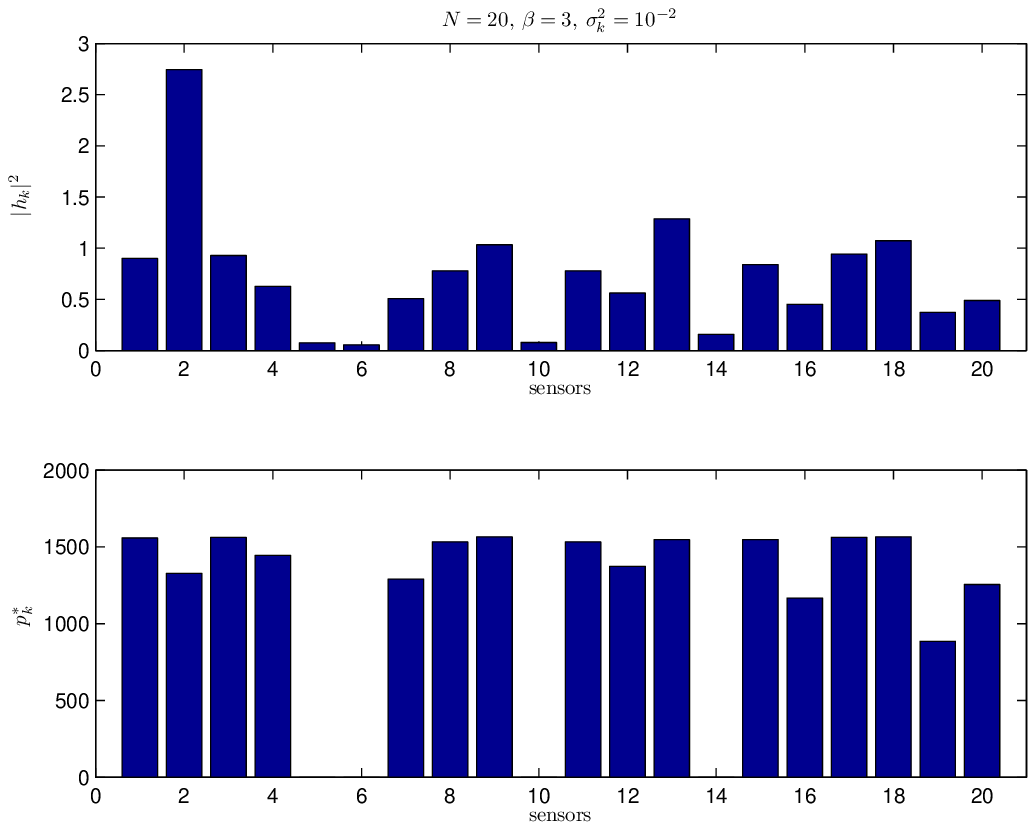}
\caption{Optimal power allocation of the sensors  for  problem $(\mathcal{P})$, fixing the per-node observation noise variance.}
\label{fig:fig4cap}
\end{figure}
\begin{figure}
\centering
\includegraphics[width=8.5cm]{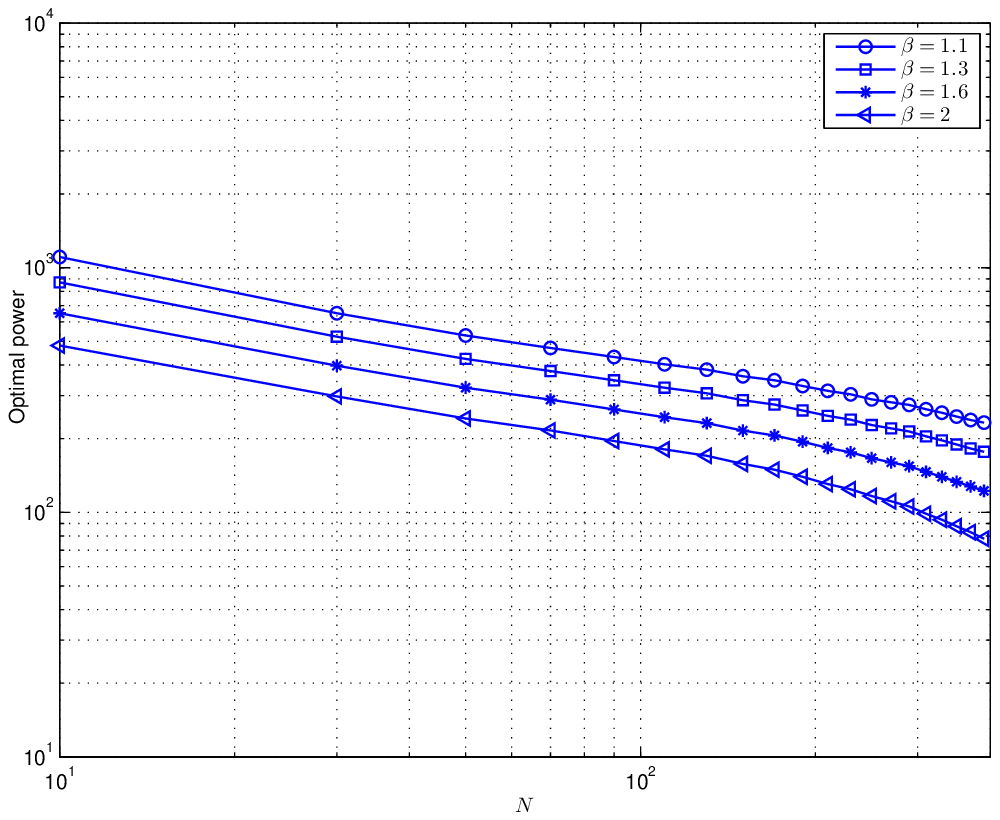}
\caption{Sum of optimal powers  versus $N$ for  several values of $\beta$.}
\label{fig:figvarN}
\end{figure}
We can  notice that  the minimum transmit power increases for smaller values of $\beta$,
i.e. when we require the realistic system to perform closer and closer to the ideal
communication case.

In the bottom subplot of Fig. \ref{fig:fig4cap}
we report an example of optimal power allocation  obtained by solving the optimization problem $(\mathcal{P})$,
corresponding to the channel realization shown in the top subplot, assuming
a constant observation noise variance $\sigma^2_k=0.01$.
We can observe that the solution is that only the nodes with the best channels coefficients
are allowed to transmit.
Finally, in Fig. \ref{fig:figvarN} we plot the sum of the optimal transmit powers  versus the number
of sensors $N$, for different  values of  $\beta$. We can see that, as $N$ increases, a lower power is necessary to achieve the desired estimation variance, as expected.

\section{Distributed detection}
\label{Decentralized detection}
The distributed detection problem is in general more difficult to handle than the estimation problem.
There is an extensive literature on distributed detection problem, but there is still
a number of open problems. According to decision theory, an ideal centralized detector having
error-free access to all the measurements collected by a set of
nodes, should form the likelihood ratio and compare it with a
suitable threshold \cite{SB-kay-II}. Denoting with ${\cal H}_0$ and
${\cal H}_1$ the two alternative hypotheses, i.e. absence or
presence of the event of interest, and with $p(\bx_1, \ldots,
\bx_N; {\cal H}_i)$ the joint probability density function of the
whole set of observed data, under the hypothesis ${\cal H}_i$,
many decision tests can be cast as threshold strategies where the likelihood
ratio (LR) is compared with a threshold $\gamma$, which depends on the decision
criterion. This is true, for example, for two important formulations leading
to the Bayes approach and to the Neyman-Pearson criterion, the only difference
between the two's being the values assumed by the threshold $\gamma$.
The detection rule decides for ${\cal H}_1$
if the threshold is exceeded or for ${\cal H}_0$, otherwise. In
formulas,
\begin{equation}
\label{LR-1}
\Lambda\left(\bx\right):=\Lambda\left(\bx_1, \ldots, \bx_N\right)
=\frac{p(\bx_1, \ldots, \bx_N; {\cal H}_1)}{p(\bx_1, \ldots, \bx_N; {\cal H}_0)}
\overset{\mathcal{H}_{1}}{\underset{\mathcal{H}_{0}}{\gtreqless}}  \gamma \;.
\end{equation}
Ideally, with no communication constraints, every node should then
send its observation vector $\bx_i$, with $i=1, \ldots, N$ to the fusion center,
which should then use all the received vectors to implement the LR test, as in (\ref{LR-1}).
In reality, there are intrinsic limitations due to, namely: a) the finite number of bits
with which every sensor has to encode the measurements before transmission; b) the maximum
latency with which the decision has to be taken; c)  the finite capacity of the channel between
sensors and fusion center. The challenging problem is then how to devise an optimum
decentralized detection strategy taking into account the limitations imposed
by the communication over realistic channels. The global problem, in the most general
setting, is still an open problem, but there are many works in the literature
addressing some specific cases. The interested reader may check the book \cite{Varshney-book}
or the excellent tutorial reviews given in
\cite{Viswanathan-Varshney-PartI}, \cite{Blum-Kassam-Poor-PartII}, \cite{Chamberland-Veeravalli07}.
The situation becomes more complicated when we take explicitly into account the
capacity bound imposed by the communication channel and we look for the
number of bits to be used to quantize the local observations before transmitting
to the fusion center. This problem was addressed in \cite{Chamberland-Veeravalli03},
\cite{Chamberland-Veeravalli04}, where it was shown that binary quantization
is optimal for the problem of detecting deterministic signals in Gaussian noise and for detecting
signals in Gaussian noise using a square-law detector. The interesting indication,
in these contexts, is that the gain offered by having more sensor nodes outperforms
the benefits of getting more detailed (nonbinary) information from each sensor.
A general framework to cast the problem of decentralized detection is the one
where the topology describing the exchange of information among sensing nodes
is not simply a tree, with all nodes sending data to a fusion center, but it is
a  graph. Each node is assumed to transmit a finite-alphabet symbol to its
neighbors and the problem is how to find out the encoding (quantization)
rule on each node. A class of problems admitting a message passing algorithm
with provable convergence properties was proposed in \cite{Kreidl-Willsky}.
The solution is a sort of distributed fusion protocol, taking explicitly into
account the limits on the communication resources.
An interesting and well motivated observation model is a correlated random field,
as in many applications the observations concern physical quantities,
like temperature or pressure, for example, which, being subject to
diffusion processes, are going to be spatially and temporally
correlated.  One of the first works addressing the detection of a known
signal embedded in a correlated Gaussian noise was \cite{Willett-Swaszek-Blum}.
Using large deviations theory, the authors of \cite{Chamberland-Veeravalli06}
study the impact of node density, assuming that observations become
increasingly correlated as sensors are in closer proximity of each other.
More recently, the detection of a Gauss-Markov Random field (GMRF) with nearest-neighbor
dependency was studied in \cite{Anandkumar-Tong-Swami}. Scaling laws for the energy consumption
of optimal and sub-optimal fusion policies were then presented in \cite{Anandkumar-Yukich-Tong-Swami}.
The problem of energy-efficient routing of sensor observations from a Markov random field
was analyzed in \cite{Anandkumar-Tong-Swami-ciss07}.\\

A  classification of the various detection algorithms depends on the
adopted criterion.
A first important classification is the following:
\begin{enumerate}
\item{Global decision is taken at the fusion center}
\begin{enumerate}
\item Nodes send data to FC; FC takes global decision
\item Nodes send local decisions to FC; FC fuses local decisions
\end{enumerate}
\item{Every node is able to take a global decision}
\begin{enumerate}
\item Nodes exchange data with their neighbors
\item Nodes exchange local decisions with their neighbors
\end{enumerate}
\end{enumerate}
In the first case, the observation is distributed across the nodes,
but the decision is centralized. This case has received most of the
attention. The interested reader may check, for example, the
book \cite{Varshney-book} or the tutorial reviews given in
\cite{Viswanathan-Varshney-PartI}, \cite{Blum-Kassam-Poor-PartII},
\cite{Chamberland-Veeravalli07}. In the second case, also the
decision is decentralized. This case has been considered only
relatively recently. Some references are, for example, \cite{Alanyali-Saligrama04},
\cite{Saligrama-06}, \cite{Aldosari-Moura}, \cite{Xiao-Boyd-Lall}, \cite{Kar-Moura-07},
\cite{Bajovic-Moura-11}, \cite{Cattivelli-Sayed-11}.

An alternative classification is between
\begin{enumerate}
\item Batch algorithms
\item Sequential algorithms
\end{enumerate}
In the first case, the network collects a given amount of data
along the time and space domains and then it takes a decision.
In the second case, the number of observations, either in time
or in terms of number of involved sensors, is not decided a priori,
but it is updated at every new measurement. The network stops
collecting information only when some performance criterion is
satisfied (typically, false alarm and detection probability) \cite{Marano-Matta-Willett-Tong},
\cite{Mei-08}, \cite{Sardellitti-SB-Pezzolo}.\\

One of the major difficulties in distributed detection comes from establishing the optimal decision thresholds
at local and global level. The main problem is how to optimize the local decisions, taking into
account that the final decisions will be only the result of the interaction
among the nodes. Taking a local decision can be
interpreted as a form of source coding. The simple (binary) hypothesis testing
can be seen in fact as a form of binary coding. Whenever the observations are conditionally
independent, given each hypothesis, the likelihood
ratio test at the sensor nodes is indeed optimal  \cite{tsitsiklis93}. However, finding the
optimal quantization levels is a difficult task. Even when the observations
are i.i.d., assuming identical decision rules is very common and apparently
well justified. Nevertheless there are counterexamples showing that nonidentical decision
rules are optimal  \cite{tsitsiklis93}.  Identical decision rules in the i.i.d. case
turns out to be optimal only asymptotically, as the number of nodes tends to infinity \cite{tsitsiklis88}.\\

\noindent A simple example may be useful to grasp some of the difficulties associated with distributed detection. For this purpose, we briefly recall the seminal work of Tenney and Sandell \cite{Tenney-Sandell}.
Let us consider two sensors, each measuring a real quantity $x_i$, with $i=1, 2$.
Based on its observation $x_i$, sensor $i$ decides whether the phenomenon of interest
is present or not. In the first case, it sets the decision variable $u_i=1$, otherwise, it sets $u_i=0$.
The question is how to implement the decision strategy, according to some optimality criterion.
The approach proposed in \cite{Tenney-Sandell} is a Bayesian approach, where the goal of each
sensor is to minimize the Bayes risk, which can
be made explicit by introducing the cost coefficients and the observation probability model.
Let us denote by $C_{ijk}$ the cost of detector $1$ deciding on ${\cal H}_i$,
detector $2$ deciding on ${\cal H}_j$, when the true hypothesis is ${\cal H}_k$.
Denoting by $P_k$ the prior probability of event ${\cal H}_k$ and by
$p(u_1, u_2, x_1, x_2, {\cal H}_k)$ the joint pdf of having ${\cal H}_k$,
observing the pair $(x_1, x_2)$ and deciding for the pair $(u_1, u_2)$,
the average risk can be written as
\begin{eqnarray}
{\cal R}&=&\sum_{i, j, k}\int C_{ijk}\,p(u_1, u_2, x_1, x_2, {\cal H}_k)\, dx_1 dx_2 \nonumber\\
&=&\sum_{i, j, k} P_k \int C_{ijk}\,p(u_1, u_2, x_1, x_2/ {\cal H}_k)\, dx_1 dx_2 \nonumber\\
&=&\sum_{i, j, k} P_k \int C_{ijk}\,p(u_1, u_2/ x_1, x_2, {\cal H}_k) p(x_1, x_2/ {\cal H}_k)\, dx_1 dx_2 \;.
\end{eqnarray}
In this case, each node observes only its own variable and takes a decision independently of
the other node. Hence, we can set
\begin{equation}
{\cal R}=\sum_{i, j, k} P_k \int C_{ijk}\,p(u_1/ x_1)\, p(u_2/ x_2)\,  p(x_1, x_2/ {\cal H}_k)\, dx_1 dx_2 \;.
\end{equation}
Expanding the right hand side by explicitly summing over index $i$, we get
\begin{equation}
{\cal R}=\sum_{j, k} P_k \int \, p(u_2/ x_2)  p(x_1, x_2/ {\cal H}_k) [C_{0jk}p(u_1=0/ x_1)+C_{1jk}p(u_1=1/ x_1)]\, dx_1 dx_2 \;.
\end{equation}
Considering that $p(u_1=1/ x_1)=1-p(u_1=0/ x_1)$ and ignoring all terms which do not contain $u_1$, we get
\begin{equation}
{\cal R}=\int p(u_1=0/ x_1) \sum_{j, k} P_k \left\{ \int \, p(u_2/ x_2)  p(x_1, x_2/ {\cal H}_k) [C_{0jk}-C_{1jk}]\, dx_2 \right\} dx_1+ {\rm const.}
\end{equation}
The average risk is minimized if $p(u_1=0/ x_1)$ is chosen as follows
\begin{equation}
\label{loc_dec}
p(u_1=0/ x_1)=
\left\{
\begin{array}{ll}0,\quad {\rm if} \quad \sum_{j, k} P_k \int \, p(u_2/ x_2)  p(x_1, x_2/ {\cal H}_k) [C_{0jk}-C_{1jk}]dx_2\ge 0 &   \\
1,\quad {\rm otherwise}. \quad &
\end{array}
\right.
\end{equation}
This expression shows that the optimal local decision rule is a {\it deterministic} rule.
After a few algebraic manipulations, (\ref{loc_dec}) can be rewritten, equivalently, as
\cite{Varshney-book}
\begin{equation}
\label{local_LRT}
\Lambda(x_1):=\frac{p(x_1/\mathcal{H}_{1})}{p(x_1/\mathcal{H}_{0})} \overset{\mathcal{H}_{1}}{\underset{\mathcal{H}_{0}}{\gtreqless}}
\frac{P_0 \sum_{j} \int \, p(u_2/ x_2)  p(x_2/ x_1, {\cal H}_0) [C_{1j0}-C_{0j0}]\,dx_2}{P_1 \sum_{j} \int \, p(u_2/ x_2)  p(x_2/ x_1, {\cal H}_1) [C_{0j1}-C_{1j1}]\,dx_2},
\end{equation}
where $\Lambda(x_1)$ is the LR at node $1$. Equation (\ref{local_LRT}) has the structure
of a LRT. However, note that the threshold on the right hand side of (\ref{local_LRT}) depends on the
observation $x_1$, through the term $p(x_2/ x_1, H_1)$, which incorporates the
statistical dependency between the observations $x_1$ and $x_2$. Hence, Equation (\ref{local_LRT})
is not a proper LRT.

The situation simplifies if the observations are conditionally independent, i.e.
$ p(x_2/ x_1, {\cal H}_k)= p(x_2/ {\cal H}_1)$. In such a case, the threshold $t_1$
can be simplified into
\begin{equation}
\label{t1}
t_1=\frac{P_0 \int p(x_2/\mathcal{H}_{0})\{p(u_2=0/x_2)[C_{100}-C_{000}]+p(u_2=1/x_2)[C_{110}-C_{010}]\}dx_2}{P_1 \int p(x_2/\mathcal{H}_{1})\{p(u_2=0/x_2)[C_{001}-C_{101}]+p(u_2=1/x_2)[C_{011}-C_{111}]\}dx_2}.
\end{equation}
Since $p(u_2=1/x_2)=1-p(u_2=0/x_2)$, (\ref{t1}) can be rewritten as
\begin{equation}
\label{t1mod}
t_1=\frac{P_0 \int p(x_2/\mathcal{H}_{0})\{[C_{110}-C_{010}]+p(u_2=0/x_2)[C_{100}-C_{000}+C_{010}-C_{110}]\}dx_2}{P_1 \int p(x_2/\mathcal{H}_{1})\{[C_{011}-C_{111}]+p(u_2=0/x_2)[C_{001}-C_{101}+C_{111}-C_{011}]\}dx_2}.
\end{equation}
Hence, the threshold $t_1$ to be used at node $1$ is a function of $p(u_2=0/x_2)$, i.e.,
on the decision taken by node $2$. At the same time, the threshold $t_2$ to be used
by node $2$ will depend on the decision rule followed by node $1$. This means that, even
if the observations are conditionally independent and the decisions
are taken autonomously by the two nodes, the decisions are still coupled
through the thresholds. This simple example shows how the detection problem
can be rather complicated, even under a very simple setting.

In the special case where $C_{000}=C_{111}=0$, $C_{010}=C_{100}=C_{011}=C_{101}=1$,
and $C_{110}=C_{001}=2$, i.e., there is no penalty if the decisions are correct, the penalty
is $1$, when there is one error, and the penalty is $2$ when there are two errors, the threshold simplifies into
\begin{equation}
\label{t1mod2}
t_1=\frac{P_0}{P_1}.
\end{equation}
Hence, in this special case, the two thresholds are independent of each other and
the two detectors become independent of each other.\\

\noindent After having pointed out through a simple example some of the problems related
to distributed detection, it is now time to consider in more detail the cases
where the nodes send their (possibly encoded) data to the FC or they take local decisions first
and send them to the FC. In both situations, there are two extreme cases: a) there is only
one FC; b) every node is a potential FC, as it is able to take a global decision.

\subsection{Nodes send data to decision center}
Let us consider for simplicity the
simple (binary) hypothesis testing problem. Given a set of vector observations $\bx:=[\bx_1, \ldots, \bx_N]$,
where $\bx_i$ is the vector collected by node $i$, $i=1, \ldots, N$, the optimal decision  rule for the simple hypothesis testing problem, under a variety of optimality criteria, amounts to compute
the likelihood ratio (LR) $\Lambda\left(\bx\right)$ and compare it with a threshold. In formulas,
\begin{equation}
\label{LR-1b}
\Lambda\left(\bx\right):=\Lambda\left(\bx_1, \ldots, \bx_N\right)
=\frac{p(\bx_1, \ldots, \bx_N; {\cal H}_1)}{p(\bx_1, \ldots, \bx_N; {\cal H}_0)}
\overset{\mathcal{H}_{1}}{\underset{\mathcal{H}_{0}}{\gtreqless}}  \gamma \;.
\end{equation}
In words, the detector decides for ${\mathcal{H}_{1}}$ if the LR exceeds
the threshold, otherwise it decides for ${\mathcal{H}_{0}}$.
In general, what changes the distributed detection problem from the standard
centralized detection is that the data are sent to the decision center after
source encoding into a discrete alphabet. The simplest form of encoding is quantization.
But also taking local decisions can be interpreted as a form
of binary coding. Clearly, source coding is going to affect the detection
performance. It is then useful to show, through a simple example, how
local quantization affects the final detection performance and how we
can benefit from the theoretical analysis to optimize the number of
bits associated to the quantization step in order to optimize performance
of the detection scheme.\\

\subsubsection{Centralized detection of deterministic signal embedded in additive noise}
Let us consider the detection of a deterministic (known) signal embedded
in additive noise. In this section, we consider the case where the decision is taken
at a FC, after having collected the data sent by the sensors. This case could refer for example to the detection
of undesired resonance phenomena in buildings, bridges, etc. The form of
the resonance is known. However, the measurements taken by the sensors
are affected by noise and then it is of interest to check the performance
as a function of the signal to noise ratio.

The measurement vector is $\bx=(x_1, \ldots, x_N)$, where $x_i$ is
the measurement taken by node $i$. Let us denote as $\bs$ the
known deterministic signal. The observation can be modeled as
\beq
\bx\sim\left\{ \begin{array}{ll}
\bv+\bw & \mbox{under } {\cal H}_0\\
\bs+\bv+\bw & \mbox{under } {\cal H}_1\\
\end{array}\right. \; ,
\eeq
where $\bv$ is the background noise, whereas $\bw$ is the quantization noise.
We assume the noise to be Gaussian with zero mean and (spatial) covariance
matrix $\bC_n$, i.e. $\bv \sim {\cal N}(\bzero, \bC_n)$.  To simplify the mathematical tractability, we consider
a dithered quantization so that the quantization error can be modeled as
a random process statistically  independent of noise. We may certainly assume that,
after dithering, the quantization noise variables over different sensors are statistically independent.
Hence, we can state that the quantization noise vector $\bw$ has zero mean
and a diagonal covariance matrix $\bC_q={\rm diag}(\sigma_{q1}^2, \ldots, \sigma_{qN}^2)$.
If the amplitude of the useful signal spans the dynamic range $[-A, A]$ and the number of bits used by
node $i$ is $n_i$, the quantum range is $q_i=2A/2^{n_i}$ so that the quantization noise variance at
node $i$ is
\begin{equation}
\label{qi}
\sigma_{qi}^2=\frac{(2A)^2}{12\,\, 2^{2 n_i}}=\frac{A^2}{3\,\, 2^{2 n_i}}.
\end{equation}
The overall noise has then a zero mean and covariance matrix $\bC=\bC_n+\bC_q$.\\

\noindent If the quantization noise is negligible, the Neyman-Pearson criterion
applied to this case leads to the following linear detector
\begin{equation}
\label{Tx}
T(\bx)={\cal R}\{\bs^H \bC^{-1}\bx\}\overset{\mathcal{H}_{1}}{\underset{\mathcal{H}_{0}}{\gtreqless}}  \gamma
\end{equation}
where ${\cal R}(x)$ denotes the real part of $x$ and the detection threshold $\gamma$
is computed in order to guarantee the desired
false alarm probability $P_{fa}$.
Unfortunately, since the quantization noise is not Gaussian,
the composite noise $\bv+\bw$ is not Gaussian and then the
detection rule in (\ref{Tx}) is no longer optimal. Nevertheless, the rule
in (\ref{Tx}) is still meaningful as it maximizes the signal to noise ratio (SNR).
Hence, it is of interest to look at the performance of this detector in the
presence of quantization noise. The exact computation of the detection probability
is not easy, at least in closed form, because it requires the computation of the
pdf of $T(\bx)$. Nevertheless, when the number of nodes is sufficiently high
(an order of a few tens can be sufficient to get a good approximation), we
can invoke the central limit theorem to state that $T(\bx)$ is approximately
Gaussian. Using this approximation, the detection probability can be written
in closed form for any fixed $P_{fa}$, following standard derivations (see, e.g. \cite{SB-kay-II}),
as
\begin{equation}
\label{Pd}
P_d=Q\left[Q^{-1}\left(P_{fa}\right)-\sqrt{\bs^H \bC^{-1} \bs}\right]=Q\left[Q^{-1}\left(P_{fa}\right)-\sqrt{\bs^H \left(\bC_n+\bC_q\right)^{-1} \bs}\,\,\right].
\end{equation}
This formula is useful to assess the detection probability as a function of the
bits allocated to each transmission. At the same time, we can also use (\ref{Pd})
as a way to find out the bit allocation that maximizes the detection probability.
This approach establishes an interesting link between the communication
and detection aspects. In practice, in fact, encoded data are transmitted over a finite capacity channel.
Hence, it is useful to relate the number of quantization bits used by  each node and
capacity of the channel between that node and the FC. For simplicity, we consider
the optimization problem under the assumption of spatially uncorrelated noise, i.e.
$\mathbf{C}_n={\rm diag}(\sigma_{n1}^2, \ldots, \sigma_{nN}^2)$. The problem we wish to solve is the maximization
of the detection probability, for a given false alarm rate and a maximum global transmit power.
To guarantee an arbitrarily low transmission error rate, we need to respect Shannon's channel coding theorem,
so that the number of bits per symbol must be less than channel capacity.
Denoting with $p_i$ the power transmitted by user $i$ and assuming flat fading channel,
with channel coefficient $h_i^2$, the capacity is given by (\ref{capacity}).
From (\ref{Pd}), maximizing $P_d$ is equivalent to maximizing $\bs^H \left(\bC_n+\bC_q\right)^{-1} \bs$.
Hence, using (\ref{qi}), the maximum $P_d$, for a given $P_{fa}$ and a given global transmit power
$P_T$, can be achieved by finding the power vector $\bp=(p_1, \ldots, p_N)$ that solves the following
constrained problem
 \begin{eqnarray}
\max_{\bp}\sum_{i=1}^N |s_i|^2 \left(\sigma_{ni}^2+\frac{A^2}{3\,(1+a_i\,p_i)}\right)^{-1}\\
{\rm s.t.} \quad \sum_{i=1}^{N}p_i \le P_T;\,\,\,p_i\ge 0, i=1, \ldots, N.
\end{eqnarray}
It is straightforward to check that this is a convex problem. Imposing the Karush-Kuhn-Tucker conditions,
the optimal powers can be expressed in closed form as:
\begin{equation}
p_i=\left[\frac{1}{\sqrt{\lambda}}\sqrt{\frac{s_i^2 A^2}{3 a_i \sigma_{ni}^4}}-\frac{A^2}{3 a_i \sigma_{ni}^2}-\frac{1}{a_i}\right]^+
\end{equation}
where the Lagrange multiplier $\lambda$ associated to the sum-power constraint can be
determined as the value that makes $\sum_{i=1}^{N}p_i=P_T$.\\

A numerical example is useful to grasp some of the properties of the proposed algorithm.
Let us consider a series of sensors placed along a bridge of length $L$. The purpose of the network
is to detect one possible spatial resonance, which we represent as the signal $s(z)=A\cos(\pi z /L)$,
where $z\in[-L/2, L/2]$ denotes the spatial coordinate.
\begin{figure}
\centering
\includegraphics[width=8.5cm]{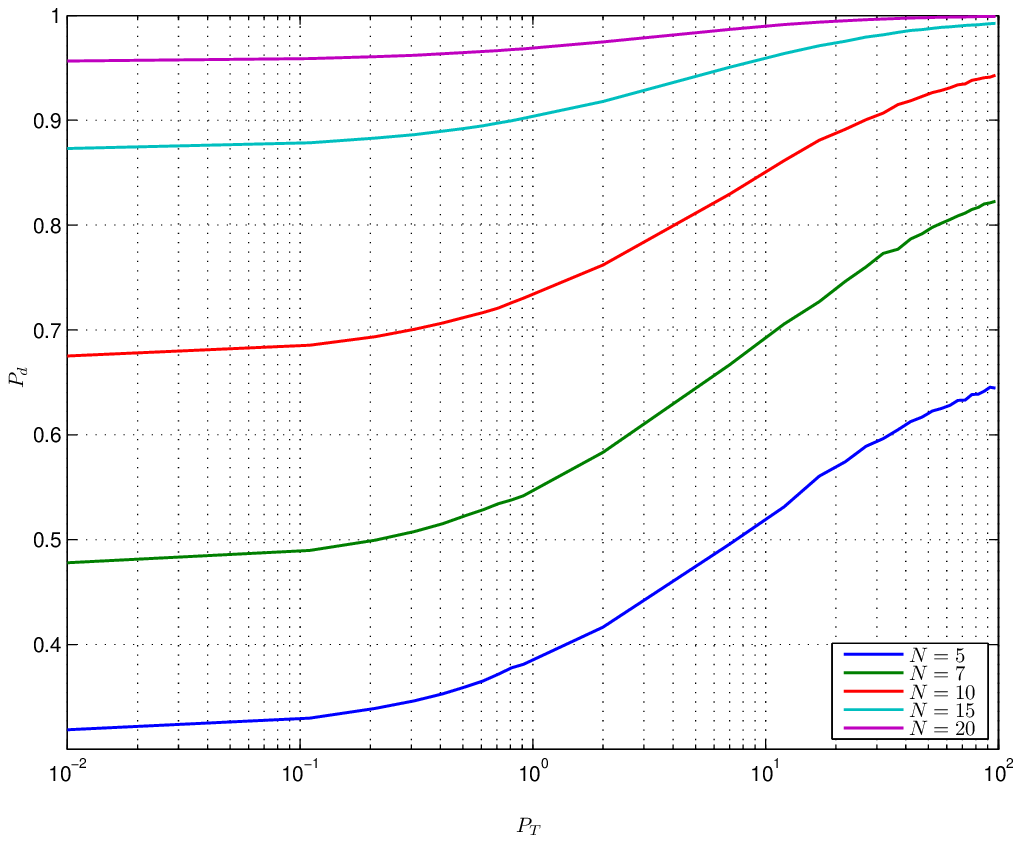}
\caption{Detection probability vs. sum transmit power, for different number of sensors.}
\label{Pd_vs_PT}
\end{figure}
The sensors are uniformly spaced along the bridge, at positions $z_i=(i-1) L/N$, with $i=1, \ldots, N$.
Every sensor measures a shift $x_i=s(z_i)+v_i$, affected by the error $v_i$. To communicate
its own measurement to the FC, every sensor has to quantize the measurement first.
The optimal number of bits to be used by every sensor can be computed by using
the previous theory. In this case, in Fig. \ref{Pd_vs_PT} we report the detection probability vs.
the sum power $P_T$ available to the whole set of sensors, for different numbers $N$ of sensors.
As expected, as the total transmit power increases, $P_d$ increases because more bits per symbol
can be transmitted and then the quantization errors become negligible. It is also important to notice
how, increasing the number of sensors, the detection probability improves, for any given transmit power.
Furthermore, in Fig. \ref{nbits_channel} we can see the optimal per channel bit allocation (bottom), together with
the channels profiles $|h_k|^2$ (top). Interestingly, we can see that the method allocates
more bits in correspondence with the best channels and the central elements of the array, where
the useful signal is expected to have the largest variations.
\begin{figure}
\centering
\includegraphics[width=8.5cm]{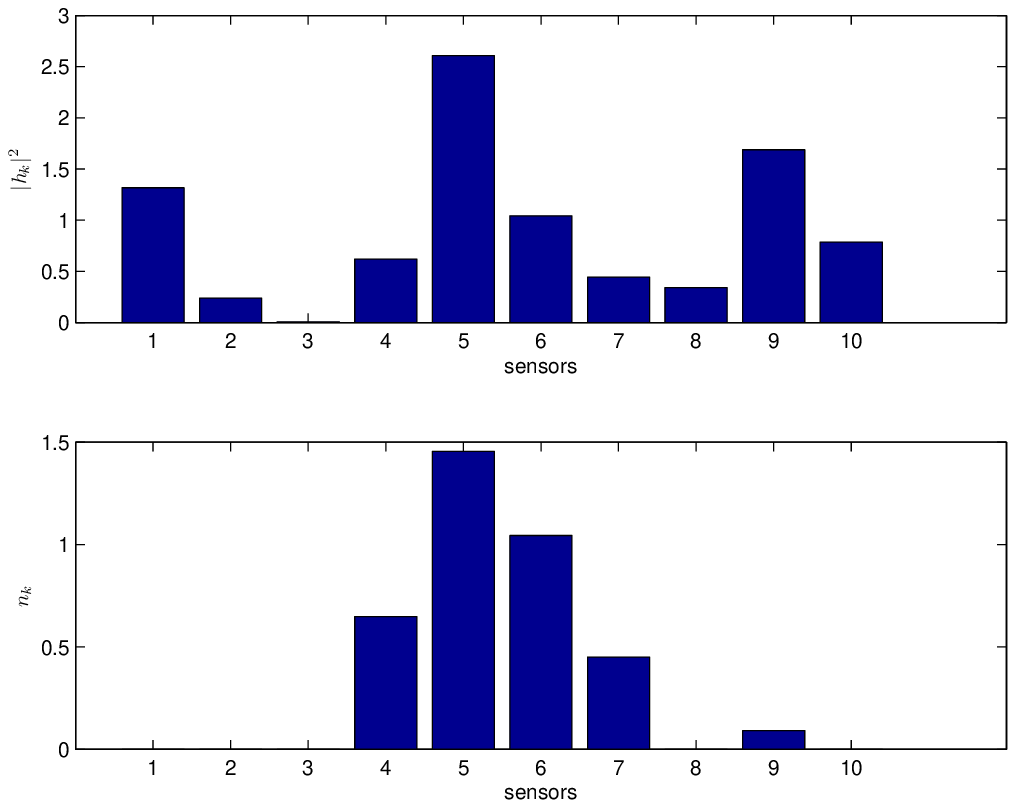}
\caption{Optimal bit allocation.}
\label{nbits_channel}
\end{figure}

\subsubsection{Decentralized detection under conditionally independent observations}
Let us consider now the case where the globally optimal decision can be taken, in principle, by any node.
To enable this possibility, every node must be able to implement the statistical test (\ref{LR-1b}).
If the measurements collected by the sensors are conditionally independent, the logarithm of the
likelihood ratio can be written as
\begin{equation}
\label{sumLR}
\log \Lambda(\bx_1, \ldots, \bx_N) =\sum_{i=1}^{N} \log \Lambda_i(\bx_i)= \sum_{i=1}^{N}
\left[\log p_{X_i}(\bx_i, {\cal H}_1)-\log p_{X_i} (\bx_i, {\cal H}_0)\right].
\end{equation}
This formula shows that, in the conditionally independent case, running a consensus algorithm is
sufficient to enable every node to compute the global LR. It is only required that every sensor
initializes its own state with the local log-LR $\log \Lambda_i(\bx_i)$ and then runs the consensus
iterations. If the network is connected, every node will end up with the average value of the local LR's.
In practice, to send the local LR, every node must quantize it first. Then, we need to refer to
the consensus algorithm in the presence of quantization errors. However, we have already seen in
Section  \ref{Consensus algorithms over realistic channels} that the consensus iterations
may be properly modified to make the algorithm robust against a series of drawbacks
coming from communications through realistic channels, as, eg. random packet drops and
quantization. Hence, a consensus algorithm, properly modified, can enable every node
to compute the global LR with controllable error.

\subsection{Nodes send local decisions to fusion center}
Consider now the case where each node $i$ takes a local decision,
according to a locally optimal criterion, and encodes the decision
into the binary variable $u_i$. Then, the node sends the variable $u_i$
to the fusion center, which is asked to take a global decision on the basis
of the vector $\bu:=(u_1, \ldots, u_N)$ containing all local decisions.
Let us consider for simplicity the binary hypothesis test.
This problem was considered in \cite{Varshney-book} and we will
now review the basic results.
This problem is distinct from the case studied in the previous section because here
the local decision thresholds are optimized according to a detection criterion,
whereas in standard quantization the decision thresholds are not optimized.\\

\noindent Under both Bayesian and Neyman-Pearson (NP) formulations, the optimal test amounts
to a likelihood ratio test, based on $\bu$, i.e.
\begin{equation}
\label{LR-2}
\frac{p(u_1, \ldots, u_N; {\cal H}_1)}{p(u_1, \ldots, u_N; {\cal H}_0)}\gtrless \eta \;.
\end{equation}
In the case of conditionally independent local decisions, the LRT
converts into
\begin{equation}
\label{fusion_rule_cond_ind}
\frac{\prod_{i=1}^{N}p(u_i; {\cal H}_1)}{\prod_{i=1}^{N}p(u_i; {\cal H}_0)}:=\prod_{i=1}^{N}\Lambda_i(u_i)\gtrless \eta \;.
\end{equation}
Since each variable $u_i$ can only assume the values $0$ or $1$, we
can group all the variables into two subsets: the subset $S_0$ containing
all variables $u_i=0$ and the subset $S_1$ containing
all variables $u_i=1$, thus yielding
\begin{equation}
\label{fusion_rule_cond_ind2}
\prod_{i\in S_0}\frac{p(u_i=0; {\cal H}_1)}{p(u_i=0; {\cal H}_0)}\prod_{i\in S_1}\frac{p(u_i=1; {\cal H}_1)}{p(u_i=1; {\cal H}_0)} \gtrless \eta \;.
\end{equation}
Denoting with $P_{Mi}=p(u_i=0; {\cal H}_1)$,  and  $P_{Fi}=p(u_i=1; {\cal H}_0)$, the probabilities of miss
and the probability of false alarm of node $i$, respectively,
 (\ref{fusion_rule_cond_ind2}) can be rewritten as
\begin{equation}
\label{fusion_rule_cond_ind3}
\prod_{i\in S_0}\frac{P_{Mi}}{1-P_{Fi}}\prod_{i\in S_1}\frac{1-P_{Mi}}{P_{Fi}} \gtrless \eta \;.
\end{equation}
Taking the logarithm of both sides and reintroducing the variables $u_i$,
the fusion rule becomes
\begin{equation}
\label{fusion_rule_cond_ind4}
\sum_{i=1}^N\left[\log\left(\frac{1-P_{Mi}}{P_{Fi}}\right)\, u_i+ \log \left(\frac{P_{Mi}}{1-P_{Fi}}\right)\, (1-u_i)\right] \gtrless \log\eta
\end{equation}
or, equivalently
\begin{equation}
\label{fusion_rule_cond_ind5}
\sum_{i=1}^N\log\left[\frac{(1-P_{Mi})(1-P_{Fi})}{P_{Mi}P_{Fi}}\right]\, u_i \gtrless \log\left[\eta\prod_{i=1}^{N}\frac{1-P_{Fi}}{P_{Mi}}\right] \;.
\end{equation}
The optimal fusion rule is then a simple weighted sum of the local decisions,
where the weights depend on the reliabilities of the local decisions: Larger weights are
assigned to the most reliable nodes.\\

If instead of having a single FC, we wish to enable every node
to implement the decision fusion rule described above, we can see that, again,
running a consensus algorithm suffices to reach the goal.
In fact, if each local state variable is initialized with a value
$x_i[0]=\log\left[\frac{(1-P_{Mi})(1-P_{Fi})}{P_{Mi}P_{Fi}}\right]\, u_i$,
running a consensus algorithm allows every node to know the function in (\ref{fusion_rule_cond_ind5}).
The only constraint is, as always, network connectivity. The drawback of this simple
approach is that running this sort of consensus algorithm requires the transmission  of
real variables, rather than the binary variables $u_i$. In fact, even if the local decision $u_i$
is binary, the coefficient multiplying $u_i$ is a real variable, which needs to
be quantized before transmission over a realistic channel. Again, the consensus
algorithm can be robustified against quantization errors by using dithered quantization and a decreasing step size,
as shown in \ref{Consensus algorithms over realistic channels}. However, it is important to
clarify that we cannot make any claim of optimality of this kind of distributed decision. In
principle, when the nodes exchange their decisions with the neighbors, the decision thresholds
should be adjusted in order to accommodate some optimality criterion. This is indeed
an interesting, yet still open, research topic.

\section{Beyond consensus: Distributed projection algorithms}
\label{Beyond consensus: Distributed projected algorithms}
In many applications, the field to be reconstructed by a sensor network is typically a smooth
function of the spatial coordinates. This happens for example, in the reconstruction
of the spatial distribution of the power radiated by a set of transmitters. The problem
is that local measurements may be corrupted by local noise or fading effects. An
important application of this scenario is given by cognitive networks. In such a case,
a secondary node would need to know the channel occupation across space, to
find out unoccupied channels, within the area of interest. This requires some sort
of spectrum sensing, but in a localized area. The problem of sensing is that
wireless propagation is typically affected by fading or shadowing effects,
so that a sensor in a shadowed location might indicate that a channel
is unoccupied, while this is not true. To avoid this kind of error, which
would lead to undue channel occupation from opportunistic users,
it is useful to resort to cooperative sensing. In such a case,  nearby
nodes exchange local measurements to counteract the effect of shadowing.

The problem with local averaging operations is that they should reduce the effect of
fading, but without destroying valuable spatial variations. In the following, we
recall a distributed algorithm proposed in \cite{SB-Scu-Bat09} to recover a
spatial map of a field, using local weighted averages where the weights are
chosen so as to improve upon local noise or fading effects, but without
destroying the spatial variation of the useful signal.

Let us consider a network composed of $N$ sensors located
at positions $(x_{i},y_{i})$, $i=1, \ldots, N$,
and denote the measurement collected by the $i$-th sensor by $g(x_{i},y_{i})=z(x_{i},y_{i})+v_{i}$,
where $z(x_{i},y_{i})$ represents the useful field while $v_{i}$ is the
observation error. Let us also denote by  $u_k(x, y)$, $k=1, \ldots, r$, a set of
linearly independent spatial functions defining a basis for the useful signal.
The useful signal can then be represented through the basis expansion model
\begin{equation}
z(x_{i},y_{i})=\sum_{k=1}^{r} s_k u_k(x_i, y_i).
\end{equation}
In vector notation, introducing the $N$-size column vector $\bg:=[g(x_{1},y_{1}), g(x_{2},y_{2}), \ldots, g(x_{N},y_{N})]^T$
and similarly for the vector $\bz$, we may write
\begin{equation}  \label{g=f+e}
\bg=\bz+\bv=\bU \bs+\mathbf{v},
\end{equation}%
where $\bU$ is the $N \times r$
matrix whose $m$-th column is $\mathbf{u}_m=(u_m(x_i, y_i), \ldots, u_m(x_N, y_N))$,
$\mathbf{s}=(s_1, \ldots, s_r)$ is an $r$-size vector of coefficients and $\bz=\bU\bs$
is the useful signal.
The spatial smoothness of the useful signal field
may be captured by choosing the functions $u_k(x, y)$ to be the low frequency
components of the Fourier basis or low-order 2D polynomials. For instance,
if the space under monitoring is a square of side $L$, we may choose
the set
\begin{equation}
\label{2DFourier}
\{u_{nm}(x, y)\}=\left\{1, \cos\left(2\pi \frac{n x+m y}{L}\right), \sin\left(2\pi \frac{n x+m y}{L}\right)\right\}_{m=0, n=0; m+n\neq 0}^{m=\infty, n=\infty}
\end{equation}
In practice, the
dimension $r$ of the useful signal subspace is typically much smaller than the
dimension $N$ of the observation space, i.e. of the number of sensors. We can
exploit this property to devise a distributed denoising algorithm.

\label{sec:format} If we use a Minimum Mean Square Error (MMSE)
strategy, the goal is to find the useful signal vector $\hat{\bs}$ that minimizes
the mean square error
\begin{equation}
{\cal E}:=E\{\|\bg-\bU \hat{\bs}\|^2\}.
\label{MSE}
\end{equation}
The solution is well known and is given by \cite{SB-kay}:
\begin{equation}
 \hat{\bs}=(\bU^T\bU)^{-1} \bU^T \mathbf{g}.
\label{MSEsol}
\end{equation}
Our goal is actually to recover the vector $\bz$, rather than $\bs$.
In such a case, the estimate of $\bz$ is
\begin{equation}
\label{lin_mod_estimator}
    \hat{\mathbf{z}}=\bU (\bU^T\bU)^{-1} \bU^T \mathbf{g}.
\end{equation}
The operation performed in (\ref{lin_mod_estimator}) corresponds to projecting
the observation vector onto the subspace spanned by the columns of $\bU$.
Assuming, without  any loss of generality (w.l.o.g.), the columns of $\bU$ to be
orthonormal,  the projector simplifies into
\begin{equation}
\label{lin_mod_estimator_orthogonal}
\hat{\bz}=\bU \bU^T \bg.
\end{equation}
The centralized solution to this problem is then very simple: The fusion center collects
all the measurements
$g(x_i, y_i)$, compute $\bU$ and then recovers $\hat{\bz}$ from (\ref{lin_mod_estimator_orthogonal}).

The previous approach is well known. The interesting point is that the MMSE solution
can be achieved with a totally decentralized approach, where every sensor
interacts only with its neighbors, with no need to send any data to a fusion center.
The proposed approach is based on a very simple iterative procedure,
where each node initializes a state variable with the local
measurement, let us say $z_i[0]=g(x_i, y_i)$, and then it updates its own state
by taking a linear combination of its neighbors' states, similarly with what
happens with consensus algorithms, but with coefficients computed in order
to solve the new problem.

More specifically, denoting by $\bz[k]$, the $N$-size vector containing the states of
all the nodes, at iteration $k$, and by $\bg$
the vector containing the initial measurements collected by all the nodes, the
vector $\bz[k]$  evolves according to the following linear state equation:
\begin{equation}\label{LTI-system}
\mathbf{z}[k+1]=\mathbf{W}\mathbf{z}[k],\quad\mathbf{z}[0]=\bg\in\mathbb{R}^{N},
\end{equation}
where $\mathbf{W}\in\mathbb{R}^{N\times N}$ is typically a {\it sparse} (not
necessarily symmetric) matrix. The network topology is reflected into the sparsity of
$\bW$. In particular, the number of nonzero entries of, let us say, the $i$-th
row is equal to the number of neighbors of node $i$.
In a WSN,  the neighbors of a node are the  nodes falling within the coverage area of that node, i.e.
within a circle centered on the location of the node, with radius dictated by the transmit
power of the node and by the power attenuation law. Our goal is to find the nonnull
coefficients of $\bW$ that allow the convergence of $\bz[k]$ to the vector $\hat{\bz}$
given in (\ref{lin_mod_estimator_orthogonal}).
In general, not every network topology guarantees the existence of a solution of this problem.
In the following, we will show that a solution exists only if each node has a number of neighbors
greater than the  dimension $r$ of the useful signal subspace.

Let us denote by $\mathbf{P}_{\mathcal{R}(\mathbf{U})}\in\mathbb{R}^{N\times N}$
the orthogonal projector onto the $r$-dimensional subspace of
$\mathbb{R}^{N}$ spanned by the columns of
$\mathcal{R}(\mathbf{U)}$, where $\mathcal{R}(\cdot)$ denotes the
range space operator and $\mathbf{U}\in\mathbb{R}^{N\times r}$ is a
full-column rank matrix, assumed, w.l.o.g., to be semi-unitary.
System  (\ref{LTI-system}) converges to the desired orthogonal
projection of the initial value vector $\mathbf{z}[0]=\bg$ onto
$\mathcal{R}(\mathbf{U)}$, for {\it any} given
$\mathbf{g}\in\mathbb{R}^{N}$, if and only if
\begin{equation}
\lim_{k\rightarrow+\infty}\bz[k]=\lim_{k\rightarrow+\infty}\bW^k
\bg=\mathbf{P}_{\mathcal{R}(\mathbf{U})} \bg, \label{eq:limit_point}
\end{equation}
i.e.,
\begin{equation}
\lim_{k\rightarrow+\infty}\mathbf{W}^{k}=\mathbf{P}_{\mathcal{R}(\mathbf{U})}.\label{eq:limit_point_2}
\end{equation}
Resorting to basic algebraic properties of discrete-time systems, it is possible
to derive immediately some basic properties of $\bW$. In particular, denoting with
OUD the Open Unit Disk, i.e. the set $\{x\in \mathbb{C}: |x|<1\}$, a matrix
$\bW$ is {\it semistable} if its spectrum ${\rm spec}(\bW)$ satisfies ${\rm spec}(\bW)\subset {\rm OUD}\cup\{1\}$
and, if $1 \in {\rm spec}(\bW)$, then $1$ is semisimple, i.e. its algebraic and geometric multiplicities coincide.
If $\bW$ is semistable, then \cite[p. 447]{bernstein}
\begin{equation}
\lim_{k\rightarrow+\infty}\mathbf{W}^{k}=\mathbf{I}-(\mathbf{I}-\mathbf{W})^{\sharp}(\mathbf{I}-\mathbf{W}),\label{semistable}
\end{equation}
where ${ }^{\sharp}$ denotes group generalized inverse \cite[p. 228]{bernstein}.
Furthermore, setting, without loss of generality, the matrix $\bW$ in the form $\bW=\bI-\epsilon \bL$,
(\ref{semistable}) can be rewritten as
\begin{equation}
\lim_{k\rightarrow+\infty}\mathbf{W}^{k}=\mathbf{I}-\mathbf{L}^{\sharp}\mathbf{L}.\label{semistableL}
\end{equation}
But $\mathbf{I}-\mathbf{L}^{\sharp}\mathbf{L}$ is the projector onto the null-space of $\bL$.
Hence, we can state the following
\begin{proposition}\label{Proposition_0}Given the dynamical system
in (\ref{LTI-system}) and the projection matrix
$\mathbf{P}_{\mathcal{R}(\mathbf{U})}$, the vector
$\mathbf{P}_{\mathcal{R}(\mathbf{U})}\mathbf{z}[0]$ is globally
asymptotically stable for \emph{any fixed}
$\mathbf{z}[0]\in\mathbb{R}^{N}$, if and only if the following
conditions are satisfied:\\
i) $\bL$ has a nullspace of dimension $r$, spanned by the columns of $\bU$;\\
ii) $\bL$ and $\epsilon$ must be chosen so that $\bW$ is semistable.
\end{proposition}
Alternatively, the previous conditions can be rewritten equivalently in the following form
\begin{proposition}\label{Proposition_1} Given the dynamical system
in (\ref{LTI-system}) and the projection matrix
$\mathbf{P}_{\mathcal{R}(\mathbf{U})}$, the vector
$\mathbf{P}_{\mathcal{R}(\mathbf{U})}\mathbf{z}[0]$ is globally
asymptotically stable for \emph{any fixed}
$\mathbf{z}[0]\in\mathbb{R}^{N}$, if and only if the following
conditions are satisfied:\begin{equation}
\mathbf{W}\,\mathbf{P}_{\mathcal{R}(\mathbf{U})}=\mathbf{P}_{\mathcal{R}(\mathbf{U})}\tag{C.1}\label{eq:C_1}\end{equation}
\begin{equation}
\mathbf{P}_{\mathcal{R}(\mathbf{U})}\,\mathbf{W}=\mathbf{P}_{\mathcal{R}(\mathbf{U})}\tag{C.2}\label{eq:C_2}\end{equation}
\begin{equation}
\rho\left(\mathbf{W}-\mathbf{P}_{\mathcal{R}(\mathbf{U})}\right)<1\tag{C.3}\label{eq:C_3}\end{equation}
where $\rho(\cdot)$ denotes the spectral radius operator
\cite{SB-horn}. \hfill $\square$
\end{proposition}

$ $

\noindent\textbf{Remark 1}: Conditions
\ref{eq:C_1}-\ref{eq:C_3} have an intuitive interpretation.
In particular, \ref{eq:C_1} and \ref{eq:C_2} state that, if system
(\ref{LTI-system}) asymptotically converges, then it is
guaranteed to converge to the desired value. In fact, \ref{eq:C_1}
guarantees that the projection of vector $\mathbf{z}[k]$ onto
$\mathcal{R}(\mathbf{U})$ is an invariant quantity for the dynamical
system, implying that the system in (\ref{LTI-system}), during its
evolution, keeps the component
$\mathbf{P}_{\mathcal{R}(\mathbf{U})}\mathbf{z}[0]$ of
$\mathbf{z}[0]$ unaltered. At the same time,  \ref{eq:C_2} makes
$\mathbf{P}_{\mathcal{R}(\mathbf{U})}\mathbf{z}[0]$ a fixed point of
matrix $\mathbf{W}$ and thus a potential accumulation point for the
sequence $\{\mathbf{z}[k]\}_{k}$. Both conditions
\ref{eq:C_1} and \ref{eq:C_2} do not state anything about the
convergence of the dynamical system. This is guaranteed by
\ref{eq:C_3}, which imposes that all the modes associated to the
eigenvectors orthogonal to $\mathcal{R}(\mathbf{U})$ are
asymptotically vanishing.

\smallskip
\noindent\textbf{Remark 2}: The conditions \ref{eq:C_1}-\ref{eq:C_3} contain,
as a special case, the convergence conditions of average consensus algorithm.
In fact, it is sufficient to set in (\ref{eq:limit_point}), $r=1$ and
$\mathbf{U}=\mathbf{u}=\frac{1}{\sqrt{N}}\mathbf{1}_{N}$, where
$\mathbf{1}_{N}$ is the $N$-length vector of all ones. In such a
case, \ref{eq:C_1}-\ref{eq:C_3} can be restated as following: the
digraph associated to the network described by $\mathbf{W}$ must be
strongly connected and balanced.\\

The previous conditions do not make any explicit reference to the sparsity
of matrix $\bW$. However, when we consider a sparse matrix, reflecting
the network topology, additional conditions are necessary to make sure that
the previous conditions are satisfied. In other words, not every network
topology is able to guarantee the asymptotic projection onto a prescribed signal subspace.
One basic question is then what network topology is able to
guarantee the convergence to a prescribed projector. We provide
now the conditions on the sparsity of $\bW$, or equivalently $\bL$,
guaranteeing the desired convergence.

From condition $i)$ of Proposition 1, given the matrix $\bU$, $\bL$ must
satisfy the equation $\bL \bU=\bzero$. Let us assume that every row of
$\bL$ has $K$ nonzero entries and let us indicate with $\{i_{j1}, \ldots, i_{jK}\}$
the set of the column indices corresponding to the nonzero entries of the $j$-th row of $\bL$.
Hence, every row of $\bL$ must satisfy the following equation
\begin{equation}
\label{ul=0}
\left(
\begin{array}{cccc}
u_{1}(i_{11}), & u_{1}(i_{12}), & \cdots & u_{1}(i_{1K})\\
u_{2}(i_{21}), & u_{2}(i_{22}), & \cdots & u_{2}(i_{2K})\\
\cdot & \cdot & \cdot & \cdot\\
u_{r}(i_{r1}), & u_{r}(i_{r2}), & \cdots & u_{r}(i_{rK})\\
\end{array}
\right)
\left(
\begin{array}{c}
l_{j1}\\
\vdots\\
l_{jK}
\end{array}
\right)=\bzero\;.
\end{equation}
To guarantee the existence of a nontrivial solution to (\ref{ul=0}), the matrix
on the left hand side must have a kernel of dimension at least one.
This requires $K$ to be strictly greater than $r$, the dimension of the signal subspace.
Since the number of nonzero entries of, let us say the $j$-th, row of $\bL$
is equal to the number of neighbors of node $j$ plus one (the coefficient
multiplying the state of node $i$ itself), this implies that the minimum
number $K$ of neighbors of each node must be at least equal to the
dimension $r$ of the signal subspace. Of course this condition is necessary
but not sufficient. It is also necessary to check that the sparse matrix
$\bL$ built with rows satisfying (\ref{ul=0}), with $j=1, \ldots, N$, had rank $N-r$.
This depends on the location of the nodes and on the specific
choice of the orthogonal basis.

\begin{figure}[t]
\centering
\includegraphics[width=11cm]{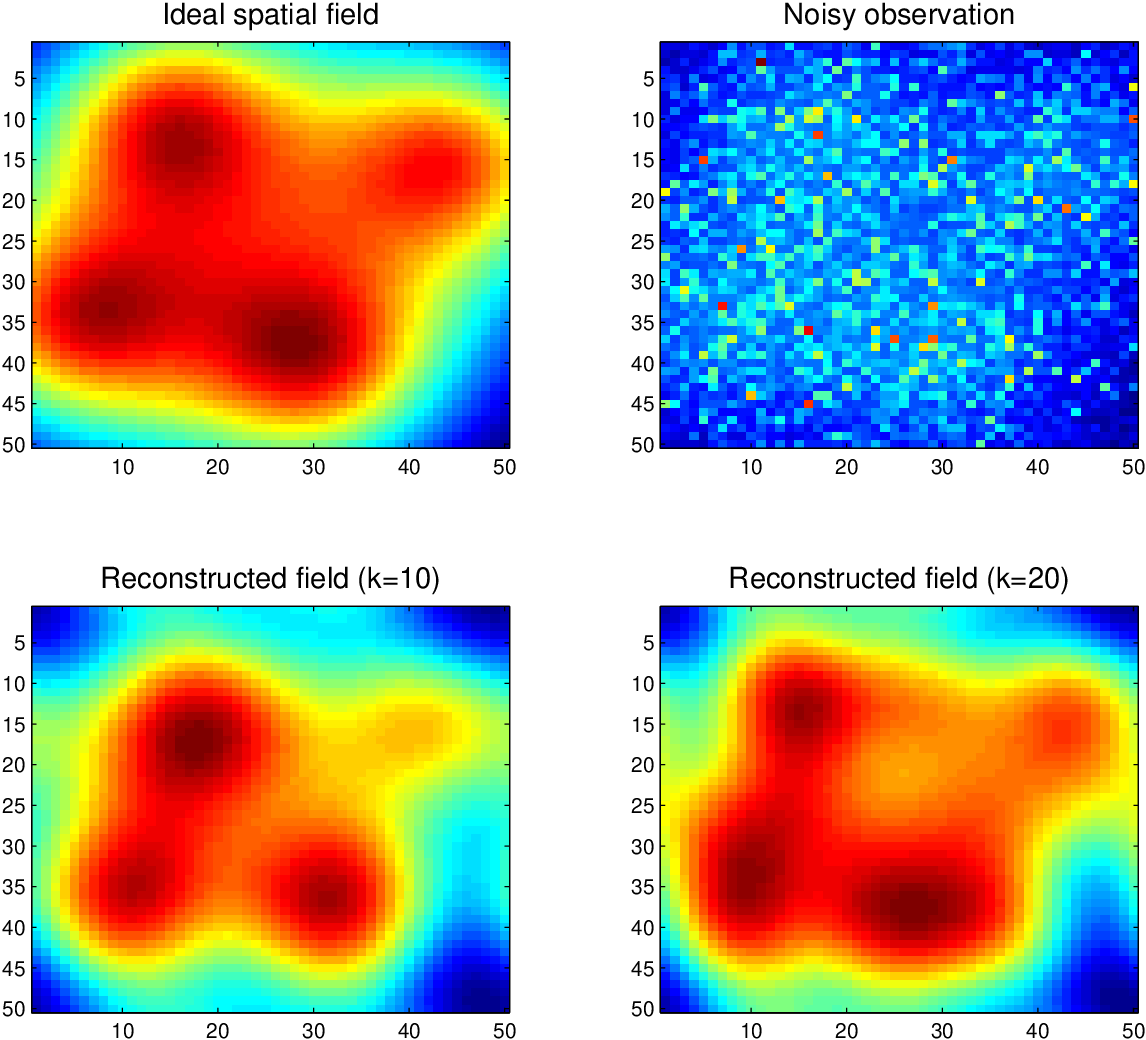}
\caption{\small Example of field reconstruction in the presence of fading: ideal spatial field (top left);
measured field (top right);  field reconstructed with order $k=10$ (bottom left) and
 $k=20$ (bottom right).}\label{Figmultnoise}
\end{figure}
An example can be useful to illustrate the benefits achievable
with the proposed technique. We consider the case where
the observation is corrupted by a  multiplicative,
spatially uncorrelated, noise, which models, for example a fading effect. Let
us denote with $P(x_i, y_i)=A(x_i, y_i)\,S(x_i, y_i)$ the measurement
carried out from node $i$, located in the point of coordinates $(x_i, y_i)$,
where $S(x_i, y_i)$ denotes the useful field, whereas $A(x_i, y_i)$
represents fading. We consider, for instance, a useful signal composed
by $N_s=4$ transmitters and we assume a polynomial power attenuation,
so that the useful signal measured at the point of coordinates $(x, y)$is
\begin{equation}
S(x, y)=\sum_{i=1}^{N_s}\frac{P_i}{1+((x-x_i)^2+(y-y_i)^2)/\sigma^2}, \quad  x\in \left[-\frac{L}{2}, \frac{L}{2}\right], \; y\in \left[-\frac{L}{2}, \frac{L}{2}\right].
\label{Cauchy}
\end{equation}
where $P_i$ is the power emitted by source $i$, located at $(x_i, y_i)$,
and $\sigma$ specifies the power spatial spread.
Furthermore, fading is modeled as a spatially uncorrelated multiplicative noise.
The sensor network is composed of $2500$ nodes uniformly
distributed over a 2D grid. All the transmitters use the same power, i.e. $P_i=P$ in (\ref{Cauchy}), and
the noise has zero mean and variance $\sigma_n^2=P$.
In this case, it is useful to apply a homomorphic filtering to the measured field.
In particular, we take  the $\log$ of the measurement, thus getting $\log(P(x_i, y_i))=\log(S(x_i, y_i))+\log(A(x_i, y_i))$.
To smooth out the undesired effect of fading, we assume a signal model composed by
the superposition of 2D sinusoids, so that the columns of the matrix $\bU$ in (\ref{g=f+e})
are composed of signals of the form $\sin(2\pi(mx+ny)/L)$, and $\cos(2\pi(mx+ny)/L)$, with $m, n=0, 1, \ldots$.
We set the initial value of the state of each node equal to $\log(P(x_i, y_i))$ and we run the distributed
projection algorithm described above. After convergence, we take the $\exp$ of the result.

Fig. \ref{Figmultnoise} shows an example of application. In particular, the spatial behavior
of the useful signal power is shown in the top left plot, while
the observation corrupted by fading is reported in the top right figure.
It is useful to consider that, in the example at hand, the useful signal would require
a Fourier series expansion with an infinite number of terms to null the
modeling error. Conversely, in our example, we  used two different orders, $k=10$ and $k=20$.
The corresponding reconstructions are shown in the bottom figures. From Fig. \ref{Figmultnoise} it is evident the capability of the proposed distributed approach to provide a significant attenuation of the fading phenomenon, without
destroying valuable signal variations.

\section{Minimum energy consensus}
\label{Minimum energy consensus}
Although distributed algorithms to achieve  consensus have received a lot of attention
because of their capability of reaching optimal decisions without the need of  a fusion center, the price paid for this simplicity is that consensus algorithms are  inherently iterative. As  a consequence the iterated exchange of data among the nodes might cause an excessive energy consumption.
Hence, to make consensus algorithms really appealing in practical applications,
it is necessary  to minimize the energy consumption necessary to reach
consensus.  The network topology plays a fundamental role
in determining the convergence rate \cite{SB-boydgossip}.
As the network connectivity increases, so does
the convergence rate. However, a highly connected
network entails a high power consumption to guarantee reliable
direct links between the nodes. On the other hand, if the network is
minimally connected, with only neighbor nodes connected to each other,
a  low power  is spent to maintain the few short range links, but, at the same time,
a large convergence time is required. Since what really matters in a WSN is the
overall energy spent to achieve consensus, in \cite{SB-Eusipco_topology, SB-SS-top}
it was considered the problem of finding the optimal network topology that minimizes
the overall energy consumption, taking into account convergence
time and transmit powers  {\it jointly}.
More specifically, in \cite{SB-SS-top} it is proposed a method  for optimizing the
network topology and the power allocation across every link in
order to minimize the energy necessary to achieve consensus.
Two  different types of networks are considered: a) deterministic topologies,
where node positions are  arbitrary, but known;  b) random geometries,
where the unknown node locations are modeled as random variables.
We will now review the methodology used in both cases.

\subsection{Optimization criterion}
By considering only the power spent to enable wireless communications, the overall energy
consumption to reach consensus can be written as the product between the sum of the power $P_{tot}$
necessary to establish the communication links among the nodes and
the number of iterations $N_{it}$ necessary to achieve consensus.
The exchange of information among the nodes is supposed to take
place in the presence of a slotted system, with a medium access
control (MAC) mechanism that prevents packet collisions. The number of
iterations can be approximated as
$N_{it}=T_c/T_s$ where
 $T_s$ denotes the duration of a time slot unit and
 $$T_c=-\frac{\log(\gamma)}{\lambda_2(\bL)}$$
 is the convergence time defined
 as the time necessary for the slowest  mode of the dynamical system (\ref{ct-consensus})
 to be reduced by a factor $\gamma \ll 1$.
 The total power spent
by the  network in each iteration is then $P_{tot}=\sum_{i, j}a_{ij}p_{ij}$
where  the coefficient $p_{ij}=p_{ji}$, $i \neq j$  denotes the power transmitted by node $i$
to node $j$, while the binary coefficients $a_{ij}$ assess the presence ($a_{ij}=1$) of a link
between nodes $i$ and $j$ or not ($a_{ij}=0$). Our goal is to minimize the energy consumption
expressed  by the following metric
\begin{equation}\label{SB-E}
{\cal E}= P_{tot} N_{it}=K\frac{ \sum_{i=1}^{N}\sum_{j=1}^{N} a_{ij}p_{i j}}{\lambda_2(\bL(\ba))},
\end{equation}
where $K$ incorporates all irrelevant constants, $N$ is the number of sensors
and $\bL(\ba)$ is the Laplacian matrix depending on the vector $\ba=\bA(:)$ containing all
the coefficients $a_{ij}$.
More specifically, we aim to find the set of active links, i.e.,
the non-zero coefficients $a_{ij}$, and the powers
$p_{ij}$ that minimize the energy consumption (\ref{SB-E}),
under the constraint of guaranteeing network connectivity,
i.e. enforcing $\lambda_2(\bL(\ba))>0$.
The problem can be formulated as follows \cite{SB-SS-top}:
\beq
    \begin{array}{llll}
     {\min}_{\ba, \bp}
      &
\ds  \frac{ \sum_{i=1}^{N}\sum_{j=1}^{N}a_{ij}p_{i j}}{\lambda_2(\bL(\ba))}\\
s.t. &  \epsilon\leq \lambda_2(\bL(\ba)) \hspace{2cm}{\rm\bf [P.0] } \\
   &   a_{ij} \in \{0,1\} \\
&   p_{ij}\geq 0 \hspace{1cm}\forall \; i,j=1,\ldots,N\\
    \end{array}
 \label{SB-combinat_problem}
    \eeq
where $\epsilon$ is an arbitrarily  small
positive constant used to ensure network connectivity and $\bp$ is the vector with entries $p_{ij}$.
Since the topology coefficients are binary variables,  ${\rm\bf [P.0] }$
is a combinatorial problem, with complexity increasing with the size
$N$ of the network as $2^{N(N-1)/2}$.
In \cite{SB-SS-top} we have modified  ${\rm\bf [P.0] }$ in order to convert it
into a convex problem, with negligible performance losses.
A first simplification comes from observing that the coefficients
$a_{ij}$ and $p_{ij}$ are dependent of each other through  the radio propagation model
so that the set of unknowns can be reduced to the set of powers $p_{ij}$.
More specifically, by assuming flat fading channel, we can assume
 that the power $p_{Rj}$  received by node $j$
when node $i$ transmits is given by
\beq
p_{Rj}=\frac{p_{i j}}{1+(r_{i j}/r_{0})^\eta} \label{SB-pr}
\eeq
where  $r_{ij}$ is the distance between  nodes $i$ and $j$,
$\eta$ is the path loss exponent, and the parameter $r_0$ corresponds to the so called Fraunhofer distance.
We have included in the denominator the unitary term to avoid the unrealistic situation in which the received
power could be greater than the transmitted one. Given the propagation model (\ref{SB-pr}), the relation between the power coefficients
$p_{ij}$ and the topology coefficients $a_{ij}$ is then
\beq a_{i j}=\left \{
\begin{array}{llll} 1 & \mbox{if} & p_{ij}> p_{min}\left[1+\left(\frac{r_{ij}}{r_0}\right)^\eta\right] \\ 0 & &
\mbox{otherwise} \end{array} \right. \label{SB-elementsAarbitrary}\;
\eeq
where $p_{\min}$ is the minimum power needed at the receiver side to establish a communication.
In \cite{SB-SS-top} we have shown how to relax this relation in order to
simplify the solution of the optimal topology control problem considering both
the deterministic and random topology.

\subsection{Optimal topology and power allocation for arbitrary networks}

In the case where the distances between the nodes are known, to find the optimal solution
of problem ${\rm\bf [P.0] }$ involves a combinatorial
strategy that makes the problem numerically very hard to solve.
In \cite{SB-SS-top} ,  we have relaxed problem  ${\rm\bf [P.0] }$ so that,
instead of requiring $a_{ij}$ to be binary, we assume  $a_{ij}$
to be a real variable belonging to the interval $[0,1]$.
This  relaxation is the first step to
transform the previous problem into a {\it convex} problem. More specifically,
we have introduced the following relationship between
the coefficients $a_{ij}$ and the distances $r_{ij}$:
 \beq
a_{ij}=\ds \frac{1}{1+(r_{i j}/r_{c_{ij}})^{\alpha}} ,
\label{SB-aijexp}
\eeq
where $\alpha$ is a positive coefficient and $r_{c_{ij}}$
is the coverage radius, which depends on the transmit power.
According to (\ref{SB-aijexp}), $a_{ij}$ is close to one when
node $j$ is within the coverage radius of node $i$, i.e., $r_{ij} \ll r_{c_{ij}}$,
whereas $a_{ij}$ is close to zero, when $r_{ij} \gg r_{c_{ij}}$. The switching
from zero to one can be made steeper by
increasing the value of $\alpha$.
In \cite{SB-SS-top} we have found   the coefficients $p_{ij}$ as  a function of $a_{ij}$
\beq
p_{i j}= q(a_{i j})= p_{min}+k_1
\left(\ds \frac{a_{ij}}{1-a_{ij}}\right)^{\eta/\alpha},
\label{SB-eqqc}
\eeq
with $k_1=p_{min}\ds \frac{r_{ij}^{\eta}}{r_0^{\eta}} $.
Consequently, we can reduce the set of  variables to the only power vector $\bp$ and problem ${\rm\bf [P.0] }$ can be relaxed into the following problem:
\beq
    \begin{array}{llll}
     {\min}_{\bp}&
\ds\frac{\bp^T \textbf{1}}{
{\lambda}_2(\bL(\bp))}\\
s.t. &  \epsilon \leq \lambda_2(\bL(\bp))\hspace{1cm}{\rm\bf [P.1] } \\
   &   p_{min}\textbf{1} \leq \bp
    \end{array} \;.
 \label{SB-costenergy}
    \eeq
The first important result proved in \cite{SB-SS-top} is that the problem ${\rm\bf [P.1] }$ is a convex-concave fractional problem  if $\eta \geq \alpha$,
so that we can use one of the methods that solve quasi-convex optimization problems, see e.g.,  \cite{SB-jeflea, SB-dinkelbach}. In \cite{SB-SS-top} we
have used the nonlinear parametric formulation proposed in
\cite{SB-dinkelbach}.
Hence we have further converted the convex-concave fractional problem ${\rm\bf [P.1] }$
into the following equivalent parametric  problem
  in terms of vector
$\ba$, i.e.
 \beq
  \begin{array}{llll}
   {\min}_{\ba} &
 \phi(\ba)-\mu
{\lambda}_2(\bL(\ba))\\
s.t. &  \epsilon \leq \lambda_2(\bL(\ba))\hspace{1cm}{\rm \bf [P.2]}\\
   & \textbf{0}\leq \ba < \textbf{1}.
    \end{array}
 \label{SB-costenergychange}
 \eeq
where
$\phi(\ba)=\ds \sum_{i=1}^N \sum_{j=1, i\neq j}^N q(a_{i j})$ and $\mu$ controls the trade-off between
total transmit power and convergence time.

The optimization problem ${\rm\bf [P.2] }$ is  a convex
parametric problem \cite{SB-SS-top} and an optimal solution can be found via efficient numerical tools.
Furthermore, using Dinkelbach's algorithm \cite{SB-dinkelbach},
we are also able to find the optimal parameter $\mu $ in $\bf [P.2] $.

\subsection{Numerical examples}
Since our optimization procedure is based on a relaxation technique, we have
evaluated  the impact of the relaxation on the final topology and performance.\\
More specifically, the topology coefficients $a_{ij}$ obtained by solving
$\bf [P.2] $ are real variables belonging
to the interval $[0, 1]$, so that to obtain the network topology, it is necessary a quantization step
 to convert them into binary values, $1$ or $0$,  by comparing
each $a_{ij}$ with a threshold $a_{th}$.
\begin{figure}[t]
\centering
\includegraphics[width=12cm]{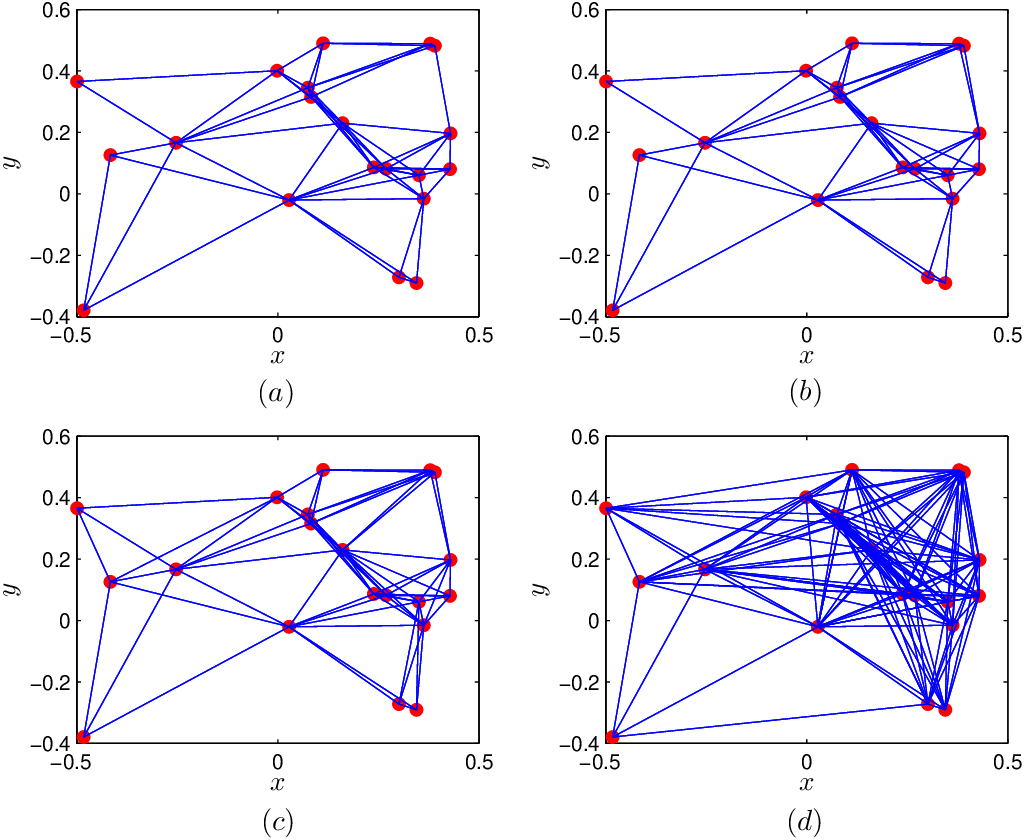}
\caption{Optimal topologies, for different threshold values and
$\eta=6$: a) $a_{th}=0.09$;
b) $a_{th}=0.05$; c) $a_{th}=10^{-4}$; d) $a_{th}=10^{-7}$.} \label{SB-Fig1}
\end{figure}
\begin{figure}[h]
\centering
\includegraphics[width=10cm]{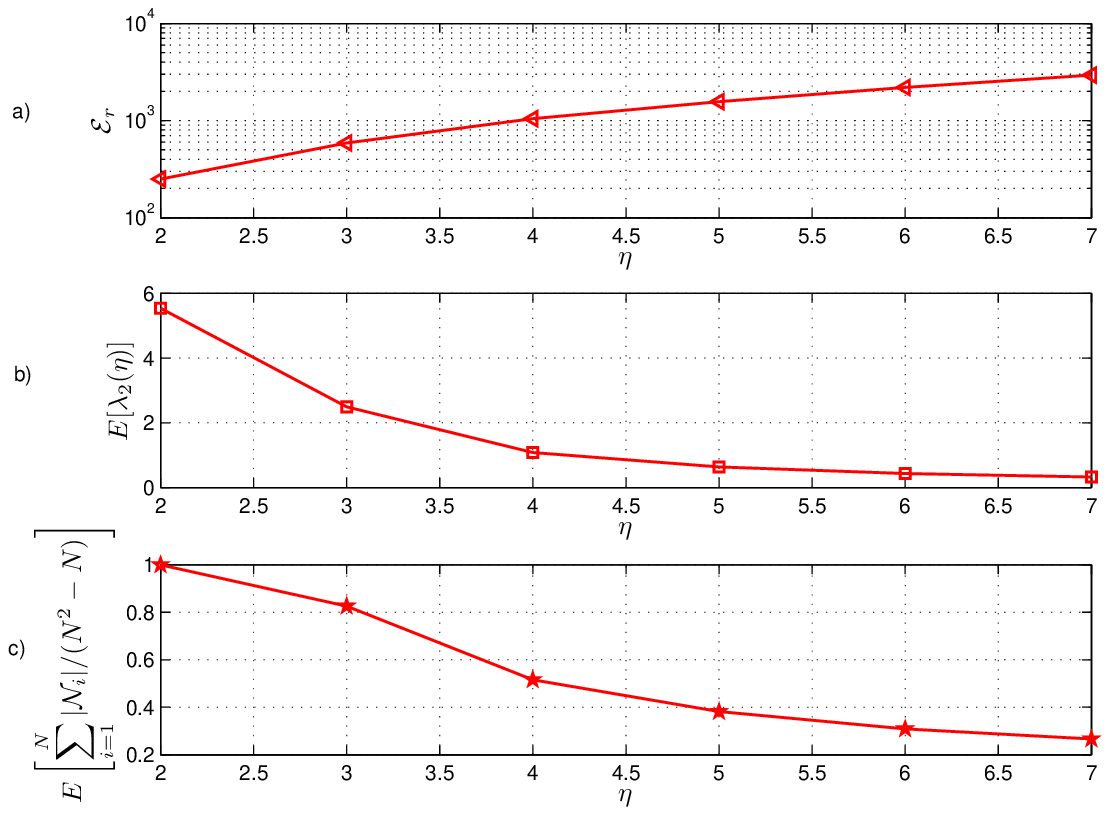}
\caption{Average value of a) energy; b) $\lambda_2(\bL)$; c) fraction of active links  vs. path loss $\eta$ for
$a_{th}=0.09$.}
  \label{SB-Fig4}
\end{figure}
It has been shown that the loss in terms of optimal energy due to the relaxation of the original
problem is negligible. To evaluate the impact of thresholding operation in Fig. \ref{SB-Fig1} we  show the  topologies
obtained by solving problem  ${\rm\bf [P.2] }$, for a network composed of $N=20$ nodes,
using different values of $a_{th}$ and assuming $\eta=6$.
Comparing the four cases reported in Fig. \ref{SB-Fig1}, we can note that
for a large range of values of  $a_{th}$, the
final topology is practically the same,
while only for very low values of the threshold (i.e., case (d)), we
can observe a sensitive change of topology.  This means that the
relaxation method is robust against the choice of the final threshold.

The previous results pertain to a specific realization of the node
locations. To provide results of more general validity, in Fig.
\ref{SB-Fig4}, we report the average value of a) the energy ($\E_r$)
b) $\lambda_2(\bL)$, and c)  fraction of active
links $\ds \frac{\sum_{i=1}^{N}|{\cal N}_i|}{N(N-1)}$, as a function of the path loss exponent $\eta$, setting $a_{th}=0.09$.
From Fig.\ \ref{SB-Fig4}, we observe that when the attenuation is high
(i.e. $\eta$ is large), reducing the number of links (making the
topology sparser) is more important than reducing convergence time.
Conversely, when the attenuation is low (i.e., $\eta$ is small),
increasing network connectivity is more important than reducing
power consumption.

\subsection{Minimization of the energy consumption over random geometric graphs}
Let us consider now the problem of minimizing  the energy consumption for a sensor network
modeled as a random geometric graph.
We will use the symbol $G(N, r)$ to indicate an RGG composed of $N$ points,
with coverage radius $r$.

In \cite{SB-boydgossip}, it has been shown that the degree of an RGG
$G(N, r)$ of points uniformly distributed over a two-dimensional unit torus\footnote{A torus geometry is
typically used to get rid of border effects.} is equal to
\beq
d(N)=\pi r^2 N \label{SB-averagdegree1}
\eeq
with high probability, i.e., with probability  $1-1/N^2$, if the radius
behaves as
$r_0(N)$ in (\ref{theor_r0(N)}). This implies that
 if the coverage radius is chosen so as to
guarantee connectivity with high probability, an RGG tends to behave,
asymptotically, as a regular graph.
In order to calculate the convergence rate
we have to derive the second eigenvalue of the Laplacian,
$\bL = \bD - \bA$, where $\bD$ is the degree matrix and $\bA$
is the adjacency matrix.
From (\ref{SB-averagdegree1}), $\bD = \pi r^2 N \bI$, so that
we only need to calculate the second {\it largest} eigenvalue of $\bA$.
In Appendix A.2, we study the asymptotic behavior of the spectrum of $\bA$
and the result is that  the second largest eigenvalue of $\bL$ tends asymptotically to
\beq
 \lambda_2(\bL)=\pi N r^2  - N r J_1(2 \pi r) \label{SB-algconnln12} \;
\eeq
where $r$ is the coverage radius of each node.\\

\subsubsection{ An analytic approach for minimizing the energy consumption }

In \cite{SB-SS-top} we studied the energy minimization problem
for RGG's, exploiting  the previous analytic expressions.
In the random topology case, since the distances are unknown, we
cannot optimize the power associated with each link. However, we can seek the
common transmit power that minimizes energy consumption. Thus, in the random setting
we assume a broadcast communication model, where each node broadcasts the value to
be shared with its neighbors. In the lack of any information about distances among the nodes,
we assume that each node uses the same transmit power.  In this case, the network  topology
can be modeled as a random graph model. In \cite{SB-mesbahi},\cite{SB-aysal} it has been shown
that the  dynamical system $\dot{\bx}(t)=-\bL \bx(t)$ converges to consensus  almost surely,
i.e. $\mbox{Pr}\left\{\ds \lim_{t\rightarrow \infty} \bx(t)=x^{*}\b1 \right\}=1$
assuming that each node  has a coverage radius so that the network is asymptotically connected with probability one.
Then the rate of convergence to consensus is given \cite{SB-mesbahi},\cite{SB-aysal} by $E[e^{-2 T_s \lambda_2(L)}]$.
In \cite{SB-SS-top} we proved that the convergence rate can be approximated as
\beq
E[e^{-2 T_s \lambda_2}]\approx e^{-2 T_s E[\lambda_2]}
\eeq
so that the energy spent to achieve consensus can now
be expressed as
\begin{equation}\label{SB-energy-consumption}
    \E=K\,\frac{N p}{2 E[\lambda_2(\bL(p))]}.
\end{equation}
This is the performance metric we wish to minimize in the
random scenario, with respect to the single unknown $p$.

In particular, using the asymptotic expression
(\ref{SB-algconnln12}) for the
algebraic connectivity,
we can introduce the following metric
\beq
\E(r)= \ds \frac{N \,p_{min} [1+(r/r_0)^{\eta}]}{N \pi r^2- r N J_1(2\pi r) }  \label{SB-enanalitic}
\eeq
 that
is a convex function of $r$, for $r_0(N)\leq r \leq 0.5$,
where $r_0(N)$, behaves as in (\ref{theor_r0(N)})
to  ensure connectivity.

\begin{figure}[ht]
\centering
\includegraphics[width=10cm]{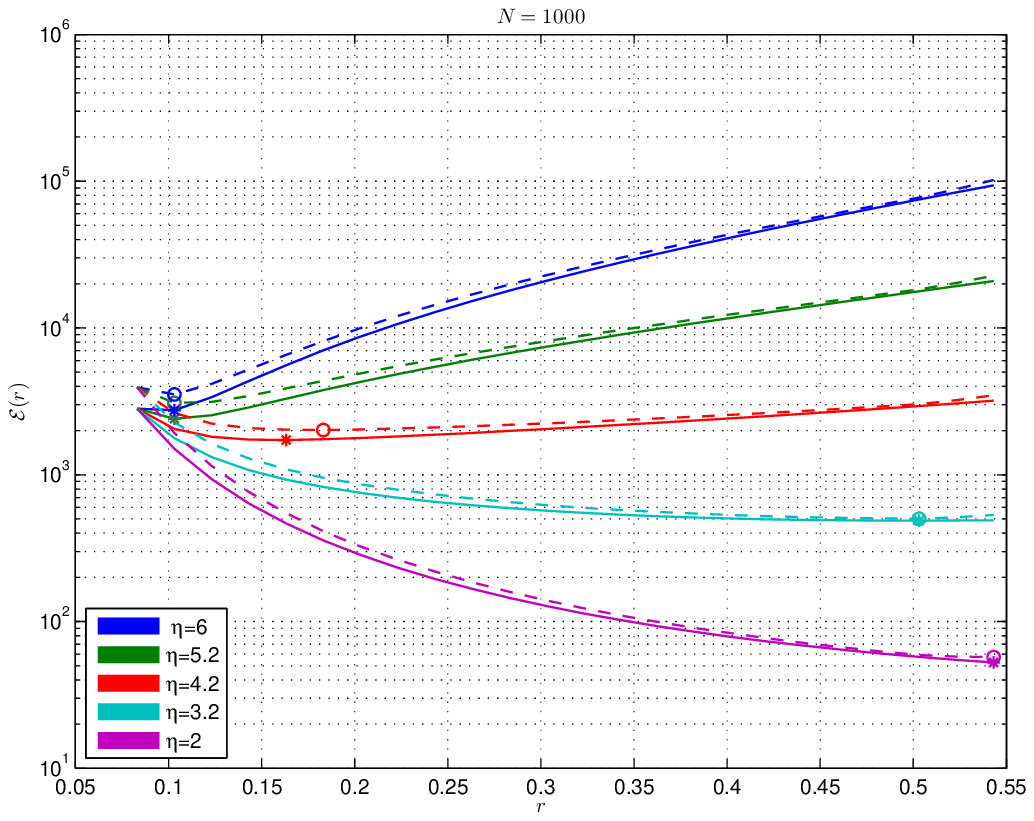}
\caption{Global energy consumption versus transmission radius for
an RGG; theoretical values (solid) and simulation results (dashed). }
\label{SB-fig:fig7}
\end{figure}

\noindent{\it Numerical examples.}
In Fig.  \ref{SB-fig:fig7}, we compare the value of $\E(r)$
obtained by our theoretical approach and by simulation,
for various values of the path loss exponent $\eta$.
The results are averaged over $100$ independent realizations of
random geometric graphs composed of $N=1000$ nodes. For each $\eta$,
we indicate the pair of radius and energy providing minimum energy consumption
by a circle (simulation) or a star (theory).  It can be noted that the theoretical
derivations provide a very good prediction of the performance achieved by simulation
and, for each $\eta$, there is a coverage radius value that minimizes energy consumption.

\section{Matching communication network topology to statistical dependency graph}
\label{Matching communication network topology to statistical dependency graph}
In Section \ref{Observation model}, we saw that the topology of a sensor network
observing a random field should depend on the structure of the graph describing
the observed field.  In this section, we recall a method proposed
in \cite{SB-SS-DSP11}, \cite{SB-SS-Journ-Match} to design the topology
of a wireless sensor network observing a Markov random field
in order to match the structure of the
dependency graph of the observed field, under constraints
on the power used to ensure the sensor network
connectivity.
As in \cite{SB-SS-DSP11}, \cite{SB-SS-Journ-Match},
our main task is to recover the sparsity of the dependency graph and to replicate it
at the sensor network level, under the constraint of limiting the transmit power
necessary to establish the link among the nodes.
Also in this case, searching for an optimal topology is a combinatorial problem.
To avoid the computational burden of solving the combinatorial problem,
we propose an ad hoc relaxation technique that allows us to
achieve the solution through efficient algorithms based on
difference of convex problems.\\
Let us assume to have a network composed of $N$ nodes, each one
observing a spatial sample of a Gaussian Markov Random Field.
We denote by $\bx:=(x_1, \ldots, x_N)$ the vector of the observations
collected by the $N$ nodes and we assume that $\bx$ has zero mean
and covariance matrix $\bC$. The statistical dependency among the
random variables $x_i$ is well captured by  the structure of the Markov graph,
whose vertices correspond to the random variables and whose
links denote statistical dependencies among the variables.
As discussed in \ref{Observation model} the main feature of a Markov graph is that it is sparse and
there is no link between  two nodes  if and only if their observations are statistically independent.
Moreover, if the random vector $\bx$ is also Gaussian, with covariance matrix
$\bC$, the sparsity of the Markov graph is completely specified by the sparsity of the
precision matrix,  which is the inverse
of the covariance matrix, i.e.  $\bB:=\bC^{-1}$.
On the other hand, the topology of the WSN can also be described by a graph,
having adjacency matrix $\bA$ such that $a_{ij}\ne 0$ only if there is
a physical link between nodes $i$ and $j$. We use a simple propagation
model such that there is a link between node $i$ and $j$ if the
power received by node $j$ exceeds a minimum power $p_{min}$.
The received power depends on the power $p_T(i, j)$ used by node $i$ to transmit
to node $j$ and
on the distance $r_{ij}$ between nodes $i$ and $j$ through the
equation
\beq
\label{recpower}
p_R(i, j)=\frac{p_T(i, j)}{1+r_{ij}^\eta}
\eeq
where $\eta$ is the path loss exponent.\\
As proposed in \cite{SB-SS-DSP11}, \cite{SB-SS-Journ-Match},
our goal is to design the topology of the WSN, and hence its
adjacency matrix $\bA$, in order to match as well as possible the
topology of the dependency graph, compatibly with the power
expenditure necessary to establish each link in the network.
Without any power constraint, we would choose $\bA$ to be equal to
$\bB=\bC^{-1}$, so as to reproduce the same sparsity of the
dependency graph. Adding the power constraints, we will end up, in
general, with a matrix $\bA$ different from  $\bB$. We measure the
difference between the two matrices $\bA$ and $\bB$ using the so
called Burg divergence, defined as \beq \label{Burgdiv} D_B(\bA,
\bB):= \frac{1}{2} \tr\left(\bA {\bB^{-1}}-\bI\right)-\frac{1}{2}
\log(\det(\bA {\bB}^{-1}))\; . \eeq
Even though the Burg
divergence does not respect all the prerequisites to be a
distance, it holds true that $D_{B}(\bA, \bB)=0$, if and only if
$\bA=\bB$, otherwise, the divergence is strictly positive. If the
matrices $\bA$ and $\bB$ are definite positive, the expression in
(\ref{Burgdiv}) coincides with the Kullback-Leibler divergence
between the probability density function (pdf) of two Gaussian
random vectors having zero mean and precision matrices $\bA$ and
$\bB$ or, equivalently, covariance matrices $\bA^{-1}$ and $\bC$.
\subsection{Encouraging sparsity by preserving total transmit power}
One of the most important tasks in wireless sensor networks
is to minimize the energy consumption for reliable data transmission. This
need can be accommodated by formulating the search for a sparse topology
incorporating a penalization for the presence of links among distant nodes.
The first strategy we propose is named Sparsity with Minimum Power (SMP) consumption.
We consider both cases where the covariance matrix is perfectly known or estimated
from the collected data.

Let us consider a wireless sensor network whose communication
graph is a geometric graph, where each node communicates only
with the nodes lying within its coverage area of radius $r$.
We assume, initially, that the covariance matrix $\bC$ of the GMRF is
 perfectly known.
Our goal is to find the optimal adjacency
matrix $\bA$ that minimizes  the divergence $D_{B}(\bA, \bB)$
given in (\ref{Burgdiv}), under the constraint of limiting the transmit power
necessary to maintain the links among the nodes of the WSN.
This constraint can be incorporated  in our optimization problem
by introducing a penalty term given by the sum of the transmit powers
over all active links, i.e.
\beq
P_N(\bA)=\sum_{i=1}^{N}\sum_{\substack{j=1\\ j\neq i}}^{N}
p_T(i,j) \delta(a_{ij})\label{eqpn}
\eeq
where $p_T(i,j)$ denotes the power used by node $i$ to transmit to node $j$
and
\beq
\delta(a_{ij})=\left\{ \begin{array}{lll}
0 & \mbox{if} \quad  a_{ij}=0\\
1 & \mbox{otherwise}
\end{array}\right. \; ,
\eeq assuming that $a_{ij}$ is different from zero only if the power $p_R(i, j)$ received by node $j$
when node $i$ transmits, as given in (\ref{recpower}), exceeds a suitable minimum level $p_{min}$, i.e.
if
\beq p_T(i,j)>(1+r_{ij}^{\eta}) p_{min}.\label{ptr}
\eeq
The optimization problem can then be formulated as \beq
\begin{array}{llll}
\underset{\bA \in \, \mbox{S}^N_{++}}{\mbox{min}} & D_{B}(\bA, \bB)+\rho P_N(\bA)\\
\end{array}\label{optprob}
\eeq where $\mbox{S}^N_{++}$ is the cone of definite positive symmetric $N\times N$-dimensional
matrices, while $\rho\geq 0$ is the penalty coefficient introduced to control
sparsity. In fact, increasing the
penalty coefficient, we assign a higher weight to power
consumption so that sparse structures are
more likely to occur.

Problem (\ref{optprob})  is indeed quite hard to solve as the penalty function is  a nonconvex discrete
function. Optimization problems with a convex penalty
 have been largely considered in several signal processing applications, for example in compressed sensing
\cite{SB-donoho} where these problems  are often formulated as  a penalized
least-square problem in which sparsity is usually induced by adding a
$l_1$-norm penalty on the coefficients, as in Lasso algorithm
\cite{SB-tibshirani}.

Indeed, non-convex penalty functions such as $l_q$-norm, with $q<1$,
are even more effective to recover sparsity than $l_1$-norm. Actually,
using the so called $l_0$ norm would be even more effective
to measure sparsity, even though the $l_0$ norm does not respect all
requisites to be a norm\footnote{The $l_0$ norm of a vector $\bx$ is
defined as the number of nonzero entries of $\bx$.}.
Here we adopt  the so called
Zhang penalty function analyzed in \cite{SB-gasso}, i.e.
\beq
z(a_{ij})=\min \left(\ds
\frac{\mid a_{ij}\mid}{\epsilon},1 \right)=\left\{
\begin{array}{lll}
 \ds \frac{\mid a_{ij} \mid}{\epsilon} & \mbox{if} \mid a_{ij}\mid \leq \epsilon\\
1 & \mbox{otherwise}
\end{array}\right. \label{zaij}
\eeq
where $\epsilon$ is an infinitesimal positive constant.  Hence, by assuming $p_T(i,j)=(1+r_{ij}^{\eta}) p_{min}$, the second term in (\ref{optprob})
can be written as
\beq
P_N(\bA)=\sum_{i=1}^{N}\sum_{j=1}^{N}d_{ij} z(a_{ij})=\tr[z(\bA) \bD]
\eeq
where $\bD$ is a $N\times N$ dimensional symmetric matrix
with entries $d_{ij}=(1+r_{ij}^{\eta})p_{min}$, $d_{i i}=0$,
$\forall\, i,j=1,\ldots,N$, while the matrix mapping $z(\bA)$
 is
defined applying the elementwise  mapping $z(a_{ij}):
\;\mathbb{R}\rightarrow \mathbb{R}^{+}$ given in (\ref{zaij}).
The
combinatorial problem in (\ref{optprob}) can then be reformulated as
\beq
\begin{array}{llll}
\underset{\bA \in \, \mbox{S}^N_{++}}{\mbox{min}} & D_{B}(\bA,\bB)+\rho \tr[z(\bA) \bD]\,.\\
\end{array}\label{optprob1}
\eeq
Unfortunately  the second term in (\ref{optprob1}),
is not convex  so
that the problem we have to solve is a nonconvex, nonsmooth
optimization problem. Nevertheless, in  \cite{SB-SS-DSP11}, \cite{SB-SS-Journ-Match} we
reformulated this problem as a difference of convex (DC) problem. Before proceeding,
we simply illustrate how to extend our approach to the case where the
covariance matrix of the observed vector is not known but estimated from the data.
In such a case, the matrix $\bC$ in (\ref{optprob1}) is substituted by
the estimated matrix $\widehat{\bC}$, whose entry $\widehat{C}_{ij}$ is
\beq \widehat{C}_{ij}=\frac 1 K \ds \sum_{k=1}^K
x_i(k)x_j(k),
\eeq where $x_i(k)$ is the observation collected by node $i$, at time $k$,
with $k=1, \ldots, K$.
The practical, relevant, difference is that while the true precision matrix
$\bB$ is sparse by hypothesis, the inverse of $\widehat{\bC}$
in general is not sparse. Also in this case, encouraging
sparsity in estimating the inverse of the covariance matrix
can be beneficial to improve the quality of the estimation itself\footnote{Provided that
the observed field is a Markov field.}.

The problem (\ref{optprob1}) can be reformulated
 as a Difference of Convex  (DC) functions problem
\cite{SB-tao}, by decomposing
the function $z(a_{ij})$ as the difference of two convex functions
$ z( a_{ij} )=g_v(a_{ij})-h( a_{ij})$
with $g_v( a_{ij})=\ds \frac{\mid  a_{ij}
\mid }{\epsilon}$ and \beq
h(a_{ij})=\left\{ \begin{array}{lll}
 0 & \mbox{if} \mid a_{ij}\mid \leq \epsilon\\
 \ds \frac{\mid a_{ij}\mid}{\epsilon}-1 & \mbox{otherwise}.
\end{array}\right.
\eeq
Hence the optimization problem in
 (\ref{optprob1})  can be rewritten as
\beq
\begin{array}{llll}
\underset{\bA \in \, \mbox{S}^N_{++}}{\mbox{min}} & D_{B}(\bA,
\bB)+\rho \tr[(g_v(\bA)-h(\bA))\bD]
 \; .
\end{array}\label{optprob2}
\eeq
To solve this problem, we have used an iterative
procedure, known as DC algorithm (DCA),
based on the duality of DC programming. The usefulness
of using DCA is that its convergence has been proved in \cite{SB-tao}
and it is simple to implement, as it  iteratively solves a convex
optimization problem.
We refer the reader to \cite{SB-SS-DSP11}, \cite{SB-SS-Journ-Match} for further
analytical details.

\subsection{Sparsification and estimation of the precision matrix  }
In this section we illustrate an alternative sparsification strategy that  improves
the estimate of the precision matrix with
respect to the SMP strategy. In this alternative formulation,
the penalty term is the sum of the absolute
values of the entries of $\bA$, weighted with the corresponding per-link
transmit power consumption. In this way, although the power consumption should not be lower
than the SMP method, we expect a sparse topology with a more
accurate estimate of the precision matrix. We call this strategy
Sparse Estimation Strategy (SES). The new problem is formulated as follows
\beq
\begin{array}{llll}
\underset{\bA \in \, \mbox{S}^N_{++}}{\mbox{min}} & D_{B}(\bA, \bB)+\rho \, \mbox{trace}[|\bA| \bD]\\
\end{array}\label{optprobpw} \;
\eeq
and it can be converted  into a
convex definite positive problem. In particular, splitting the matrix $\bA$ into the difference of two nonnegative matrices representing its positive and negative part, i.e.
$\bA=\bA^{+}-\bA^{-}$, we can rewrite (\ref{optprobpw}) as
 \beq
\begin{array}{llll}
\underset{{\bA}^{+},{\bA}^{-}\in \, \mbox{S}^{N}}{\mbox{min}} &  D_{B}(\bA^{+}-\bA^{-}, \bB)+\rho \,\mbox{trace}[(\bA^{+}+\bA^{-}) \bD]\\
s.t. & \bA^{+}-\bA^{-} \succ 0 \\
  & \bA^{+}\geq \mathbf{0} \\
    & \bA^{-}\geq \mathbf{0}\; .
 \end{array}\label{secondoalgortimo}
 \eeq
This problem can be solved using standard numerical tools or  by applying  a projected
gradient algorithm \cite{SB-zdunek}.

\subsection{Numerical results}
\begin{figure}[t]
\centering
\includegraphics[width=10cm]{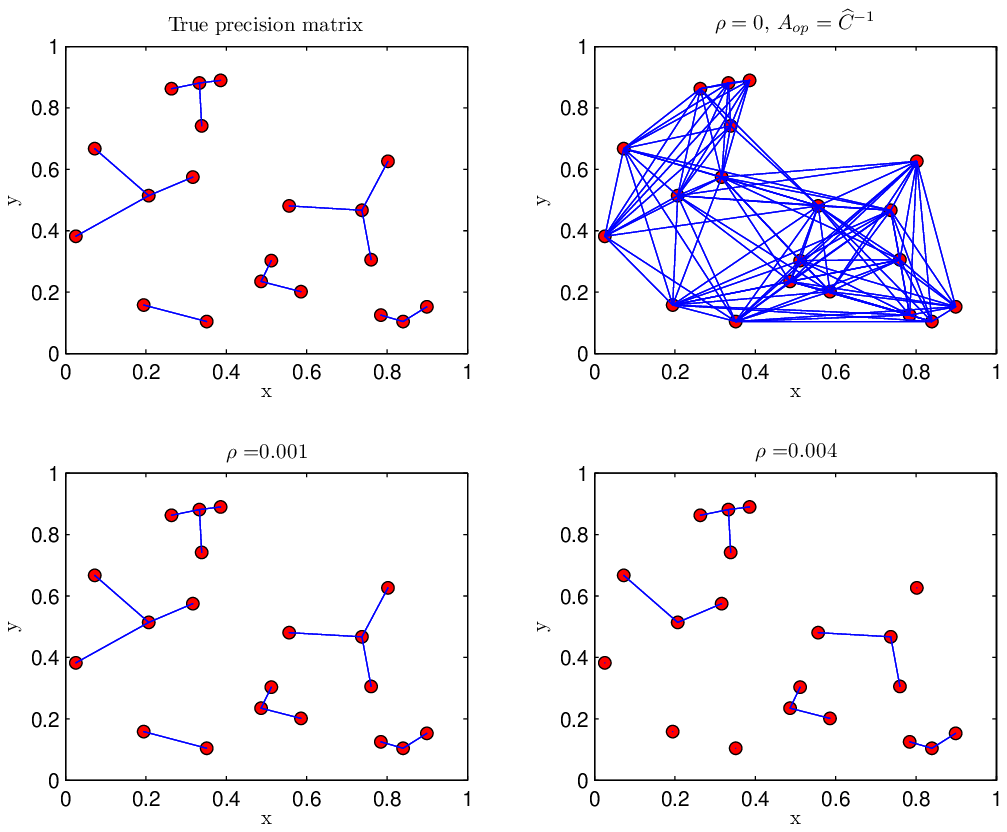}
\caption{Optimal links configurations for the SMP strategy  using
the data estimated covariance matrix $\widehat{\bC}$.}
\label{fig:fig5-estimated}
\end{figure}

In this section we report some simulation results considering a
sensors network composed of $N=20$ nodes, uniformly deployed over a unit
area square and observing correlated data from a GMRF.
We adopt the Markov model proposed in \cite{Anandkumar-Tong-Swami}, where
the correlation between neighboring nodes is a decreasing function of their
 distance and the entries of the covariance matrix can be derived in closed form.
In \cite{SB-SS-DSP11, SB-SS-Journ-Match} we have shown that  even though the problem is not convex,
the numerical results seem to indicate that the
method always converges to the same value of the
precision matrix entries, irrespective of the initializations.
In Fig. \ref{fig:fig5-estimated} we report the final optimal network topology
referring to the case of a matrix
estimated from the data. In particular, the top left plot of Fig. \ref{fig:fig5-estimated} shows the true
dependency graph and all other plots depict the network topologies
obtained using the proposed SMP algorithm, with different penalty coefficients. More specifically,
the network topologies shown in  Fig. \ref{fig:fig5-estimated} are obtained
by thresholding the values of the matrix $\bA_{op}$ obtained through our SMP algorithm, i.e.
the coefficients of matrix $\bA_{op}$ are set to zero if $|a_{op}(i,j)|<10^{-4}$.
In the top right plot
of Fig. \ref{fig:fig5-estimated},
 it can be noted that
the precision matrix achieved with a null penalty can be quite dense because of estimation errors.  Nevertheless, it is interesting to observe that, as the
penalty coefficient increases, the proposed method is not only able to recover the
desired topology, but also to correct most of the errors due to estimation.
We can say that the introduction of the penalty induces a robustness
against estimation errors.
Let us now compare the SMP method with the  SES
strategy, by considering a data estimated covariance matrix
in the divergence term and averaging the simulation results over
$100$ independent realizations of the nodes deployment.

\begin{figure}[h]
\centering
\includegraphics[width=9cm]{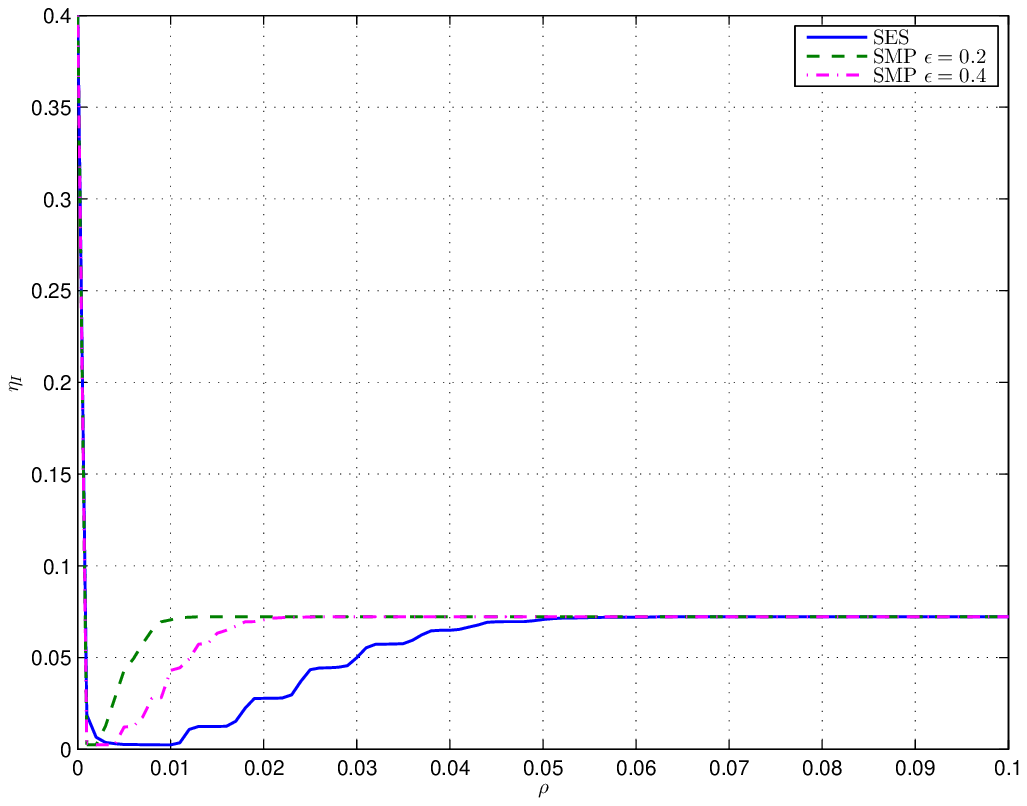}
\caption{Fraction of incorrect links versus $\rho$ for the SMP and
SES strategies.} \label{fig:fig7}
\end{figure}

Let us now evaluate the mismatch between the network topology
and the dependency graph, as a function of the penalty coefficient, obtained using the two proposed strategies.
We assess the mismatch by counting the number of links appearing in the
network topology, which do not appear in the dependency graph.
To this end, Fig. \ref{fig:fig7} shows  the fraction  of
incorrect links $\eta_{I}$, normalized to the total number of
links $N_T=N(N-1)/2$, versus $\rho$. More specifically, considering the true and
optimal precision matrices ($\bA_{t}$ and $\bA_{op}$ respectively), we can define $\eta_{I}=\ds \frac{\sum_{i=1}^N
\sum_{j=1, j> i}^N q_{i j}}{N_T}$ where
 $q_{i j}=1$ if $\delta(a_{t}(i, j))$ and
$\delta(a_{op}(i,j))$ are not equal, assuming $\delta(a_{ij})=1$ if $a_{ij}>0$ and zero otherwise. From Fig. \ref{fig:fig7}, we can deduce that SES
provides more correct links than SMP, as it achieves lower
values of  $\eta_I$.

\section{Conclusions and further developments}
\label{Conclusions}
In this article we have provided a general framework to show how an efficient design
of a wireless sensor networks requires a joint combination of in-network processing and communication.
In particular, we have shown that inferring the structure of the graph describing the statistical
dependencies among the observed data can provide important information on how to build
the sensor network topology and how to design the flow of information through the network.
We have illustrated several  possible network architectures where the global decisions,
either estimation or hypothesis testing, are taken by a central node or in a totally decentralized way.
In particular, various forms of consensus have been shown to be instrumental to achieve globally
optimal performance through local interactions only. Consensus algorithms have then been generalized to more sophisticated signal processing techniques able to provide a cartography of the observed field.
In a decentralized framework, the network topology plays an important role in terms of convergence
time as well as structure of the final consensus value. Considering that most sensor networks
exchange information through a wireless channel, we  have addressed the problem of
finding the network topology that minimizes the energy consumption required to reach consensus.
Finally, we have showed how to match the network topology to the Markov graph describing
the observed variables, under constraints imposed by the power consumption
necessary to establish direct links among the sensor nodes.\\

Even though the field of distributed detection and estimation
has accumulated an enormous amount of research works, there are still many
open problems, both in the theoretical as well as in the application sides. In the
following we make a short list of possible topics of future interest.\\
\begin{enumerate}
\item The general {\it multi-terminal source/channel coding} problem  is
still an open issue. The conventional paradigm established by the source/channel coding
separation  theorem does not hold for the multi-terminal case. This means that
source coding should be studied jointly with channel coding.
\item Distributed decision establishes a {\it strict link between statistical signal processing
and graph theory}. In particular, the network topology plays a fundamental role
in the design of an efficient sensor network. In this article, we have shown some
simple techniques aimed to matching the network topology to the statistical dependency graph
of the observed variables, but significant improvements may be expected from
cross-fertilization of methods from graph theory and statistical signal processing.
\item The design of fully decentralized detection algorithms has already received
important contributions. Nevertheless, there are many open issues concerning
the refinements of the local decision thresholds as a function of both local
observations and the decisions taken from neighbors. In a more general
setting, {\it social learning} is expected to play an important role in future
sensor networks.
\item An efficient design of wireless sensor networks requires a {\it strict
relation between radio resource allocation and decision aspects}, under
physical constraints dictated by energy limitations or channel noise and interference.
Some preliminary results have been achieved in the many-to-one setting,
but the general many-to-many case needs to be thoroughly studied.
\item The application of wireless sensor networks to new fields may be easily expected.
The important remark is that, to improve the efficiency of the network at
various levels, it is necessary to take the application needs strictly
into account in the network design. In other words, a {\it cross-layer design
incorporating all layers} from the application down to the physical layer
is especially required in sensor networks. Clearly, handling the complexity of the
network will require some sort of layering, but this layering will not necessarily
be the same as in telecommunication networks, because the requirements
and constraints in the two fields are completely different.
\end{enumerate}

\section*{Appendix A}
  In this appendix we briefly review  some important notations and basic concepts of graph  theory
  that have been adopted in the previous sections  (for a more detailed introduction to this field see \cite{SB-godsil}).\\

\emph{ \textbf{ A.1 Algebraic graph theory}}\\
Given $N$ nodes let us define a  directed graph or \emph{digraph}
$\mathcal{G} = \{\mathcal{V} , \mathcal{E}\}$  as a set of nodes  $\mathcal{V}=\{v_i\}_{i=1}^{N}$ and a set of edges or links
$\mathcal{E} \subseteq \mathcal{V} \times \mathcal{V}$
where the links $e_{ij} \in \mathcal{E}$ connect the ordered pair of nodes $(v_i,v_j)$, with the convention  that the information flows from
$v_j$ to $v_i$.  In the case where a positive weight $a_{ij}$ is associated
to each edge, the digraph is called \emph{weighted}. Let us assume that there are no loops, i.e. $a_{ii}=0$.\\
The graph is called \emph{undirected} if $e_{ij} \in \mathcal{E} \Leftrightarrow e_{ji} \in \mathcal{E}$.
 The in-degree and out-degree of node $v_i$ are, respectively,
defined as $\mbox{deg}_{in}\triangleq \sum_{j=1}^{N} a_{ij}$ and $\mbox{deg}_{out}\triangleq \sum_{j=1}^{N} a_{j i}$.
In the case of undirected graphs $\mbox{deg}_{in}=\mbox{deg}_{out}$.
Let ${\cal N}_i$
denote the set of neighbors of node $i$, so that $|{\cal N}_i| \triangleq \mbox{deg}_{in}(v_i)$.\\
 The node $v_i$ of a digraph is
said to be \emph{balanced} if and only if its in-degree and out-degree
coincide, while a digraph is
called \emph{balanced} if and only if all its nodes are balanced.\\
We recall now the basic properties of the matrices associated
to a digraph, as they play a fundamental role in the
study of the connectivity of the network associated to the graph. Given a
digraph  $\mathcal{G}$, we introduce the following matrices associated
with  $\mathcal{G}$: 1) The $N \times N$ adjacency matrix $\bA$
whose entries $a_{ij}$  are equal to the weight associated to the edge
$e_{ij}$,  or equal to zero, otherwise; 2) the degree
matrix $\bD$ which is  the diagonal matrix whose diagonal entries are
$d_{ii} =\mbox{deg}_{in}(v_i)=\sum_{j=1}^N a_{ij}$; 3) the weighted Laplacian matrix  $\bL$, defined as
$\bL = \bD-\bA$
whose entries are
 \beq
\ell_{ij} =\left\{
\begin{array}{lll} \mbox{deg}_{in}(v_i) & \mbox{if} & j=i \\
-a_{ij} & \mbox{if} & j \neq i
\end{array}\right. \; . \label{SB-eqL}
\eeq
According to this definition $\bL$ has the following properties: a) its diagonal elements are positive;
b) it has zero row sum; c) it is a diagonally row dominant matrix.
It can be easily verified that $\bL \mathbf{1}=\mathbf{0}$\footnote{ We denote by $\mathbf{1}$ and $\mathbf{0}$ the vectors of all ones or zeros,
respectively.},  i.e.
 zero is an eigenvalue of $\bL$ corresponding to a right eigenvector
$\mathbf{1}$ in the $\mbox{Null}\{\bL\}\supseteq \mbox{span}\{\mathbf{1}\}$,
 and all the other eigenvalues have positive real parts.
 Furthermore a digraph is balanced if and only if $\mathbf{1}$ is
also a left eigenvector of $\bL$ associated with the zero eigenvalue or $\mathbf{1}^T \bL=\mathbf{0}^T$.
Note that for undirected graph the Laplacian matrix is  a symmetric and then balanced matrix   with non negative real eigenvalues.\\
The algebraic multiplicity of the zero eigenvalue of $\bL$ is equal
to the number of connected components contained in $\mathcal{G}$.
For undirected graphs $\mathcal{G}$  is connected
if and only if the algebraic multiplicity of the zero eigenvalue is $1$,
or, equivalently,  $\mbox{rank}(\bL)=N-1$ if and only if $\mathcal{G}$ is connected.
Hence if an undirected graph is connected the eigenvector associated with the zero eigenvalue is $\mathbf{1}$,
and the second smallest eigenvalue of $\bL$, denoted as $\lambda_2(\bL)$ and called algebraic connectivity  \cite{SB-fiedler} of
   $\mathcal{G}$,
 is strictly positive.\\

\emph{\textbf{A.1.1 Forms of connectivity for digraphs}}\\
Before to introduce several forms of graph connectivity \cite{SB-barbarossa_Pesc}
we have to define some useful   concepts.
A
\emph{strong path} of a digraph $\mathcal{G}$ is a sequence of distinct nodes
$v_1, v_2, \ldots, v_p \in \mathcal{V}$ such that $(v_{j-1}, v_j) \in \mathcal{E}$, for $j = 2, \ldots, p$. If
$v_1 \equiv v_p$, the path is said to be closed. A \emph{weak path} is a
sequence of distinct nodes $v_1, v_2, \ldots, v_p \in \mathcal{V}$  such that either
$(v_{j-1}, v_j) \in \mathcal{E}$ or $(v_j, v_{j-1}) \in \mathcal{E}$, for $j = 2, \ldots, p$.
A closed strong path is said a strong \emph{cycle}. A digraph with $N$
 nodes is a
\emph{directed tree} if it has $N-1$ edges and there exists a
node, called the root node, which can reach all the
other nodes through an unique strong path.  As a consequence a directed
tree contains no cycles and every node, except the root, has
one and only one incoming edge. A digraph is a  \emph{forest}
if it consists of one or more directed trees. A subgraph
$\mathcal{G}_s = \{\mathcal{V}_s, \mathcal{E}_s\}$ of a digraph $\mathcal{G}$,
with $\mathcal{V}_s \subseteq \mathcal{V}$ and $\mathcal{E}_s \subseteq \mathcal{E}$, is a
directed spanning tree (or a spanning forest) if it is a directed
tree (or a directed forest) and it has the same node set as $\mathcal{G}$.\\
According to this definition we can define many forms
 of connectivity \cite{SB-barbarossa_Pesc}: a) a
digraph is {\it strongly connected} (SC) if any ordered pair of distinct
nodes can be joined by a strong path; b) a digraph is
{\it quasi strongly connected} (QSC) if, for every ordered pair of
nodes $v_i$ and $v_j$, there exists a node $r$ that can reach both $v_i$
and $v_j$ via a strong path; c) a digraph is {\it weakly connected} (WC)
if any ordered pair of distinct nodes can be joined by a weak
path; d) a digraph is disconnected if it is not weakly connected.
Note that for undirected graphs, the above notions of connectivity are equivalent.
Moreover, it is easy to check
that the quasi strong connectivity of a digraph is equivalent to
the existence of a directed spanning tree in the graph.\\

\emph{\textbf{A.1.2 Connectivity study from the condensation digraph}}\\
When a digraph $\mathcal{G}$ is WC, it may still contain strongly connected subgraphs.
A maximal subgraph of $\mathcal{G}$, which is also SC, is called a
{\it strongly connected component} (SCC) of $\mathcal{G}$ \cite{SB-brualdi}, \cite{SB-barbarossa_Pesc}.
Any digraph $\mathcal{G}$  can be partitioned into SCCs, let us
say $\mathcal{G}_k = \{\mathcal{V}_k, \mathcal{E}_k\}$ where $\mathcal{V}_k \subseteq \mathcal{V}$ and
$\mathcal{E}_k \subseteq \mathcal{E}$ for $k=1,\ldots, r$.
The connectivity properties of a digraph may be better
studied by referring to its corresponding {\it condensation
digraph}. We may reduce the original digraph $\mathcal{G}$ to the condensation
digraph $\mathcal{G}^{*}= \{\mathcal{V}^{*},\mathcal{E}^{*}\}$ by associating the node set $\mathcal{V}_k$ of
each SCC $\mathcal{G}_k$ of $\mathcal{G}$ to a single distinct node $v^{*}_k \in \mathcal{V}^{*}_k $
of $\mathcal{G}^{*}$ and
introducing an edge in $\mathcal{G}^{*}$ from $v_i^{*}$ to $v_j^{*}$, if and only if there
exists some edges from the SCC $\mathcal{G}_i$ and the SCC $\mathcal{G}_j$ of the original
graph. An SCC that is reduced to the root of a
directed spanning tree of the condensation digraph is called
the {\it root SCC} (RSCC). Looking at the condensation graph, we may
identify the following topologies of the original graph: 1) $\mathcal{G}$
is SC if and only if $\mathcal{G}^{*}$ is composed by a single node; 2) $\mathcal{G}$ is
QSC if and only if $\mathcal{G}^{*}$ contains a directed spanning tree; 3) if $\mathcal{G}$
is WC, then $\mathcal{G}^{*}$ contains either a spanning tree or a (weakly)
connected forest.\\

\begin{figure}
\centering
\includegraphics[scale=0.4]{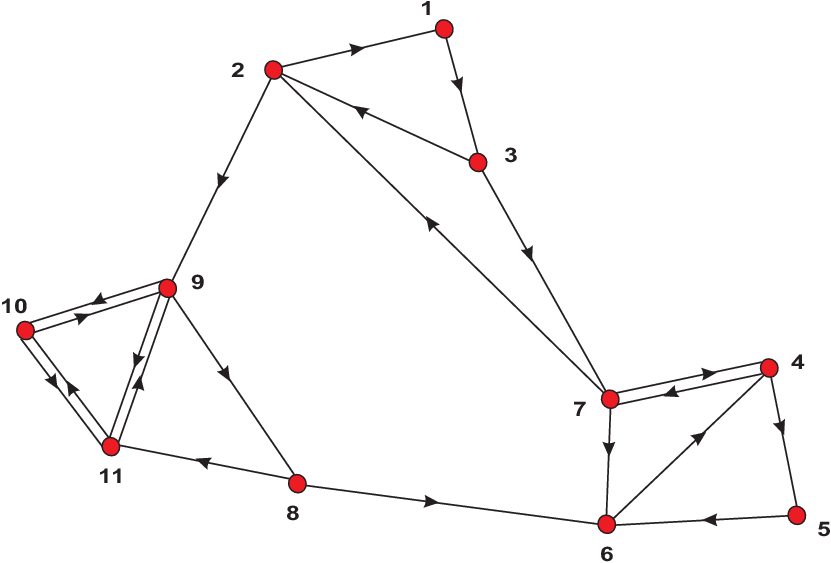}\hspace{1.3cm}
\includegraphics[scale=0.4]{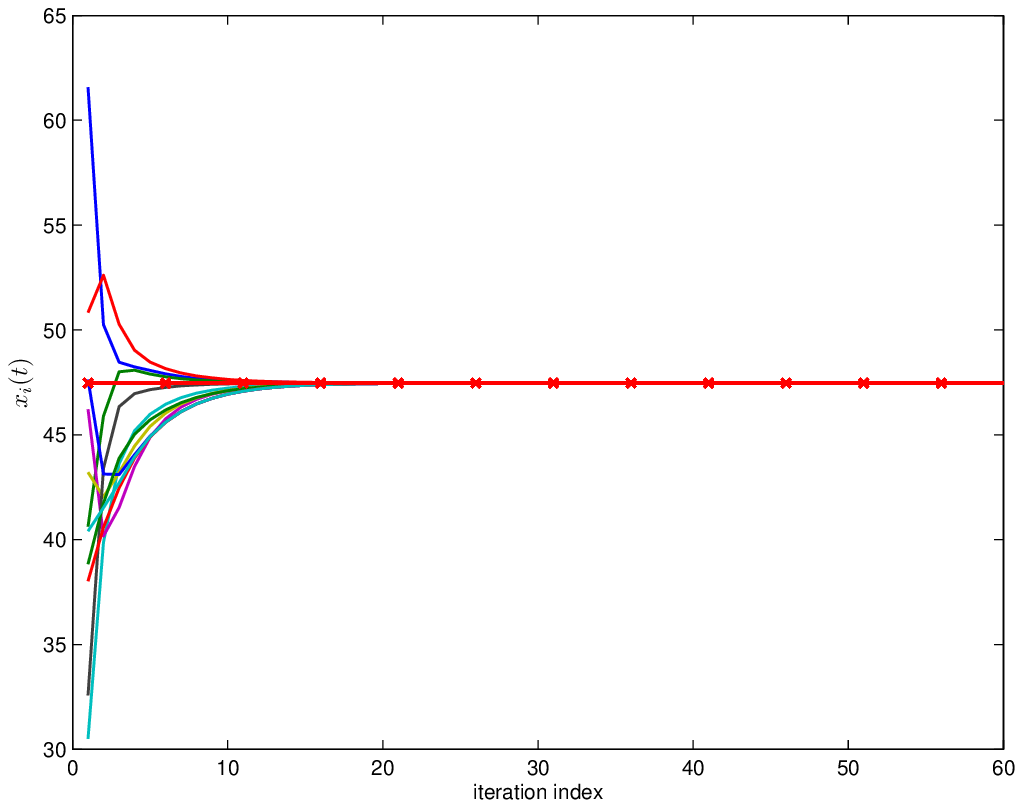}\hspace{1.3cm}\\
  a) \\
  \includegraphics[scale=0.4]{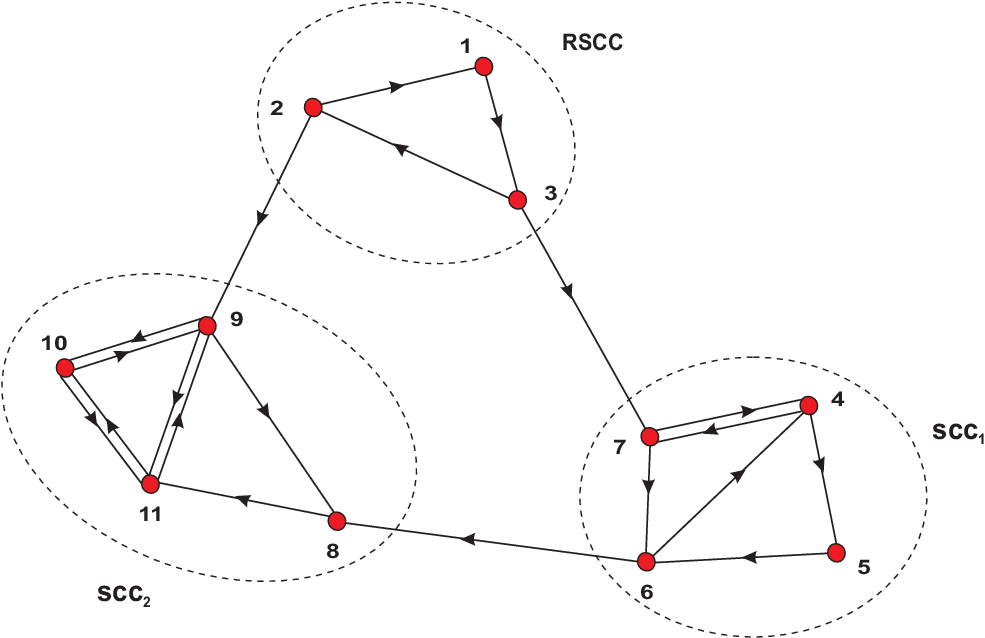}\hspace{1.3cm}
\includegraphics[scale=0.4]{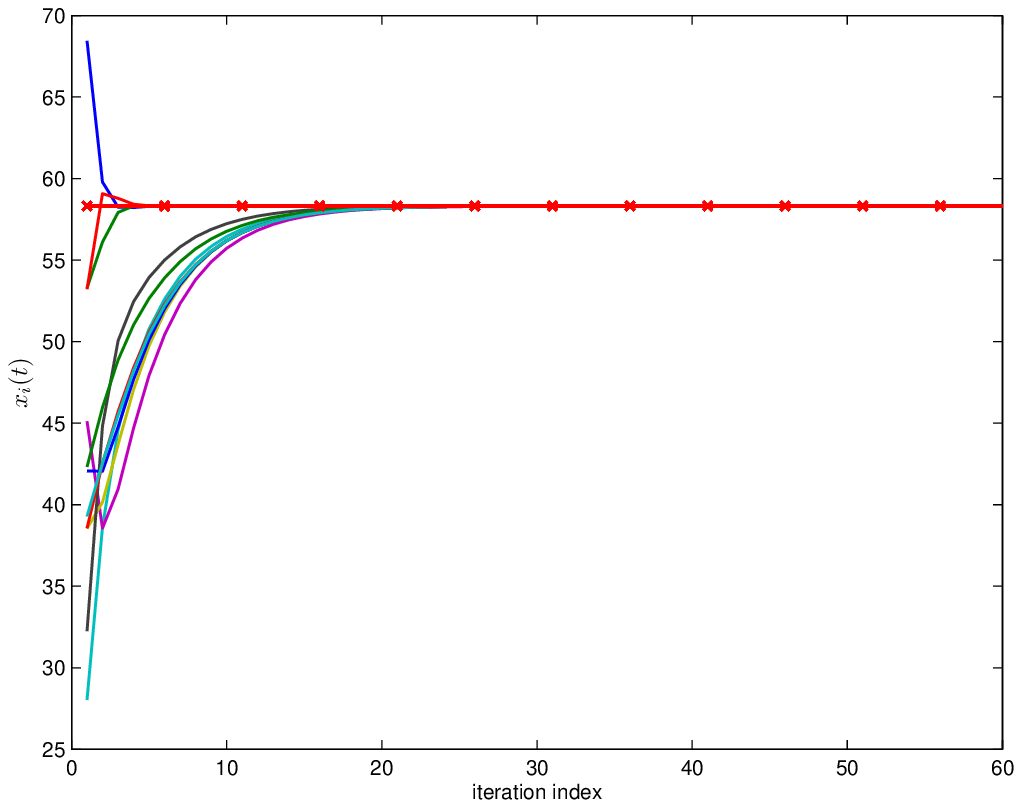}\hspace{1.3cm}\\
  b) \\
  \includegraphics[scale=0.4]{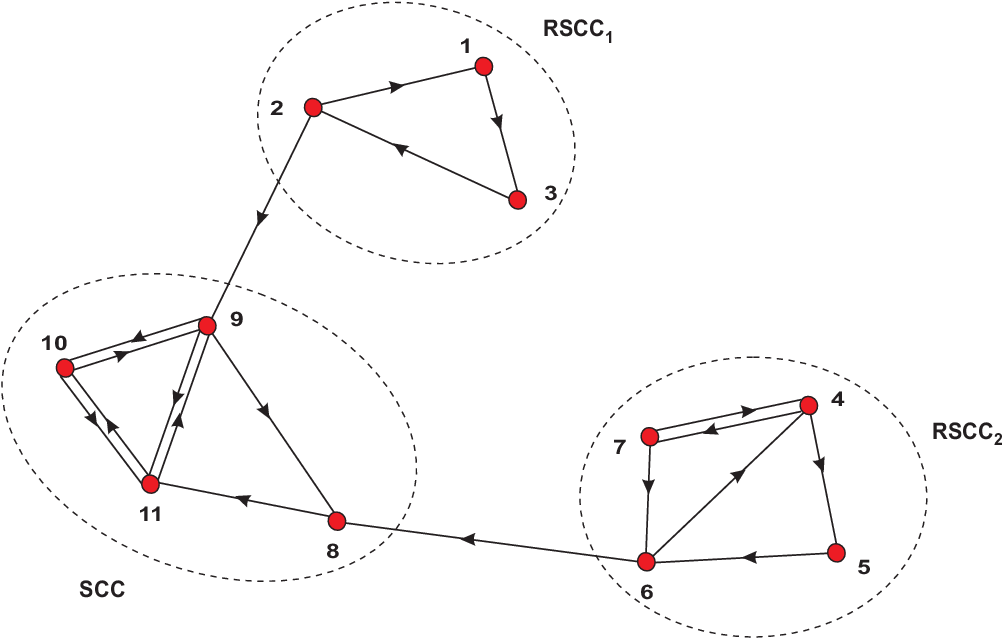}\hspace{1.3cm}
\includegraphics[scale=0.4]{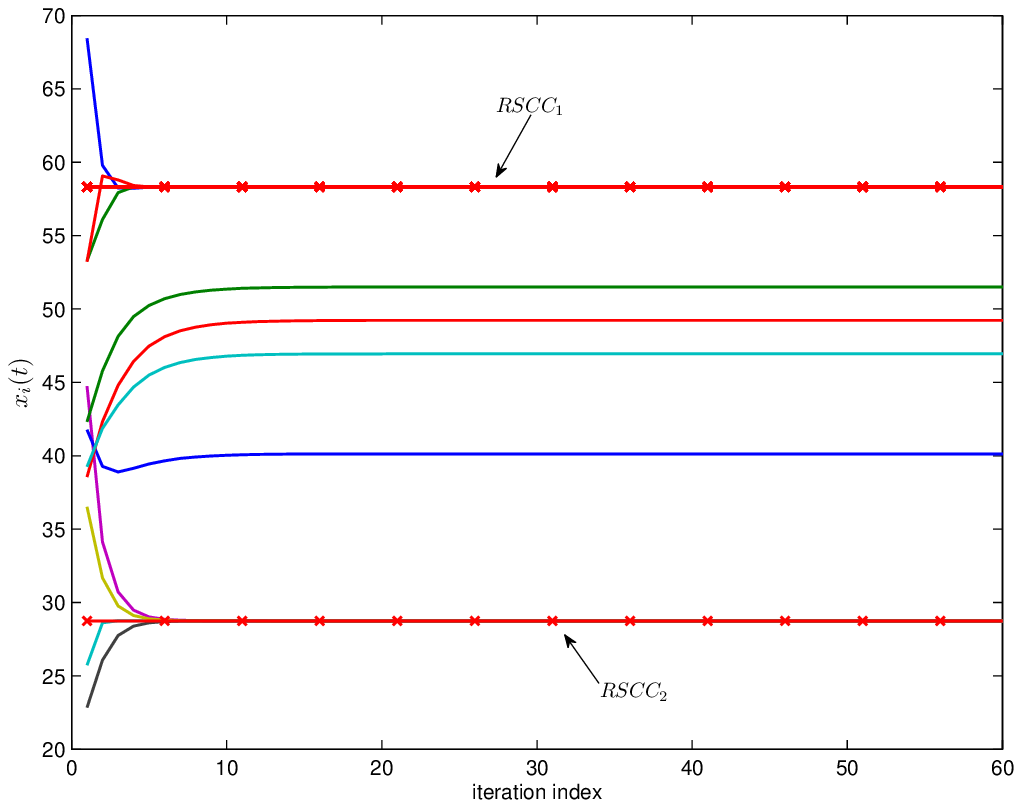}\hspace{1.3cm}\\
  c) \\
  \caption{Consensus for different network topologies: a) SC digraph; b) QSC digraph with three SCCs; c) WC digraph
  with a forest.}
  \label{fig_graph}
\end{figure}

The multiplicity of the zero eigenvalue of $\bL$ is equal
to the minimum number of directed trees contained in a
directed spanning forest of $\mathcal{G}$. Moreover, the zero eigenvalue
of $\bL$ is simple if and only if $\mathcal{G}$ contains a spanning directed
tree or, equivalently, $\mathcal{G}$ is QSC.
If $\mathcal{G}$ is SC then $\bL$ has a simple zero eigenvalue and
positive left-eigenvector associated to the zero eigenvalue.
If $\mathcal{G}$ is QSC with $Q\geq 1$ strongly connected components
$\mathcal{G}_i \triangleq \{\mathcal{V}_i, \mathcal{E}_i \}$ with $\mathcal{V}_i \subseteq \mathcal{V}$,
$\mathcal{E}_i \subseteq \mathcal{E}$ for $i=1,\ldots,Q$, $\mid \mathcal{V}_i \mid  =r_i$ and $\sum_{i} r_i=N$, numbered w.l.o.g.
so that $\mathcal{G}_1$ coincides with the root SCC of $\mathcal{G}$,
then the left-eigenvector $\bgamma=[\gamma_1, \ldots, \gamma_N]^T$
of $\bL$  associated to the zero eigenvalue has entries $\gamma_i>0$ iff $v_i \in \mathcal{V}_1$ and zero
otherwise. If $\mathcal{G}_1$ is balanced then $\bgamma_{r_1}=[\gamma_1, \ldots, \gamma_{r_1}]^T \in \mbox{span}\{\mathbf{1}_{r_1}\}$ where $r_1\triangleq \mid \mathcal{V}_1\mid$.\\
As a numerical example,  in Fig. \ref{fig_graph} we report three network topologies: a) a SC digraph; b) a QSC digraph with three SCCs; c) a WC digraph  with a two-trees forest. We have also depicted for each digraph its decomposition into SCCs corresponding to the nodes
of the associated condensation digraph; RSCC denotes the root SCC.
For each network topology, we have also reported the dynamical evolution of the
consensus algorithm in (\ref{ct-consensus}) versus time.
It can be observed that the dynamical system in Fig. \ref{fig_graph}a) achieves a global consensus
since the underlying digraph is SC. For the QSC digraph in  Fig. \ref{fig_graph}b), instead, there is  a set of nodes in the RSCC
component that is able to reach all  other nodes so that the dynamical system can achieve a global consensus.
Finally, in Fig. \ref{fig_graph}c), the system cannot achieve a global consensus since there is no node that can
reach all the others. Although we can observe two disjoint clusters corresponding
to the two RSCC components, the nodes of the SCC component (middle lines) are affected by the consensus
in the two RSCC components but are not able to influence them.\\

\noindent \emph{ \textbf{ A.2 RGG adjacency matrix}}\\
A random graph is obtained by distributing $N$ points randomly over the
$d$-dimensional space $\mathbb{R}^d$ and connecting the nodes
according to a given rule.
The graph topology is captured by the adjacency matrix $\bA$
which, in this case, is a random matrix.  An important class of random matrices,
is the so called Euclidean Random Matrix (ERM) class, introduced in  \cite{SB-parisi}.
Given a set of $N$ points located at positions $\bx_i, i=1, \ldots, N,$
an $N \times N$ adjacency matrix $\bA$ is an ERM if its generic $(i, j)$
entry depends only on the difference $\bx_i-\bx_j$,
i.e., $a_{i j}=F(\bx_i-\bx_j)$, where $F$  is a measurable mapping from $\mathbb{R}^d$ to
 $\mathbb{R}$.
An important subclass of ERM is given by the adjacency matrices of the
so called Random Geometric Graphs (RGG). In such a case, the entries $a_{ij}$ of the adjacency matrix
are either zero or one depending only on the distance between
nodes $i$ and $j$, i.e.,
\beq
a_{i j}=F(\bx_i-\bx_j)
=\left \{
\begin{array}{llll} 1 & \mbox{if} & \|\bx_i-\bx_j\|\leq r \\ 0 & &
\mbox{otherwise} \end{array} \right. \label{SB-elementsF}\; ,
\eeq
where $r$ is the coverage radius.
Next we discuss some important properties of the spectrum of the adjacency matrix of a random
geometric graph.\\

\emph{ \textbf{ A.2.2 Spectrum of a random geometric graph}}\\
Assuming that the  RGG $G(N, r)$ is connected with high probability, we have derived in \cite{SB-SS-top} an analytical
expression for the algebraic connectivity of the graph, i.e., the second eigenvalue of the symmetric Laplacian,
$\bL = \bD - \bA$, where $\bD$ is the degree matrix and $\bA$
is the adjacency matrix. From (\ref{SB-averagdegree1}), $\bD = \pi r^2 N \bI$, so that
we only need to investigate the second {\it largest} eigenvalue of $\bA$.
Hence, let us start  by studying  the spectrum of $\bA$ as discussed in \cite{SB-SS-top}.
In  \cite{SB-bordenave, SB-rai}, it is shown that the eigenvalues of the
adjacency matrix tend to be concentrated, as the
number of nodes tend to infinity.
In particular, in  \cite{SB-bordenave} it is shown that
the eigenvalues of the normalized adjacency matrix $\bA_N=\bA/N$ of an RGG $G(N, r)$, composed of points
uniformly distributed over a unitary two-dimensional torus, tend to the Fourier series coefficients
of the function $F$ defined in (\ref{SB-elementsF}),
\beq
\hat{F}(\bz)=\int_{\Omega_r} \exp{(- 2 \pi j \bz^T \bx)} d\bx
\eeq
almost surely, for all $\bz=[z_1, z_2] \in \mathbb{Z}^2$,
where $\Omega_r=\{\bx=[x_1,x_2]^T \in \mathbb{R}^2 \; :\; \|\bx\|\leq r\}$.
Using polar coordinates, i.e., $x_1= \rho \sin{\theta}$ and $x_2= \rho \cos{\theta}$,
 with $0\leq \rho \leq r$ and $0\leq \theta\leq 2\pi$ , we obtain
\beq \nonumber
\hat{F}(\bz)=\int_{0}^{r} \int_{0}^{2 \pi} \exp{(- 2 \pi j \rho (z_1 \sin{\theta}+z_2 \cos{\theta}))} \rho d\rho d\theta \; . \label{SB-integr}
\eeq
This integral can be computed in closed form. Setting $z_1=A \sin{\phi}$ and $z_2=A \cos{\phi}$, we have
\beq \nonumber
\hat{F}(A,\phi)=\int_{0}^{r} \int_{-\phi}^{2 \pi-\phi} \exp{(- 2 \pi j \rho A \cos(\xi))} \rho d\rho d\xi \label{SB-integr1}
\eeq
with $\xi=\theta-\phi$.
Furthermore, using the integral expression for the  Bessel function of the first kind of order $k$,
$J_{k}(x)= \ds \frac{1}{2 \pi} \int_{-\pi}^{\pi} \exp{(j x \sin(\xi) -
j k\xi)} d\xi$,
we get
\beq \nonumber
\hat{F}(A,\phi)= \hat{F}(A) = 2 \pi \int_{0}^{r}  J_{0}(2 \pi \rho A) \rho d\rho  \; .
\eeq
Finally, using the identity $\int_{0}^{u} v J_{0}(v) dv= u J_{1}(u)$,
we can make explicit the dependence of $\hat{F}(A)$ on the index pair $[z_1, z_2]$
\beq
\hat{F}(z_1, z_2)=\ds \frac{r}{\sqrt{z_1^2+z_2^2}}\, J_{1}\left(2 \pi r \sqrt{z_1^2+z_2^2}\right) \; .
\eeq
This formula allows us to rank the eigenvalues of $\bA_N=\bA/N$. In particular,
we are interested in the second largest eigenvalue of $\bA_N$.
Considering that the minimum coverage radius ensuring connectivity behaves as
$r(N) \sim \sqrt{ \frac{\log(N)}{N}}$, i.e., it is a vanishing function of $N$, we can use the
Taylor series expansion of $\hat{F}(z_1, z_2)$, for small $r$. Recalling that, for small $x$,
$J_1(x)= x/2-x^3/16+o(x^5)$, we can approximate the eigenvalues as
\begin{equation}
\hat{F}(z_1, z_2)= \pi r^2-\frac{\pi^3 (z_1^2+ z_2^2)\, r^4}{2}+o(r^6).
\end{equation}
This expansion shows that, at least for small $r$, the largest eigenvalue
equals $ \pi r^2$ and occurs at $z_1=z_2=0$,
whereas the second largest eigenvalue corresponds to the cases $(z_1=1, z_2=0)$ and
$(z_1=0, z_2=1)$.
More generally, we can check numerically that,
for $r\le 1/2$ and $A\geq 1$,  the following inequalities hold true:
\beq
\pi r^2 \geq  r J_1(2 \pi r) \geq \ds \frac{r}{A} \left| J_1(2 \pi r A)\right| \label{SB-inequalities} \;.
\eeq

In summary, denoting the spectral radius of $\bA_N$ as
$\zeta_1(\bA_N)=\ds \max_{1\leq i \leq N} \ds \frac{ \mid \lambda_i(N)\mid}{N}$, where $\{\lambda_i(N)\}_{i=1}^{N}$
is the set of eigenvalues of $\bA$, it follows that
\beq
\ds \lim_{N\rightarrow \infty} \zeta_1(\bA_N)=\ds \max_{\bz \in \mathbb{Z}^2 } \mid \hat{F}(\bz)\mid=\hat{F}(0, 0)=\pi r^2 \;, \label{SB-radspec}
\eeq
while the second largest eigenvalue of $\bA_N$, $\zeta_2(\bA_N)$, converges
 to
\beq
 \lim_{N\rightarrow \infty}\zeta_2(\bA_N)= \hat{F}(1, 0)= \hat{F}(0, 1)= r J_1(2\pi r)\; . \label{SB-secondlar}
\eeq

We are now able to derive the asymptotic expression for the second largest eigenvalue of
the normalized Laplacian $\bL_N=\bD_N-\bA_N$, where $\bD_N:=\bD/N$ is the normalized
degree matrix. Because of the asymptotic property of the degree of an RGG, shown in
(\ref{SB-averagdegree1}),  the second largest eigenvalue of $\bL_N$ tends asymptotically to
 \beq
 \lambda_2(\bL_N)=\pi r^2 - \zeta_2(\bA_N)  \label{SB-algconnln} \; .
 \eeq
Thus, the algebraic connectivity of the graph can be approximated, asymptotically, as
\beq
 \lambda_2(\bL)=\pi N r^2  - N r J_1(2 \pi r) \label{SB-algconnln1} \; .
\eeq

\end{document}